\documentclass[11pt]{article}
\pdfoutput=1
\usepackage{jcapmod}
\usepackage{dsdshorthand}

\usepackage{mathtools}
\usepackage{booktabs}
\usepackage[english]{babel}
\usepackage{amsmath,amssymb,amsbsy,amstext, amsthm, simplewick, amsfonts}
\usepackage[hang, flushmargin]{footmisc}
\usepackage{graphicx}
\usepackage[small]{caption}
\usepackage{siunitx}
\usepackage{upgreek}
\usepackage{framed}
\usepackage{wrapfig}
\usepackage{multirow}
\usepackage{bbm}
\usepackage[svgnames,dvipsnames,x11names]{xcolor}
\usepackage{selinput}
\usepackage{bm}
\usepackage{float}
\usepackage{geometry}
\usepackage{yfonts}
\usepackage{caption}
\usepackage{subcaption}
\usepackage{sidecap}
\usepackage{longtable}
\usepackage{anyfontsize}
\usepackage{tcolorbox}
\usepackage{bbm}
\usepackage{placeins}
\usepackage{tikz-cd}
\usepackage{setspace}

\makeatletter
\def\leftrightharpoon{%
  \@ifnextchar[{\@lrharp}{\@lrharp[]}%
}
\def\@lrharp[#1]{%
  \arrow[#1, leftharpoonup  ,yshift= 0.225ex]%
  \arrow[#1,rightharpoondown,yshift=-0.225ex]%
}
\makeatother

\usepackage{colortbl}
\definecolor{lightgreen}{cmyk}{0.2, 0, 0.2, 0.2}
\definecolor{lightgray}{cmyk}{0.1,0.2,0,0.1}
\definecolor{lightgray2}{cmyk}{0.1,0.1,0,0.1}
\definecolor{greyish}{rgb}{.90,.90,.90}
\definecolor{greyish2}{rgb}{.96,.96,.96}

\definecolor{lightgray}{gray}{0.9}

\usepackage[framemethod=TikZ]{mdframed}
\mdfdefinestyle{boxed}{%
linecolor=red,
roundcorner=5pt,
innerleftmargin=0,%
innerrightmargin=0,
backgroundcolor=gray!10,%
}

\usepackage{tikz}
\tikzset{every picture/.style={line width=0.75pt}} 
\usepackage{tikz-cd, tikz-feynman} 
\tikzfeynmanset{warn luatex=false}

\def\beq{\begin{equation}}
\def\eeq{\end{equation}}
\def\bea{\begin{eqnarray}}
\def\eea{\end{eqnarray}}

\newcommand{\bfx}{\mathbf{x}}

\newcommand{\bfk}{\mathbf{k}}
\newcommand{\bfp}{\mathbf{p}}
\newcommand{\bfq}{\mathbf{q}}
\newcommand{\bfK}{\mathbf{K}}

\newcommand{\col}[2]{ \begin{array}{c} #1 \\ #2  \end{array} }

\newcommand{\SBD}{\mathcal{S}} 
\newcommand{\m}{\mu}  
\renewcommand{\G}{\mathcal{G}} 

\begin{document}

\title{A de Sitter $S$-matrix from amputated cosmological correlators}

\author[a]{Scott Melville}
\author[b]{and Guilherme L. Pimentel}
\affiliation[a]{Queen Mary University of London, Mile End Road, London, E1 4NS, U.K.}
\affiliation[b]{Scuola Normale Superiore and INFN, Piazza dei Cavalieri 7, Pisa, 56126, Italy}


\abstract{
Extending scattering to states with unphysical mass values (particles ``off their mass shell'') has been instrumental in developing modern amplitude technology for Minkowski spacetime. 
Here, we study the off-shell correlators which underpin the recently proposed $S$-matrix for scattering on de Sitter spacetime. 
By labelling each particle with both a spatial momentum and an independent ``energy'' variable (the de Sitter analogue of a 4-momentum), we find that the practical computation of these correlators is greatly simplified. 
This allows us to derive compact expressions for all 3- and 4-particle $S$-matrices at tree-level for scalar fields coupled through any derivative interactions. 
As on Minkowski, we find that the 3-particle and exchange part of the 4-particle $S$-matrices are unique (up to crossing).
The remaining contact part of the 4-particle $S$-matrix is an analytic function of just two differential operators, which become the usual Mandelstam variables in the Minkowski limit. 
Finally, we introduce a spectral decomposition for the tree-level exchange of a heavy field responsible for a cosmological collider signal. 
Once projected onto physical mass eigenstates, these $S$-matrix elements encode the statistical properties of the early inflationary perturbations. 
}

\maketitle

\section{Introduction}

The $S$-matrix has played a central role in the development of quantum field theory \cite{ChewBook,Eden:1966dnq,Martin:1969book}. 
It is how we connect theories, such as the Standard Model Lagrangian, to phenomenological predictions for observables, such as cross-sections. 
Since $S$-matrix elements are free from the gauge and other redundancies that affect the Lagrangian or Hamiltonian, they are the natural objects to compare with data or to compare when discriminating between different models.
Most effective field theories are built on the idea that their effective parameters can be fixed by matching their low-energy $S$-matrix elements onto the $S$-matrix of the underlying high-energy theory. 
Fundamental physical principles, such as causality and unitarity, are encoded in the $S$-matrix as concrete mathematical properties, such as analyticity and positivity. These properties underpin many modern amplitude techniques (see e.g.~\cite{Cutkosky:1960sp,Bern:1994zx,Bern:1994cg} and \cite{Elvang:2013cua, Bern:2022jnl, Buonanno:2022pgc} for modern reviews) and can be used to  bound the space of consistent quantum field theories (see e.g. \cite{Adams:2006sv, Manohar:2008tc, Komargodski:2011vj, Luty:2012ww} and \cite{Kruczenski:2022lot, deRham:2022hpx} for modern reviews).   \\

Historically, it was data from colliders that drove much of the early progress in establishing $S$-matrix theory. 
Today, the cosmological collider programme aims to use the inflationary expansion of the early Universe to probe the new degrees of freedom that become important at very high energy / high spacetime curvature \cite{Chen:2009zp,Arkani-Hamed:2015bza}.
Due to the strong gravitational fields, quanta of these new fields can be excited from an initial vacuum state and this particle production leaves a characteristic correlation in fluctuations of the Cosmic Microwave Background and the large scale structure of galaxies---signals that could be within the reach of upcoming sky surveys; for reviews see \cite{Meerburg:2019qqi,Achucarro:2022qrl}, and for a recent analysis see \cite{Cabass:2024wob}.
By now there are numerous studies of the cosmological correlators produced by inflation (see \cite{Baumann:2022jpr, Benincasa:2022gtd} for modern reviews). 
The earliest focus was on computing in-in correlation functions, which are the field-theoretic objects most directly connected to what we observe in the CMB. 
More recently, the wavefunction of the Universe has emerged as an economical way to capture these observables, as it has many nice properties that are less manifest at the correlator level.       
However, despite many interesting connections to flat-space scattering amplitudes \cite{Bonifacio:2021azc,Lee:2023kno, Chowdhury:2023arc}, the wavefunction leaves much to be desired. 
For instance, the cutting rules which encode perturbative unitarity are more involved than on Minkowski \cite{Goodhew:2020hob,Cespedes:2020xqq,Melville:2021lst,Goodhew:2021oqg,Baumann:2021fxj,Jazayeri:2021fvk,Albayrak:2023hie,Agui-Salcedo:2023wlq}, the spectral density is not guaranteed to be positive \cite{Salcedo:2022aal}, and there is no manifest double copy structure \cite{Armstrong:2023phb} (but see \cite{Lee:2022fgr} for progress in this direction). In most formulations, the wavefunction coefficients themselves are not invariant under field redefinitions and are sensitive to total derivative (boundary) terms in the action.  \\

Recently, in an effort to describe cosmological collider phenomenology in a language that more closely parallels flat space amplitudes, we introduced an $S$-matrix for the inflationary patch of de Sitter spacetime \cite{Melville:2023kgd}. This object is naturally insensitive to field redefinitions and total derivatives, since it is defined using particle eigenstates rather than field eigenstates. Once a particular field basis is specified, the wavefunction and cosmological correlators can be constructed from this $S$-matrix, so it encodes all of the physically relevant information about the dynamics.  \\

One challenge, however, is to explicitly compute these $S$-matrix elements. 
Like its wavefunction counterpart, the naive construction in terms of $(\tau, \bfk)$ co-ordinates (conformal time and spatial momentum) leads to nested integrals over products of Hankel functions: frightening special functions whose analytic structure is often far from clear. Rather than an inevitable feature of de Sitter, we would describe this as an inevitable feature of the co-ordinates. A similar situation would be encountered on Minkowski spacetime if one attempted to compute amplitudes in position space by performing time-integrals over products of Feynman propagators (which are also Hankel functions). Of course, on Minkowski we know the resolution---rather than work in position space, the propagator and corresponding scattering amplitudes become remarkably simple once expressed in terms of covariant momenta $p_\mu$. Crucially, the original position space correlator corresponds to an integral over all values of the energy $p_0$, despite the fact that on-shell particles propagate with $p_0 = \sqrt{ |\bfp|^2 + m^2}$. The extension to off-shell values of the energy is central to how we simplify and analyse Minkowski amplitudes. \\

In this work, we extend the de Sitter $S$-matrix to off-shell kinematics and identify a particularly convenient set of ``energy'' variables which mirror the 4-momenta of Minkowski spacetime.
This allows us to explicitly compute a much larger class of Feynman diagrams, some of which enjoy a surprisingly simple analytic structure.
Since all equal-time correlation functions can be extracted from the de Sitter $S$-matrix, these technical advances provide another tool with which to analyse inflationary correlation functions.  \\

For example, suppose we have a simple interaction like $(\nabla \phi )^2 \sigma$, where $\sigma$ is a heavy scalar field with mass $m^2_\sigma = \mu_\sigma^2 + \frac{d^2}{4}$, and we want to answer a question such as: 
\begin{quotation}
\noindent \emph{Find the $S$-matrix for $2 \to 2$ scattering mediated by $\sigma$, and show that it has a positive real part in the forward limit.}
\end{quotation}
On Minkowski, the answer is straightforward: focussing on the $s$-channel contribution\footnote{
We omit the energy and momentum conserving $\delta$ functions. 
},
\begin{align}
i S_{12 \to 34}  =  \frac{ p_1 \cdot p_2  \, \times \,  p_3 \cdot p_4 }{ m^2_\sigma - s - i \epsilon }  \qquad\qquad \text{(Minkowski)}
\end{align}
we see indeed that the discontinuity 
\begin{align}
 S_{12 \to 34} + S_{34 \to 12}^*  =  \left( p_1 \cdot p_2  \times p_3 \cdot p_4 \right) \, 2 \pi  \, \delta \left(  m_\sigma^2 - s  \right)  \qquad\qquad \text{(Minkowski)}
\end{align}
implies a positive imaginary part in the forward limit ($p_4 = p_2$ and $p_3 = p_1$), as required by unitarity\footnote{
On Minkowski we often define the scattering amplitude $A$ via $S = i A  \, (2 \pi)^{d+1} \delta^{d+1} ( p_{\rm total} )$, in which case unitarity implies that the imaginary part of $A$ is positive. 
}.
The goal of this paper is to develop the technology needed to write down the analogous answers in de Sitter just as simply.
In the above example, our results lead to
\begin{align}
 \SBD_{12 \to 34} = \int_{\nu} \,  \frac{ \left[ \hat{p}_1 \cdot \hat{p}_2 \, F_{j_2} \right]  \times \left[ \hat{p}_3 \cdot \hat{p}_4 \, F_{j_2} \right]  }{ ( \mu_\sigma^2 - \nu^2 )_{i \epsilon} }\,, 
 \qquad\qquad \text{(de Sitter)}
\end{align}
where $\hat{p}_b^\mu$ is a known differential operator and $F_{j_2}$ is the two-particle mode function analogous to $e^{i (p_1 + p_2 ) \cdot x}$ (both of which are uniquely fixed by the de Sitter isometries). $\int_\nu$ represents an integral over all mass values in the principal series and the $( \mu_\sigma^2 - \nu^2 )_{i \epsilon}$ is a particular combination of $(\mu_\sigma^2 - \nu^2 \pm i \epsilon)$ that accounts for particle production. 
The discontinuity is then 
\begin{align}
 \SBD_{12 \to 34} + \SBD_{34 \to 12}^*=   \SBD_{12 \to \mu_\sigma}  \; \SBD_{34 \to \mu_\sigma}^*  +  \SBD_{12 \mu_\sigma \to 0} \; \SBD_{34 \mu_\sigma \to 0}^*\, ,
\end{align}
where $\SBD_{12 \to \mu_\sigma} = \left[ \hat{p}_1 \cdot \hat{p}_2 \, F_{j_2} \right]$, and we find it is also manifestly positive in the forward limit. 
Since derivative interactions like this one are expected to dominate the cosmological collider signal from inflation, and since the $S$-matrix and its positivity have played such an important role in how we analyse signals from terrestrial colliders, we see this machinery as an important theoretical step towards understanding the early Universe.  \\

We end this introduction with a short summary of our main results and conventions. 
In the following section~\ref{sec:observables}, we review the $S$-matrix proposal in \cite{Melville:2023kgd} and provide an LSZ reduction formula from time-ordered correlators in both the time and the energy domains. 
Then through sections~\ref{sec:mandelstam} and~\ref{sec:spectral} we systematically describe all tree-level contact and exchange diagram contributions to this $S$-matrix.   
We conclude in section~\ref{sec:disc} with some discussion of future directions and provide an appendix containing useful identities for our mode functions.

\subsection{Executive summary}

For perturbative interactions between quantum fields on a fixed Minkowski background, there is a clear and well-established path: we compute in-out $S$-matrix elements (i.e. scattering amplitudes) as a function of the energy and momenta of each particle (i.e. the covariant $p_\mu$). 
These objects are phenomenologically useful because they are closely related to the observable cross section, and yet mathematically tractable since the principles of causality/analyticity and unitarity ensure that they are ``simple'' functions of the Mandelstam variables.

When describing interacting quantum fields on a curved spacetime background, the best way forward is far less clear. Asymptotic scattering states are more difficult to define (as the vacuum is no longer unique) and Mandelstam variables are no longer sufficient to capture the complete kinematics (as energy is no longer conserved). One must therefore decide
\begin{itemize}

\item[(i)] which object to consider?

\item[(ii)] which kinematic variables to use? 

\end{itemize}
In the context of cosmological collider physics, possible answers to (i) include correlations functions, wavefunction coefficients or other notions of scattering amplitudes on de Sitter. 
Possible answers to (ii) include labelling each field by its bulk position $(\tau, \bfx)$, or by its conformal time and spatial momentum $(\tau, \bfk )$, or by angular momentum quantum numbers adapted to the de Sitter isometry group. Each option has advantages and disadvantages: some properties which are manifest in one object with certain variables may appear very non-trivial in another object or variables\footnote{
We see this already to some extent on Minkowski, where causality corresponds to branch cuts in position space with a clear physical meaning (reflecting the ordering ambiguity of time-like separated operators), but this translated into an analytic structure in momentum space which can be very subtle, particularly beyond $4$-particle scattering. 
}.

In this work, we have chosen to

\begin{itemize}

\item[(i)] focus on \emph{amputated} time-ordered correlators, which become $S$-matrix elements once we put the external particles on-shell. These are analogous to \emph{off-shell} scattering amplitudes on Minkowski, where the $p_\mu$ are no longer constrained by the condition $p^2 = - m^2$. 

\item[(ii)] label each field by a \emph{conformal energy} $\omega$ and spatial momentum $\bfk$, which are the Fourier conjugates to $\tau$ and $\bfx$. In the flat-space limit, $(\omega, \bfk)$ reduces to $p_\mu$.  

\end{itemize}
This choice seems to occupy the same sweet spot between being observationally useful and being mathematically tractable. 
The on-shell limit of these correlators, in which $\omega$ is fixed in a mass-dependent way, determines the in-in correlators that we ultimately measure in primordial non-Gaussianity. 
And yet, they have a simple analytic structure, crossing 
and perturbative Feynman rules that allow them to be computed and studied with relatively few special functions.  

%
%
%
%
%

\paragraph{Main results.}
We derive a number of interesting features enjoyed by the amputated correlator (off-shell $S$-matrix) when written in terms of the energy and momenta $(\omega, \bfk)$ of each field, including: 

\begin{itemize} 

\item \emph{Contact seed solutions}. 
A quartic $\lambda \phi^4$ interaction gives a simple contact-diagram contribution to the $4$-point amputated correlator,
\begin{align}
    \begin{tikzpicture}[scale=0.6,baseline=0.0cm]
			\draw (-2,1) -- (-1,0);
			\draw (-2,-1) -- (-1,0);
			\draw (0,1) -- (-1,0);
			\draw (0,-1) -- (-1,0);
                \draw (-2,1) node [anchor=east][inner sep=2.0pt] {$\omega_1$};
                \draw (-2,-1) node [anchor=east][inner sep=2.0pt] {$\omega_2$};
                \draw (0,1) node [anchor=west][inner sep=2.0pt] {$\omega_3$};
			\draw (0,-1) node [anchor=west][inner sep=2.0pt] {$\omega_4$};
             \draw (-1,-0.1) node [anchor=north][inner sep=3.0pt] {\small \color{gray} $\phi^4$};
        \end{tikzpicture} 
        =
       \frac{ i \lambda \, \Gamma (d)}{ ( i \omega_1 + i \omega_2 + i \omega_3 + i \omega_4 )^d }  \equiv   \lambda \;  \G_4^{\rm con} 
\label{eqn:intro_G4_seed}
\end{align}
This basic structure, $\G_4^{\rm con}$ is the analogue of the energy conserving $\delta$-function in flat space.  
Since time translations have been replaced by dilation invariance, the isometries now require that the total energy appears with a fixed power (to ensure the correlator has the correct scaling dimension) instead of requiring energy conservation.

\item \emph{Covariant momenta}.
Using the de Sitter isometry generators, we construct a differential operator $\hat{p}_\mu$ which corresponds to the Fourier transform of the covariant derivative $\nabla_\mu$. The Feynman rules for derivative interactions are then simple: for instance the interaction $\phi_1 \phi_2 \nabla_\mu \phi_3 \nabla^\mu \phi_4$ produces an amputated correlator of $- \hat{p}_3 \cdot \hat{p}_4 \, \G_4^{\rm con}$. 
These momenta obey the useful relations that $\hat{p}^2$ is fixed in terms of the particle's mass when on-shell, and the total momentum operator $\sum_{b=1}^4 \hat{p}_b$ annihilates $\G_4^{\rm con}$.

\item \emph{All contact diagrams}.
From this seed solution and covariant momenta, we can express \emph{any} contact-diagram (i.e. the contribution of any quartic interaction with arbitrarily many derivatives) as
\begin{align}
    \begin{tikzpicture}[scale=0.6,baseline=0.0cm]
			\draw (-2,1) -- (-1,0);
			\draw (-2,-1) -- (-1,0);
			\draw (0,1) -- (-1,0);
			\draw (0,-1) -- (-1,0);
                \draw (-2,1) node [anchor=east][inner sep=2.0pt] {$\omega_1$};
                \draw (-2,-1) node [anchor=east][inner sep=2.0pt] {$\omega_2$};
                \draw (0,1) node [anchor=west][inner sep=2.0pt] {$\omega_3$};
			\draw (0,-1) node [anchor=west][inner sep=2.0pt] {$\omega_4$};
        \end{tikzpicture} 
        =
      \sum_{a, b = 0} c_{ab} \; \hat{t}^b \; \hat{s}^a  \;  \G^{\rm con}_4 \; , 
\end{align}
up to terms that vanish on-shell and do not contribute to any $S$-matrix element. $\hat{s} = - ( \hat{p}_1 + \hat{p}_2 )^2$ and $\hat{t} = - ( \hat{p}_1 + \hat{p}_3 )^2$ are differential operators that are given explicitly in \eqref{eqn:shat_def_3} and which reduce to the usual Mandelstam $s$ and $t$ in the flat-space limit. 
For example, the interaction $\phi_1 \phi_2 \nabla^2  \phi_3 \phi_4$ gives
\begin{align}
\hat{s} \;  \G^{\rm con}_4 = \left( -\omega_{12} \omega_{34} - k^2_s  \right) \frac{ 
 i  \, \Gamma (d+2 ) }{ ( i \omega_T )^{d+2} }
\end{align}
where $\omega_{ij} = \omega_i + \omega_j$ and $k_{s} = | \bfk_1 + \bfk_2 | = | \bfk_3 + \bfk_4 |$. The total energy $\omega_T = \omega_{12} + \omega_{34} \to 0$ in the flat-space limit, which produces the usual $s = \omega^2_s - k_s^2$ multiplying a divergence that plays the role of the $2 \pi \delta ( \omega_T  )$ in the Minkowski $S$-matrix.

\item \emph{Multi-particle mode functions.}  
$\hat{s}$ and $\hat{t}$ do not commute, so it is not possible to diagonalise both simultaneously. 
However, interactions that depend only on $\hat{s}$ can be expressed very simply using the eigenfunctions of this operator,
\begin{align}
  \hat{s}^a \,   \G^{\rm con}_4 = \frac{ 1 }{ ( i k_s )^d } \int_\nu \; \left(  \nu^{2} + \tfrac{d^2}{4} \right)^a \; F_{j_2} \left(  \frac{ \omega_{12} }{  k_{s} }  , \nu \right) F_{j_2} \left( \frac{ \omega_{34} }{  k_{s} }  , \nu \right) 
  \label{eqn:intro_saG4}
\end{align} 
where $\int_\nu = \tfrac{2 i}{\pi} \int_0^{\infty} d \nu \, \nu \sinh( \pi \nu) $ is an integral over all mass values in the principal series (weighted by the appropriate de Sitter density of states). 
$F_{j_2}$ has the interpretation of a \emph{two-particle} mode function and is the analogue of $e^{ i (p_1 + p_2 )_\mu x^\mu}$ on Minkowski. It is given explicitly in \eqref{eqn:Fmode} in terms of Legendre functions. 
\eqref{eqn:intro_saG4} also makes manifest the fact that $ \left( \hat{p}_1 + \hat{p}_2 \right)^2 =  \left( \hat{p}_3 + \hat{p}_4 \right)^2 $ when acting on $\G^{\rm con}_4$, since it is symmetric in $12 \leftrightarrow 34$.

\item \emph{Exchange seed solutions.} 
A cubic $g \phi^3$ interaction gives the following exchange-diagram contribution to the $4$-point amputated correlator,
\begin{align}
    \begin{tikzpicture}[scale=0.6,baseline=0.0cm]
			\draw (-4,1) -- (-3,0);
			\draw (-4,-1) -- (-3,0);
			\draw (-3,0) -- (-1,0);
			\draw (0,1) -- (-1,0);
			\draw (0,-1) -- (-1,0);
                \draw (-4,1) node [anchor=east][inner sep=2.0pt] {$\omega_1$};
                \draw (-4,-1) node [anchor=east][inner sep=2.0pt] {$\omega_2$};
                \draw (-2,0) node [anchor=south][inner sep=2.0pt] {$\mu$};
                \draw (-1.1,-0.1) node [anchor=north][inner sep=2.0pt] {\small \color{gray} $\phi^3$};
                \draw (-2.9,-0.1) node [anchor=north][inner sep=2.0pt] {\small \color{gray} $\phi^3$};
                \draw (0,1) node [anchor=west][inner sep=2.0pt] {$\omega_3$};
			\draw (0,-1) node [anchor=west][inner sep=2.0pt] {$\omega_4$};
        \end{tikzpicture} 
        =
 \frac{ (i g)^2 }{ ( i k_s)^d }   \int_{\nu}\, \frac{  F_{j_2} \left( \frac{ \omega_{12} }{  k_{s} }   , \nu  \right) F_{j_2} \left(  \frac{ \omega_{34} }{  k_{s} }  , \nu   \right)   }{  \nu^2 - \mu^2  }  \equiv    g^2  \;  \G_4^{\mu}
 \label{eqn:intro_G4_exch}
\end{align}
where $\mu$ is related to the mass of the exchange field by $m^2 = \mu^2 + \frac{d^2}{4}$ and the relevant $i \epsilon$ prescription for the pole is given in \eqref{eqn:our_prescription}. 
This exchange structure obeys the equation
\begin{align}
 \left(  m^2 - \hat{s}  \right) \G_4^{\mu} =  \G^{\rm con}_4 \, ,
 \label{eqn:intro_exch_eqn}
\end{align} 
and is therefore the analogue of $1/(m^2 - s)$ on Minkowski.

\item \emph{Any exchange diagram.}
\eqref{eqn:intro_G4_exch} is the \emph{unique} kinematic structure allowed by local interactions compatible with de Sitter isometries, such that any exchange diagram can be written as 
\begin{align}
    \begin{tikzpicture}[scale=0.6,baseline=0.0cm]
			\draw (-4,1) -- (-3,0);
			\draw (-4,-1) -- (-3,0);
			\draw (-3,0) -- (-1,0);
			\draw (0,1) -- (-1,0);
			\draw (0,-1) -- (-1,0);
                \draw (-4,1) node [anchor=east][inner sep=2.0pt] {$\omega_1$};
                \draw (-4,-1) node [anchor=east][inner sep=2.0pt] {$\omega_2$};
                \draw (-2,0) node [anchor=south][inner sep=2.0pt] {$\mu$};
                \draw (0,1) node [anchor=west][inner sep=2.0pt] {$\omega_3$};
			\draw (0,-1) node [anchor=west][inner sep=2.0pt] {$\omega_4$};
        \end{tikzpicture} 
        =
    g^2 \,  \G_4^{\mu}  +  \sum_{a, b = 0} c_{ab} \;  \hat{t}^b \; \hat{s}^a \;  \G^{\rm con}_4\, ,
\end{align}
where again $g$ and the $c_{ab}$ correspond to constant model-dependent couplings. 
The $t$ and $u$ channels are related to this $s$-channel diagram by a simple relabelling of the energies.

\item \emph{Going on-shell.} 
We derive the LSZ projection of these amputated correlators onto the on-shell $S$-matrix. On Minkowski this is a simple algebraic replacement of $\pm \omega = \sqrt{| \bfk|^2 + m^2}$ for outgoing/ingoing particles, but on de Sitter going on-shell requires integrating over all $\pm \omega  >  | \bfk|$ for outgoing/ingoing particles in the following way:  
\begin{gather}
{}_{\rm out} \langle \bfk_3 \bfk_4   |  \bfk_1 \bfk_2   \rangle_{\rm in}  \propto   \int d^4 \rho \; e^{i \mu \cdot \rho }   \; \G_4   \label{eqn:intro_LSZ} \\ 
\text{with } \omega_1 = - | \bfk_1| \cosh \rho_1 \; , \;\; \omega_2 = - | \bfk_2 | \cosh \rho_2 \; , \;\; \omega_3 =  + | \bfk_3 | \cosh \rho_3 \; , \;\; \omega_4 = + | \bfk_4| \cosh \rho_4 \; .  \nonumber 
\end{gather}
Note that $\mu \cdot \rho = \mu_1 \rho_1 + \mu_2 \rho_2 + \mu_3 \rho_3 + \mu_4 \rho_4$, where each $\mu_b$ is related to the mass of particle $b$ by $m_b^2 = \mu_b^2 + \frac{d^2}{4}$. 
Remarkably, while going from the time-domain $n$-point correlator to the $n$-particle $S$-matrix requires doing $n$ integrals over Hankel mode functions, this operation can be replaced by doing a pair of Fourier transforms (first $n$ Fourier transforms from $\tau$ to $\omega = \pm | \bfk| \cosh \rho$, and then a further $n$ Fourier transforms from $\rho$ to $\mu$). 
The first of these is straightforward, producing simple expressions like \eqref{eqn:intro_G4_seed} for $\G_4$. The second transform, for generic mass parameter $\mu$, converts these simple expressions into complicated multi-variate hypergeometric functions (for \eqref{eqn:intro_G4_seed} specifically, the result of \eqref{eqn:intro_LSZ} is a Lauricella function). 
All of the dynamical information is encoded in the simpler $\G_4$---the projection onto an $S$-matrix element is purely kinematics (since \eqref{eqn:intro_LSZ} is the universal integral transform that would be applied to any model, regardless of its detailed field content or interactions).  
This demonstrates that much of the complexity of cosmological correlators stems from how we treat the asymptotic (free) propagation of the fields!

\item \emph{Conformally coupled scalars.}
For conformally coupled external masses, going on-shell via \eqref{eqn:intro_LSZ} is simple and amounts to setting $\omega = \pm | \bfk|$ for outgoing/ingoing and lowering the spacetime dimension by 2. 
An off-shell correlator in $d$ dimensions therefore corresponds to a conformally coupled $S$-matrix element in $d+2$ dimensions, at least for 4-particle scattering at tree-level. This immediately implies that all previous literature that has computed such wavefunction coefficients or $S$-matrices for conformally coupled scalars can be applied to the study of the underlying off-shell correlator.
The slogan would be: an off-shell $S$-matrix element (for fields of any mass) is as simple as an on-shell object with conformally coupled masses\footnote{
The shift in the dimension typically also reduces complexity, for instance the exchange diagram with all fields conformally coupled gives the dilogarithmic expression \eqref{eqn:cc_dilogs} for the on-shell $S$-matirx in $d=3$ dimensions, but the underlying off-shell correlator is a simple rational function (see \eqref{eqn:intro_cc}) that coincides with the $S$-matrix in $d=5$ with each momentum magnitude $k_b \to \omega_b$. 
}.

\item \emph{Exchange bootstrap.}
One example of conformally coupled technology that can be readily applied to amputated correlators is the bootstrap approach of \cite{Arkani-Hamed:2018kmz,Baumann:2019oyu,Baumann:2020dch}, which solves an exchange equation analogous to \eqref{eqn:intro_exch_eqn} by demanding only physical singularities. 
The same approach can be applied to the off-shell correlators with two adjustments. First, the two $s$-channel variables $u$ and $v$ are now functions of the off-shell energy
\begin{align}
 u = \frac{| \bfk_{12}|}{\omega_1 + \omega_2} \; , \qquad v =  \frac{| \bfk_{34}|}{\omega_3 + \omega_4} \; .  
\end{align}
Solving \eqref{eqn:intro_exch_eqn} will produce functions of $u$ and $v$ with undetermined constants of integration: fixing these corresponds to deciding which object one is considering (wavefunction, $S$-matrix or correlator). 
The relevant boundary conditions for the $S$-matrix (with all particles outgoing) are that it is regular at $u ,v \to +1$ and that any singularity at $u, v \to -1$ must be regular as the other variable $v, u \to -1$ (i.e. no singularities in overlapping channels, a consequence of factorisation). 
For instance, for the exchange of a conformally coupled scalar in $d=3$, this procedure gives
\begin{align}
k_s^3 \;  \G_4^{\mu = i/2}  = - \frac{ u v (  u + v  +  2 u v) }{ 2  (1 + u) (1 + v) (u + v) }  \; .
 \label{eqn:intro_cc}
\end{align}
which agrees with the explicit bulk time integral. 
The on-shell $S$-matrix can then be found by performing the LSZ integrals~\eqref{eqn:intro_LSZ_rho} to project from $\omega_b = k_b \cosh \rho_b$ onto any desired mass state. 
The same bootstrap approach can also be used to find the on-shell $S$-matrix elements directly, in which case we set $\omega_b = k_b$ in $u$ and $v$ so they reduce to the variables of \cite{Arkani-Hamed:2018kmz}, but retain the adjusted boundary conditions relevant for the $S$-matrix. 
For four conformally coupled scalars exchanging a generic massive field, solving the exchange equation with this factorisation condition gives \eqref{eqn:Fuv_def}.

\item \emph{Unitarity cuts.}
The ``discontinuity'' of our exchange solution \eqref{eqn:Fuv_def} is
\begin{align}
 \SBD_{0 \to 1234} + \SBD_{ 1234 \to 0}^*    =   
 \SBD_{0 \to 12 \mu} \, \SBD_{\mu \to 34} 
 +
 \SBD_{\mu \to 12}  \, \SBD_{0 \to 34 \mu} \, .
 \label{eqn:intro_cut}
\end{align}
This reflects the simplicity of perturbative cutting rules for the $S$-matrix, which require fewer terms than their wavefunction counterparts. 
In fact, the integral representation~\eqref{eqn:intro_G4_exch} is particularly well-suited for proving that \eqref{eqn:intro_cut} holds for any tree-level diagram.  
Going beyond individual tree-level diagrams, we derive a simple optical theorem for the on-shell $S$-matrix elements which closely parallels the usual Minkowski result: the only conceptual difference is that terms which were forbidden in flat space due to energy conservation (such as $\SBD_{ 0 \to 12 \mu}$) can contribute to the de Sitter discontinuity. In spite of these additional terms, we find that unitarity connects the discontinuity $\SBD_{12 \to 12} + \SBD_{12 \to 12}^*$ to a sum of positive terms ($\SBD_{12 \to I} \SBD_{12 \to I}^*$) in the forward limit. Positivity of the forward discontinuity is a crucial ingredient of the Minkowski $S$-matrix, underpinning the $S$-matrix bootstrap and positivity bounds for effective field theories, and we will explore its consequences further in future work. 

\item \emph{Higher-point interactions.} 
Finally, all of the above generalises straightforwardly to more than 4 particles, and we describe how to use $n$-point seed solutions and general multi-particle mode functions to express \emph{any} tree-level Feynman diagram for scalar fields in terms of a small number of unique exchange structures (the analogues of $1/(m^2 - p^2)$ on Minkowski) plus a derivative expansion in powers of Mandelstam operators.

\end{itemize}

\noindent Overall, we see these results as forming the basis of future investigations into the mathematical structure of amputated correlators and their use as a well-defined QFT observable that can be computed and constrained in inflationary field theories.

\subsection{Notation and conventions}

Throughout we consider the expanding Poincare patch of a fixed de Sitter spacetime with $d$ spatial dimensions, $ds^2 = (-\tau)^{-2} \left( -d \tau^2 + d \bfx^2 \right)$, where $-\infty < \tau \leq 0$ and we work in units where the Hubble rate $H=1$. 
For a real scalar field $\phi (\tau, \bfx)$ of mass $m^2 = (d/2)^2 + \mu^2$, we split the action $S = S_{\rm free} + S_{\rm int}$, where the free quadratic action is
\begin{align}
S_{\rm free} [ \phi ] = - \tfrac{1}{2} \int d\tau d^d \bfx \; \sqrt{-g} \; \left( g^{\alpha \beta} \partial_\alpha \phi \partial_\beta \phi +  m^2 \phi^2 \right)  
\label{eqn:Sfree}
\end{align}
and $S_{\rm int}$ contains all the non-linear interactions. 
Fields in the so-called ``principal series'' have sufficiently large masses that the parameter $\mu = +\sqrt{  m^2 - d^2 / 4 }$ is real---our discussion will begin with these fields and treat light fields as an analytic continuation\footnote{
The $S$-matrix elements, like the wavefunction coefficients, can develop late-time divergences for particular imaginary values of the mass parameter $\mu$. We restrict our attention to IR finite interactions, for instance between only heavy fields or between light fields with derivative interactions.  
} from real to imaginary $\mu$. 
The canonically normalised field is
\begin{align}
	\varphi (\tau , \bfx ) = (- \tau )^{-d/2} \phi (\tau , \bfx ) \; ,
\end{align}
so that solutions to the free equations of motion for $\varphi$ can be written as $f^{\pm} ( k\tau )$ with Wronskian $f^- i \tau \partial_\tau f^+ - f^+ i \tau \partial_\tau f^- = 1$. 
Its spatial Fourier transform is $\varphi (\tau, \bfk) = \int d^d \bfx \, e^{ - i \bfk \cdot \bfx} \varphi (\tau ,  \bfx )$ where $\bfk \cdot \bfx = \sum_{i=1}^{3} k_i x^i$ is $\tau$-independent\footnote{
We use the same symbol to denote functions in different representations: for instance $\varphi (\tau, \bfx)$ and $\varphi ( \tau, \bfk)$ are of course different functions (one is the field in position space, the other is the field in momentum space). 
}.  
Since both our background and interactions will be invariant under spatial translations, we will often factor out the overall momentum-conserving $\delta$ function which is guaranteed by this symmetry. This is denoted by a prime like so: 
\begin{align}
 \langle \mathcal{O} ( \bfk ) \rangle = ( 2\pi)^d \delta^d \left( \bfk \right) \langle \mathcal{O} (\bfk ) \rangle'  \; . 
\end{align}


In \cite{Melville:2023kgd}, we showed how the time-ordered correlator of $n$ canonically normalised scalar fields could be projected onto a set of ``$S$-matrix elements'' by amputating the external legs and going on-shell---analogous to the familiar LSZ procedure on Minkowski.
This work is an extension of those ideas, and we will use the following taxonomy to describe the different kinds of correlator:  
\begin{align}
\begin{array}{r r}
\text{Correlator:}  & {}_{\rm out} \langle 0 
 | \, T \, \hat{\varphi}_1 ...  \hat{\varphi}_n  | 0 \rangle_{\rm in}   \nonumber \\ 
\text{Amputated correlator:}  &\left( \prod_{j=1}^n i \mathcal{E}_j  \right) {}_{\rm out} \langle 0 
 | \, T \, \hat{\varphi}_1   ...  \hat{\varphi}_n   | 0 \rangle_{\rm in} \nonumber \\ 
\text{On-shell amputated correlator:}  &\quad \left( \prod_{j=1}^n \int_{\tau_j}  f_j  \right) \left( \prod_{j=1}^n  i \mathcal{E}_j  \right) {}_{\rm out} \langle 0 
 | \, T \, \hat{\varphi}_1   ...  \hat{\varphi}_n   | 0 \rangle_{\rm in}   
\end{array}
\end{align}
where $| 0 \rangle_{\rm in}$ ($| 0 \rangle_{\rm out}$) is the state that coincides with the free Bunch-Davies vacuum in the far past (future), $\mathcal{E}_j$ is the free equation of motion for $\varphi_j$ given in \eqref{eqn:E_def} and $f_j$ is the corresponding mode function ($f^+$ for outgoing particles and $f^-$ for ingoing particles). 
Each field will always be labelled by a spatial momentum $\bfk$, together with a single additional variable which can be either $\tau$ or $\omega$, and we collect these kinematic variables into a column argument like so: 
\begin{align}
 G_n \left( \begin{array}{c}
	\tau_1  \\
	\bfk_1 
\end{array}  , \cdots , \begin{array}{c}
	\tau_n  \\
	\bfk_n 
\end{array} \right) &= \;  {}_{\rm out} \langle 0 |   \, T\, \hat{\varphi} (\tau_1, \bfk_1) ...    \hat{\varphi} (\tau_n , \bfk_n)  \,  | 0 \rangle'_{\rm in}     \\ 
 \G_n \left( \begin{array}{c}
	\tau_1  \\
	\bfk_1 
\end{array}  , \cdots , \begin{array}{c}
	\tau_n  \\
	\bfk_n 
\end{array} \right) &=  \left( \prod_{j=1}^n i \mathcal{E}_{\mu_j} [ k_j \tau_j ]  \right) \;  {}_{\rm out} \langle 0 |   \, T\, \hat{\varphi} (\tau_1, \bfk_1) ...    \hat{\varphi} (\tau_n , \bfk_n)  \,  | 0 \rangle'_{\rm in}    \nonumber  
\end{align}
where the subscript on $G_n$ or $\G_n$ is implicitly a list of all other quantum numbers (e.g. the masses of the fields) and $T$ is time-ordering in $\tau$. 
We use the same notation for (amputated) correlators written in terms of $\omega$, which are related to those in the time-domain by a Fourier transform in conformal time,
\begin{align}
\G_n \left( \begin{array}{c}
	\omega_1  \\
	\bfk_1 
\end{array}  , \cdots , \begin{array}{c}
	\omega_n  \\
	\bfk_n 
\end{array} \right)  =  \left[ \prod_{j=1}^n \int_{-\infty}^0 \frac{d \tau_j}{- \tau_j} e^{i \omega_j \tau_j}  \right] \G_n \left( \begin{array}{c}
	\tau_1  \\
	\bfk_1 
\end{array}  , \cdots , \begin{array}{c}
	\tau_n  \\
	\bfk_n 
\end{array} \right) \; . 
\label{eqn:intro_FT}
\end{align}

We will adopt a condensed notation when the same integral/differential operator $\hat{\mathcal{O}}$ (which can depend on the mass, momentum, time, etc.) acts on every field of a correlation function, writing such expressions as
\begin{align}
  \left[   \prod_{b=1}^n \hat{\mathcal{O}} \left( \mu_b, k_b, \tau_b, ... \right) \right]   G_n \left( \begin{array}{c}
	\tau_1  \\
	\bfk_1 
\end{array} , ... , \begin{array}{c}
	\tau_n  \\
	\bfk_n 
\end{array}    \right) 
\equiv   
\left[ \hat{\mathcal{O}} \left( \mu, k, \tau, ... \right) \right]   G_n \left( \begin{array}{c}
	\tau  \\
	\bfk 
\end{array}   \right)  \;\; . 
\label{eqn:condensed_integrals}
\end{align}
For instance the Fourier transform \eqref{eqn:intro_FT} is then
\begin{align}
 \G_n \left(  \begin{array}{c}
	\omega  \\
	\bfk
\end{array}  \right) = \left[ \int_{-\infty}^0 \frac{d \tau}{-\tau} e^{i \omega \tau}  \right]  \G_n \left( \begin{array}{c}
	\tau  \\
	\bfk 
\end{array} \right) \; . 
\end{align}

We use the following notation for sums of energies and momenta,
\begin{align}
 \omega_{ab ... z} &= \omega_a + \omega_b + ... + \omega_z \; , &\bfk_{ab...z} &= \bfk_a + \bfk_b + ... + \bfk_z \; . 
\end{align}
$k$ (and $k_{ab...z}$) denotes the magnitude of $\bfk$ (and $\bfk_{ab...z}$). When discussing 4-particle kinematics, we will use $k_s$, $k_t$ and $k_u$ for the total spatial momentum in each channel:
\begin{align}
 k_s &= k_{12} = k_{34} , &k_t &= k_{13} = k_{24} , &k_u&=k_{14} = k_{23}
\end{align}
It will also be useful to introduce the following shorthand:
\begin{align}
 j_n &= \frac{d}{2} ( n- 1 ) \; , &J_n = \frac{d}{2} ( n -2 )
\end{align}
such that $j_{n} + j_{n'} = J_{n + n'}$ and where $d$ is the number of spatial dimensions.

Finally, we would like to make a comment regarding our choice of sign/phase conventions. 
The reader may wonder why some later expressions seem to contain inconvenient factors of $-1$, $i$ and $\sqrt{\pi}$. We suspect that a sufficiently prescient normalisation for the states and mode functions could remove these. We have chosen to work with the conventions detailed in Appendix~\ref{app:mode}, which are fixed by having:
\begin{itemize}

 \item[(a)] ${}_{\rm out}\langle \bfk' | \bfk \rangle_{\rm in} = (2 \pi )^d \delta ( \bfk - \bfk')$, which fixes the normalisation of the state (up to a phase), 

 \item[(b)] the simple LSZ formula \eqref{eqn:LSZ_intro}, which together with $f^+ = (f^- )^*$ fixes the normalisation of the mode function in the time-domain,

  \item[(c)] the simple integral transform \eqref{eqn:f_to_F}, which fixes the normalisation of the mode function in the energy-domain. 
   
\end{itemize}

\section{Observables from amputated correlators}
\label{sec:observables}

In this section, we describe how amputated correlators, in both $(\tau, \bfk)$ and $(\omega, \bfk)$ variables, can be projected onto $S$-matrix elements, and how these in turn can be used to construct any desired cosmological correlator (which encode the primordial non-Gaussianity produced by inflation).

\subsection{$S$-matrix from time-dependent correlators}

We begin with a brief description of the de Sitter $S$-matrix recently introduced in \cite{Melville:2023kgd}.
%
%
These $S$-matrix elements 
are defined through the usual adiabatic hypothesis that interactions turn off sufficiently quickly in the far past/future that asymptotic states can be defined by matching onto states of the free theory (which can be quantised exactly). 
In particular, we define $| n \rangle_{\rm in}$ ($| n \rangle_{\rm out}$) to be the states which coincide in the far past (future) with the free theory state that has $n$ particles in the far past (i.e. definite energy excitations on top of the Bunch-Davies vacuum).
The overlap between these different asymptotic states, 
\begin{align}
   \SBD_{n' \to n} =   {}_{\rm out} \langle n | n' \rangle_{\rm in}' \; ,
   \label{eqn:S_def_intro}
\end{align}
was dubbed the ``Bunch-Davies'' $S$-matrix in \cite{Melville:2023kgd}.

\paragraph{States in the free theory.}
To extract \eqref{eqn:S_def_intro} from the field theory $S[\phi]$, we first define the instantaneous vacua and $n$-particle states of the free theory defined by the quadratic action \eqref{eqn:Sfree}. 
In the Heisenberg picture, the time evolution of the $\varphi$ operator can be written as
\begin{align}
\hat{\varphi} ( \tau , \bfk )= f^- ( k \tau ,  \m ) \hat{a}_{ - \bfk}  + f^{+} ( k \tau  , \m  ) \hat{a}_{\bfk} ^{ \dagger} \, , 
\end{align}
where the mode functions $f^{\pm} (  k\tau , \m )$ satisfy the free equation of motion,
\begin{align}
	\mathcal{E}_\mu [ k \tau ]  f^{\pm} ( k \tau , \m )  \equiv  \left( 
	\left( \tau \partial_\tau \right)^2 + k^2 \tau^2 + \m^2 \right) f^{\pm} ( k \tau , \m ) = 0 \; , 
 \label{eqn:E_def}
\end{align}
with the boundary condition $f^{\pm} \sim e^{\pm i k \tau}$ imposed at $\tau \to - \infty$, which ensures that $\hat{a}_{\bfk}$ diagonalises the instantaneous Hamiltonian in the far past. 
Consequently, $ \hat{a}_{\bfk}   | 0 \rangle  =  0$ defines the Bunch-Davies vacuum state: the state with the lowest energy at time $\tau \to - \infty$. 
A complete basis of states for the Hilbert space is then provided by
\begin{align}
	| n  \rangle = \hat{a}^\dagger_{n} ... \hat{a}^\dagger_{1} | 0  \rangle  
 \label{eqn:n_def}
\end{align}
where here the general label on $\hat{a}^\dagger_b$ denotes the momenta $\bfk_b$ of each particle together with its other quantum numbers (e.g. mass $\m_b$) and $| n \rangle$ denotes the complete list of $n$-particle data. 

\paragraph{States in the interacting theory.}
For the interacting theory, we use the adiabatic hypothesis, assuming that interactions turn off sufficiently quickly in the far past/future. This allows us to construct a corresponding basis of states in the interacting theory.
Although the operator $\hat{\varphi}$ is no longer a linear function of $\hat{a}$ and $\hat{a}^\dagger$, the hypothesis is that the states $| 0 \rangle_{\rm out}$ and $|\bfk \rangle_{\rm out}$, defined by
\begin{align}
0 &=\lim_{\tau \to 0} i f^+ ( k \tau , \mu ) ( \overset{\leftrightarrow}{ \tau \partial_\tau} ) \hat{\varphi} ( \tau, \bfk )  | 0 \rangle_{\rm out}   \nonumber \\
   | \bfk \rangle_{\rm out} &= \lim_{\tau \to 0} i f^- ( k \tau , \mu ) ( \overset{\leftrightarrow}{ \tau \partial_\tau} )  \hat{\varphi} ( \tau, \bfk )  | 0 \rangle_{\rm in} \; ,   
\end{align}
coincide with the free theory state $| 0 \rangle$ and $| \bfk \rangle$ in the $\tau \to 0$ limit. Similarly, the states $|0\rangle_{\rm in}$ and $|\bfk \rangle_{\rm in}$, defined by 
\begin{align}
0 &=\lim_{\tau \to -\infty} i f^+ ( k \tau , \mu ) ( \overset{\leftrightarrow}{ \tau \partial_\tau} ) \hat{\varphi} ( \tau, \bfk )  | 0 \rangle_{\rm in}   \nonumber \\
| \bfk \rangle_{\rm in} &=  \lim_{\tau \to -\infty} i f^- ( k \tau , \mu ) ( \overset{\leftrightarrow}{ \tau \partial_\tau} ) \hat{\varphi} (\tau ,\bfk )  | 0 \rangle  \; , 
\end{align}
coincide with the free theory state $| 0 \rangle$ and $| \bfk \rangle$ in the $\tau \to -\infty$ limit.
This can be demonstrated explicitly in perturbation theory for all fields with masses $m^2 > d^2/4$ (and for lighter fields with IR finite interactions).

\paragraph{LSZ reduction.}
With this definition of asymptotic states, the in-out overlap \eqref{eqn:S_def_intro} can be obtained from the time-ordered correlators of the interacting theory via the following LSZ reduction formula \cite{Melville:2023kgd}, 
\begin{align}
   \SBD_{0 \to n} ( \bfk )  &=  \left[  \int_{-\infty}^0 \frac{d \tau}{- \tau} \;  f^+ ( k \tau , \mu  ) \,  i \mathcal{E}_\mu [ k \tau ]  \right] \,  G_n \left( \begin{array}{c}
	\tau  \\
	\bfk 
\end{array}   \right)  
=   \left[  \int_{-\infty}^0 \frac{d \tau}{- \tau} \;  f^+ ( k \tau , \mu  )  \right] \,  \G_n \left( \begin{array}{c}
	\tau  \\
	\bfk 
\end{array}   \right)
\label{eqn:LSZ_intro}
\end{align}
where we have used the condensed notation~\eqref{eqn:condensed_integrals}. This corresponds to amputating each external propagator (acting with $i \mathcal{E}$) and putting each external particle ``on-shell'' (integrating against the free mode function). 
The $S$-matrix elements with ingoing particles are given by the analogous expression with an $f^-$ mode function used for each ingoing particle.

\paragraph{Crossing.}
As on Minkowski, the CPT properties of the field can be used to ``cross'' particles from ingoing to outgoing. 
In particular, thanks to the reality of $\varphi$ in position space (which implies $\hat{\varphi} ( \tau , -\bfk ) =  \hat{\varphi}^{\dagger} ( \tau , \bfk ) $) and the CPT relation \eqref{eqn:CPT_f}, we see that crossing a particle corresponds to sending its $\bfk \to \bar{\bfk}$, where $\bar{\bfk}$ denotes a suitable analytic continuation which gives $- \bfk$ when appearing as vector and $| \bar{\bfk} | = e^{-i \pi} k$ when appearing as a magnitude. 
This procedure first appeared in \cite{Goodhew:2020hob, Cespedes:2020xqq} and has recently been discussed in various works \cite{Goodhew:2020hob, Cespedes:2020xqq, Melville:2021lst, Goodhew:2021oqg, Baumann:2021fxj, Albayrak:2023hie, Agui-Salcedo:2023wlq}.
In practice, this means that once one $n$-particle $S$-matrix is known as a function of the $n$ individual momenta, all other channels can be derived using this analytic continuation, e.g.
\begin{align}
 \SBD_{0 \to n} (\bfk_1, ... , \bfk_n ) \big|_{\bfk_n \to \bar{\bfk}_n}  = \SBD_{1 \to n-1} \left( \bfk_n ; \bfk_1 , ... , \bfk_{n-1} \right)  \; . 
 \label{eqn:cross_kbar}
\end{align}

\paragraph{Going off-shell.}
To make crossing manifest, \cite{Melville:2023kgd} defined an ``off-shell'' $S$-matrix, 
\begin{align}
\tilde{\SBD}_n \left( \col{ \tilde{k}}{ \bfk }  \right)   \equiv    
  \left[ \int_{-\infty}^0 \frac{d \tau}{- \tau} \; f^+ (  \tilde{k} \tau , \mu  )  \right] 
\G_n \left(  \col{\tau}{\bfk} \right)
\label{eqn:S_offshell_time}
\end{align}
which is a function of $d+1$ variables per particle. 
This coincides with the all-outgoing channel $\SBD_{0 \to n}$ when every $\tilde{k}_b \to + | \bfk_b|$, and coincides with the other channels when $\tilde{k}_b \to e^{-i \pi} | \bfk_b|$ for each ingoing particle. 
At the level of this off-shell $S$-matrix, crossing corresponds to the simple operation $( \tilde{k}, \bfk ) \to ( - \tilde{k} , -\bfk )$, and therefore the function $\tilde{\SBD}_{n}$ efficiently encodes all $\SBD_{n_1 \to n_2}$ matrix elements with $n_1 + n_2 = n$. 
Another advantage of the off-shell extension \eqref{eqn:S_offshell_time} is that it is precisely analogous to the off-shell wavefunction defined in \cite{Salcedo:2022aal}---it is an analytic function of each $\tilde{k}_b$ in the lower-half of the complex plane, satisfies interesting UV/IR dispersion relations and has a discontinuity in complex $\tilde{k}^2$ that is fixed by the cutting rules of \cite{Melville:2021lst, Goodhew:2021oqg}.  
However, in practice the bulk time integrals in \eqref{eqn:S_offshell_time} are difficult to perform explicitly and can lead to complicated special functions. 
In this work, we will therefore explore a different way of making crossing manifest: by studying the amputated correlator $\G_n$ itself. We will occasionally refer to $\G_n$ as an ``off-shell'' $S$-matrix, since it is the integration over the mode function that sets the particles on-shell (sets $\omega = \pm \sqrt{ k^2 + m^2}$) on Minkowski. In either case, $\tilde{S}_n$ or $\G_n$ are effectively introducing one additional variable for each field in order to conveniently unify all $n$-particle channels into a single object with a simpler mathematical structure. 

%
%
%
%

\subsection{$S$-matrix from energy-dependent correlators}

We will find below that the Fourier transform \eqref{eqn:intro_FT} from conformal time $\tau$ to a conformal energy $\omega$ will simplify many perturbative computations. 
Before proceeding, it is worthwhile to check how these energy-dependent correlators are related to the $S$-matrix.

\paragraph{LSZ reduction.}
The key observation is that the Hankel mode functions $f^{\pm} ( k\tau , \mu )$ can be written as\footnote{
The $-\sqrt{\pi}$ is an unfortunate consequence of how we have normalised the mode functions, see Appendix~\ref{app:mode}. 
}
\begin{align}
 f^{\pm} ( k \tau , \mu ) =  \left[  - \sqrt{\pi} \int_{-\infty}^{+\infty} \frac{d \rho}{2 \pi } \, e^{ i  \mu \rho } \right] e^{ \pm i k \tau \cosh \rho} \;  
 \label{eqn:f1_rho_int_rep}
\end{align}
and therefore a Fourier transform from conformal time to $\omega = \pm k \cosh \rho$ will replace the $f^{\pm}$ mode functions with simple factors of $e^{ i \mu \rho}$. 
Substituting \eqref{eqn:f1_rho_int_rep} into the LSZ reduction formula \eqref{eqn:intro_LSZ} immediately gives,
\begin{align}
  \SBD_{0 \to n} ( \bfk ) =  \left[ - \sqrt{\pi} \int_{-\infty}^{+\infty} \frac{d \rho}{2 \pi}  \, e^{i \mu \rho} \right] \G_n \left(  \col{\omega}{\bfk}  \right)  \Bigg|_{\omega = + k \cosh \rho}  \equiv  \text{LSZ} \left[   \G_n \left( \col{\omega}{\bfk} \right)  \right]\; . 
  \label{eqn:intro_LSZ_rho}
\end{align}
On Minkowski, going on-shell is a simple algebraic procedure (i.e. set $\omega$ equal to $+ \sqrt{k^2 + m^2}$ for each outgoing particle). On de Sitter, this is replaced by the above integral over all $\omega \geq k$. 
This reflects the fact that $\omega$ is not related to any conserved quantity on de Sitter. In fact, $\rho$ appears naturally as a Schwinger proper time for each particle, and integrating over $\omega = k \cosh \rho$ accounts for the redshifting of the particle energy as it propagates through this expanding de Sitter spacetime. 
The $\omega$ variables are computationally useful but since $e^{i \omega \tau}$ is not a solution of the free equations of motion (not an eigenfunction of the de Sitter Casimir) it does not represent a propagating particle. 
We will therefore refer to \eqref{eqn:intro_LSZ_rho} as ``projecting onto mass eigenstates'', since this $\rho$ integral effectively projects $e^{i \omega \tau}$ onto $f^{\pm} ( k \tau , \mu )$, which does represent a freely propagating particle (is an eigenfunction of the de Sitter Casimir).

\paragraph{Crossing.}
To project a field onto an ingoing mass eigenstate, we must use $f^-$ in place of $f^+$ in the LSZ formula. Since they are related by complex conjugation, we see from \eqref{eqn:f1_rho_int_rep} that this amounts to setting $\omega = - k \cosh \rho$ in \eqref{eqn:intro_LSZ_rho}.  
The same LSZ reduction applied to $\G_n$ with $(\omega_b ,\bfk_b)$ replaced by $(-\omega_b, -\bfk_b)$ would therefore produce the $S$-matrix with particle $b$ incoming. At the level of $\G_n$, crossing is therefore very simple and does not require any analytic continuation of $|\bfk|$. 
For instance, the 3-particle $S$-matrix elements are given by the following LSZ reductions,
\begin{align}
 \langle \bfk_1 \bfk_2 \bfk_3 | 0 \rangle   &= 
  \text{LSZ} \left[ \G_3 \left( \col{+ \omega_1}{+\bfk_1} , \col{+ \omega_2}{+\bfk_2} , \col{+ \omega_3}{+\bfk_3} \right) \right]  ,   
&\langle  \bfk_2 \bfk_3 | \bfk_1 \rangle  &=   \text{LSZ} \left[ \G_3 \left( \col{- \omega_1}{-\bfk_1} , \col{+ \omega_2}{+\bfk_2} , \col{+ \omega_3}{+\bfk_3} \right) \right] ,  \nonumber \\ 
 \langle \bfk_3 | \bfk_1 \bfk_2 \rangle &=  \text{LSZ} \left[ \G_3 \left( \col{- \omega_1}{-\bfk_1} , \col{- \omega_2}{-\bfk_2} , \col{+ \omega_3}{+\bfk_3} \right) \right]   ,   
& \langle 0 |  \bfk_1 \bfk_2 \bfk_3 \rangle  &=   \text{LSZ} \left[ \G_3 \left( \col{- \omega_1}{-\bfk_1} , \col{- \omega_2}{-\bfk_2} , \col{- \omega_3}{-\bfk_3} \right) \right]   .
 \label{eqn:S3}
\end{align}

\paragraph{Conformally coupled fields.}
When the external particles have the ``conformally coupled'' mass $\mu = i/2$ (e.g. $m^2 = 2 H^2$ in $d=3$ dimensions), then the LSZ integrals are very simple. 
In fact, comparing the time-domain LSZ formula \eqref{eqn:intro_LSZ} with the definition of $\G_n$ in \eqref{eqn:intro_FT}, we see that the replacement $f^{\pm} ( k \tau, \mu ) \to e^{\pm i k \tau}$ for the external mode functions would transform an $S$-matrix element into an amputated correlator with $\omega = k$. 
Since $f^{\pm} = e^{\pm i k \tau} / \sqrt{\mp 2 i k \tau}$ for conformally coupled scalars, we see that the $S$-matrix for scattering such fields is very nearly the amputated correlator with $\omega = \pm k$, but with some additional factors of $\tau$. One way to account for these factors is to shift the spacetime dimension. 
At a vertex with $n_V$ legs, of which $n_E$ are external conformally coupled legs, the replacement
\begin{align}
 d \to d - \frac{n_E}{n_V - 2}
 \label{eqn:d_shift}
\end{align}
would account for the extra $\tau$ factors. 
For the 4-particle $S$-matrix, there are only two possible diagram topologies at tree-level---the contact diagram (with $n_E = n_V = 4$) and the exchange diagram (with $n_E = 2$ and $n_V =3$ at each vertex)---and in both cases \eqref{eqn:d_shift} yields $d \to d-2$. 
We therefore have the neat relation
\begin{align}
    \G_4 \left( \col{\omega = k}{\bfk}   \right)  \Bigg|_{d-2}  =  \left[ \prod_{b=1}^4 \sqrt{2 i k_b}  \right] \SBD_{0 \to 4} \big|_{d}  \;\; \qquad \left(  \text{conformally coupled at tree-level} \right)
    \label{eqn:cc_d_shift}
\end{align}
For instance, the $S$-matrix for four conformally coupled scalars exchanging a conformally coupled scalar is \cite{Melville:2023kgd}:
\begin{align}
\left[ \prod_{b=1}^4 \sqrt{2 i k_b}  \right] \SBD_{0 \to 4} \big|_{d=3} &= \frac{- 1}{2 k_s} \left( 
  \text{Li}_2 \left( \frac{k_{1} + k_2 - k_s}{k_T} \right) + \text{Li}_2 \left( \frac{k_3 + k_4 - k_s }{ k_T } \right) - \frac{\pi^2}{6}  \right) \; ,    \label{eqn:cc_dilogs} \\
 \left[ \prod_{b=1}^4 \sqrt{2 i k_b}  \right]  \SBD_{0 \to 4} \big|_{d=5} &= - \frac{ k_T + 2 k_s }{ 2 k_s k_T ( k_1 + k_2 + k_s ) ( k_3 + k_4 + k_s )  }  \;  
    \label{eqn:cc_dilogs_2}
\end{align}
while the underlying off-shell correlator is
\begin{align}
 \G_4 \left( \col{\omega}{\bfk} \right) \Bigg|_{d=3} &= 
 - \frac{ \omega_T + 2 k_s }{ 2 k_s \omega_T ( \omega_{12} + k_s ) ( \omega_{34} + k_s )  } 
 \label{eqn:cc_dilogs_3}
\end{align}
We immediately see that \eqref{eqn:cc_dilogs_3} and \eqref{eqn:cc_dilogs_2} indeed satisfy \eqref{eqn:cc_d_shift}, and it can be shown that the Fourier transform of \eqref{eqn:cc_dilogs_3} from $\rho$ to $\mu = i/2$ indeed produces \eqref{eqn:cc_dilogs}.

\paragraph{Mass as an angular momentum.}
The simplicity of \eqref{eqn:intro_LSZ_rho} comes about because the Fourier transform from $\tau$ to $\omega$ transforms the equation-of-motion operator \eqref{eqn:E_def} into a simpler operator
\begin{align}
\left[ \int_{-\infty}^0 \frac{d \tau}{-\tau} \,  e^{i \omega \tau} \right] \mathcal{E}_\mu \left[ k \tau \right] = \left(   \hat{p}^2  + m^2 \right)   \left[ \int_{-\infty}^0 \frac{d \tau}{-\tau} \,  e^{i \omega \tau} \right] 
\end{align}
where\footnote{
There is a choice to be made here about whether $\mathcal{E} \to \hat{p}^2 + m^2$ or $\hat{p}^2 + \mu^2$. With the former choice, $\hat{p}$ corresponds to the momentum of $\phi$, and we will see below that this gives very simple Feynman rules for interactions like $(\nabla \phi )^4$. With the latter choice, $\hat{p}$ corresponds to the momentum of $\varphi$. This leads to some simplifications, such as $\hat{p}^2 = \partial_\rho^2 \approx -\mu^2$ when on-shell, however these come at the cost of more complicated Feynman rules. 
} 
\begin{align}
    \hat{p}^2  &=  ( \omega \partial_\omega )^2 - k^2 \partial_{\omega}^2 - \tfrac{d^2}{4}  = \partial_{\rho}^2  - \tfrac{d^2}{4} \;\; \text{ when } \; \omega = \pm  k \cosh \rho  \; . 
\end{align}
Since the $f^{\pm}$ mode functions are eigenfunctions of $\mathcal{E}$, in the energy domain they are eigenfunctions of $\hat{p}^2$, with eigenvalue $-m^2$. 
The mass parameter $\mu^2 = m^2 - d^2/4$ corresponds to an ``angular momentum'' of the de Sitter isometry group $SO(d+1,1)$.   
Physically, the change of variables from $\tau$ to $\omega$ makes this manifest since now $\rho$ is an angle (rapidity) conjugate to this angular momentum. 
While $(\tau, \bfk)$ is a conceptually convenient set of variables---since we measure correlators imprinted in the CMB after the end of inflation (fixed time) as a function of their wavenumber (fixed $\bfk$)---they obscure much of the mathematical structure because $\tau$ and $\bfk$ are on very different footing. Just as the simplicity of Minkowski correlators only becomes manifest in terms of $(\omega, \bfk)$, here too we find that transforming both $\tau$ and $\bfx$ to conjugate variables can lead to simpler results.

\paragraph{Going on-shell.}
One important simplification is that, in terms of $( \omega, \bfk )$, it is straightforward to see whether a given correlator $\G_n$ will contribute to the $S$-matrix or whether it will vanish when going on-shell. 
On Minkowski, this is simply the observation that $p^2 = -m^2$ when on-shell, and so terms proportional to $p^2 + m^2$ in the amputated correlator will not contribute to the $S$-matrix. 
On de Sitter, the analogous observation is that
\begin{align}
\left[  \int_{-\infty}^{+\infty} \frac{d \rho}{2 \pi}  \, e^{i \mu \rho} \right]  \left( \hat{p}^2  + m^2 \right)   F \left(  \omega, \bfk  \right) |_{\omega = \pm k \cosh \rho}
 &= 0 \; , 
\end{align}
upon integration by parts in $\rho$. 
In anticipation of this integral, which is required to project $\G_n$ onto an $S$-matrix element, we will make use of the relation 
\begin{align}
\hat{p}^2 \approx - m^2
\label{eqn:on-shell}
\end{align}
where $\approx$ should be understood as ``equal up to terms which do not contribute to the on-shell $S$-matrix''. 
In particular, a term in $\G_n$ that is proportional to $(\hat{p}^2 + m^2)$ acting on any function $F$ will vanish when going on-shell.

\paragraph{Field redefinitions.}
A useful feature of the de Sitter $S$-matrix elements is that they are invariant under perturbative field redefinitions, unlike the correlators $G_n$ which are sensitive to our choice of field basis. 
However, while the amputated correlators $\G_n$ are generally not invariant under a field redefinition, they are guaranteed to shift by terms proportional to $\hat{p}^2 + m^2$ which vanish once we set $\hat{p}^2 \approx - m ^2$. This lets us enjoy the best of both worlds: an amputated correlator which is computationally simpler than the $S$-matrix, and which retains the independence under field redefinitions. 
To see how this happens, note that the free action for $\varphi$ can be written as
\begin{align}
S_{\rm free} =  \frac{1}{2} \int \frac{d^d \bfk}{(2\pi)^d} \int_{-\infty}^0 \frac{ d \tau} { \tau}  \, \varphi (\tau, - \bfk ) \, \mathcal{E}_\mu [ k \tau ] \, \varphi (\tau , \bfk )
\end{align}
up to a boundary term that affects only the power spectrum. 
A perturbative redefinition $\varphi \to \varphi + F [ \varphi ]$ will therefore shift the action by
\begin{align}
 S \to S +  \frac{\delta S_{\rm int} }{\delta \phi} F + \frac{1}{2} \int \frac{d^d \bfk}{(2\pi)^d} \int_{-\infty}^0 \frac{d \tau}{ \tau } \, F   \mathcal{E}_\mu \, \varphi   + \mathcal{O} \left( F^2 \right) \; . 
\end{align}
Now, suppose $F[\varphi]$ contains a term with $n$ fields and $S_{\rm int}$ contains a term with $n'$ fields. Since the contact contribution of $F \mathcal{E} \varphi$ to the $(n+1)$-point correlator $\G_{n+1}$ is proportional to $\left( \hat{p}^2 + m^2 \right)$, this is trivial once we go on-shell using \eqref{eqn:intro_LSZ_rho} because $\left( \partial_\rho^2 + \mu^2 \right)$ becomes a total derivative. 
The $\frac{\delta S_{\rm int}}{\delta \varphi}  F$ term first contributes to the $(n+n'-1)$-point correlator, but so does an exchange diagram in which one vertex is the original $S_{\rm int}$ and the other is the $F \mathcal{E} \varphi$ term. In this exchange diagram, the $\mathcal{E}$ will either act on an external line to give a trivial total derivative as before, or it will act on the internal line: this collapses the propagator and produces a contact contribution that exactly cancels with the $\frac{\delta}{\delta \varphi} S_{\rm int} F$ term. 
This argument can be generalised to any desired order in $F$, and shows that under perturbative field redefinitions $\G_n$ is invariant up to terms $\propto ( \hat{p}^2 + m^2)$ that vanish on-shell.

\paragraph{Uniqueness.}
Another result presented in \cite{Melville:2023kgd} is that invariance under field redefinitions and total derivatives implies a unique $3$-particle $S$-matrix (up to crossing). 
The perturbative argument\footnote{
The non-perturbative argument is that there are four possible solutions to the de Sitter Ward identities at three points \cite{Bzowski:2013sza, Bzowski:2015pba}, corresponding to the four scattering channels \eqref{eqn:S3}.
} is that any cubic interaction can be integrated by parts into the form $\phi^2 \Box^n \phi$ (summed over different $n$), which contributes to $\SBD_{0 \to 3}$ in the same way as $m^{2n} \phi^3$. 
This is the analogue of the Minkowski result that the 3-point amplitude is uniquely fixed to be a (possibly mass-dependent) constant. 
Since the amputated 3-point correlator is an off-shell object, additional structures are possible (we describe them explicitly below), however these all must vanish when $\hat{p}^2 \approx - m^2$ and we go on-shell. $\G_3$ is therefore unique up to terms proportional to $\hat{p}^2 + m^2$. 

Furthermore, for the on-shell $S$-matrix there is a unique 4-point exchange structure which describes the interaction of two $\phi$'s via the exchange of a field $\sigma$ (again up to crossing). 
The argument is essentially the same: since the most general cubic vertex is equivalent to $\phi^2 \sigma$ after integration by parts and a field redefinition, there can be only one independent exchange contribution (from $\phi^2 \sigma \times \phi^2 \sigma$) in each channel for the 4-particle $S$-matrix. 
For the amputated correlator, we again find that any additional terms it may contain must vanish when $\hat{p}^2 \approx - m^2$.
~\\

The conclusion is that invariance under field redefinitions (and the uniqueness of the 3-point and 4-point exchange structures) is a consequence of the on-shell relation~\eqref{eqn:on-shell}, which can be imposed without having to perform the LSZ integral \eqref{eqn:intro_LSZ_rho} that projects onto mass eigenstates. We shall therefore focus on amputated correlators $\G_3$ and $\G_4$ with the condition that  $\hat{p}^2 \approx - m^2$, as these are what determine the 3- and 4-particle $S$-matrix.

\subsection{Cosmological correlators from the $S$-matrix}

Our goal is to develop some machinery for computing the amputated correlators $\G_n (\omega, \bfk)$, since these can then be taken on-shell using \eqref{eqn:on-shell} and projected onto mass eigenstates using \eqref{eqn:intro_LSZ_rho}. 
Before doing this, let us briefly recap how the resulting $S$-matrix elements are related to the inflationary correlators that ultimately seed primordial non-Gaussianities in the Cosmic Microwave Background and large-scale structure of galaxies.

\paragraph{Cosmological correlators.}
The late-time in-in correlation functions, often referred to as simply ``cosmological correlators'', are\footnote{
If one used the original field $\phi$ in place of the canonical normalised field $\varphi$, then the correlator would decay like $\tau^{n d/2}$ at late times and require renormalisation to extract a meaningful prediction as $\tau \to 0$. 
}$^{,}$\footnote{
Notice that we have assumed $\int d \varphi \, | \Psi [ \varphi ] |^2 = 1$, otherwise this denominator should be explicitly included: its role is to cancel vacuum bubble contributions. 
}
\begin{align}
\lim_{\tau \to 0} \;  {}_{\rm in} \langle 0 |  \, \hat{\varphi} ( \tau, \bfk_1 )  ... \hat{\varphi} ( \tau, \bfk_n ) | 0 \rangle_{\rm in} \;  = \; \int d \varphi \;  \varphi ( \bfk_1 )  ... \varphi ( \bfk_n ) \, | \Psi [ \varphi ] |^2
\end{align}
where $\Psi$ is the late-time Bunch-Davies wavefunction and this integral is over all field configurations on the $\tau = 0$ spatial slice\footnote{
While inflation is not eternal and in reality this de Sitter expansion should terminate at some time before the conformal boundary at $\tau =0$, since the correlator is dominated by the much earlier time at which the modes cross the horizon using $\tau = 0$ as our final time is a good approximation.  
}.
The correlations between these small perturbations produced during inflation can be related to those of the curvature perturbation $\zeta$: these are conserved on large-scales until they re-enter the horizon and imprint density perturbations on the primordial plasma of the hot Big Bang cosmology (see e.g.  \cite{Baumann:2022mni}).

\paragraph{Wavefunction coefficients.}
The wavefunction is often parameterised in terms of ``wavefunction coefficients'' $\psi_n$,
\begin{align}
 \Psi [ \varphi ] = \lim_{\tau \to 0} \,  \exp \left[  
\sum_{n}^{\infty} \left[ \prod_{b=1}^n \int \frac{d^d \bfk_b}{(2\pi)^d} \,  \frac{ \varphi ( \bfk_b ) }{ f^+ ( k_b \tau ) } \right]  \, \frac{  \psi_{ \bfk_1 ... \bfk_n} ( \tau )  }{n!}  ( 2 \pi )^d \delta^d \left( \bfk_T \right)
\right] \;  ,
\end{align}
where our normalisation for $\psi_n$ ensures they remain finite as $\tau \to 0$ (this is the ``interaction picture'' of \cite{Cespedes:2020xqq}).  
In \cite{Melville:2023kgd}, we showed that the coefficients of the late-time wavefunction are related to the $S$-matrix elements by
\begin{align}
\lim_{\tau \to 0} \psi_n (\tau )  = 
\lim_{\tau \to 0} {\textstyle \sum_j}  B^{j}_{n} ( \tau ) \;\;   \SBD_{0 \to j} \; . 
\end{align} 
where the $B_n^j$ are known model-independent coefficients. 
For instance, the first two terms that contribute to $\psi_4$ are
\begin{align}
\lim_{\tau \to 0} \psi_{\bfk_1 \bfk_2 \bfk_3  \bfk_4}  (\tau )
=
  \SBD_{0 \to \bfk_1 \bfk_2 \bfk_3 \bfk_4}
- \int_{\bfq}  \frac{f^-}{f^+}   \SBD_{0 \to \bfk_1 \bfk_2 \bfk_3 \bfk_4 \bfq  -\bfq}
\end{align}
where $f^-/f^+$ is the late-time limit of $f^- ( q \tau) / f^+ ( q \tau)$ and captures the oscillatory behaviour of $\psi_4$ at late-times.
The six-particle $S$-matrix contains both connected and disconnected components: in perturbation theory its leading contribution is
\begin{align}
 \SBD_{0 \to \bfk_1 \bfk_2 \bfk_3 \bfk_4 \bfq \bfq'} \supset \SBD_{0 \to \bfk_1 \bfk_2 \bfq } \SBD_{0 \to \bfk_3 \bfk_4  \bfq' } + \text{perm.}
 \label{eqn:S6}
\end{align}
At a fixed order in perturbation theory, it is straightforward to invert this relation to express the $S$-matrix elements in terms of wavefunction coefficients. For instance, at leading order
\begin{align}
    \SBD_{0 \to \bfk_1 \bfk_2 \bfk_3 \bfk_4} = 
    \lim_{\tau \to 0} \left[ 
    \psi_{\bfk_1 \bfk_2 \bfk_3  \bfk_4} ( \tau ) 
    + \int_{\bfq} \frac{f^- (q \tau )}{f^+ ( q \tau )} \, \psi_{\bfk_1 \bfk_2 \bfq} ( \tau ) \psi_{\bfk_3 \bfk_4 -\bfq} ( \tau) + \text{2 perm.}
    \right]
    \label{eqn:S4_from_psi}
\end{align}
where the $\psi_3 \times \psi_3$ term arises from replacing \eqref{eqn:S6} with wavefunction coefficients.
This means that previous results in the literature for the wavefunction coefficients can be translated into properties of the $S$-matrix (at least in perturbation theory).

\paragraph{Born rule for correlators.}
On Minkowski, the magnitude-squared of the scattering amplitude corresponds to an interaction probability, and the observable cross-section corresponds to a phase space integral over this probability. 
On de Sitter, we have an analogous relation between the in-in observable correlator and the square of the in-out scattering matrix,
\begin{align}
\lim_{\tau \to 0}  {}_{\rm in} \langle 0 |  \hat{\varphi} ( \tau, \bfk_1)  ... \hat{\varphi} (\tau, \bfk_n) | 0 \rangle_{\rm in}'   = \lim_{\tau \to 0}  {\textstyle \sum_{j, j'}}  \; \SBD_{0 \to j}^* \; C_n^{j j'} ( \tau)  \; \SBD_{0 \to j'} \; , 
\end{align}
where the $C_n^{j j'}$ are known model-independent coefficients. 
For instance, separating each $S$-matrix element into connected and disconnected parts, the primordial bispectrum is given simply by,
\begin{align}
\lim_{\tau \to 0}  \frac{  \langle \hat{\varphi} ( \tau, \bfk_1)  \hat{\varphi} (\tau, \bfk_2) \hat{\varphi} (\tau, \bfk_3) \rangle' 
}{ \prod_{b=1}^3  \langle \hat{\varphi} ( \tau, \bfk_b) \hat{\varphi} ( \tau, -\bfk_b)  \rangle'  }
= 2 \text{Re} \left[  \SBD_{0 \to \bfk_1 \bfk_2 \bfk_3}  \right] \; . 
\end{align}

~\\

So once the amputated correlators $\G_n$ have been projected onto $S$-matrix elements, these objects can be related to the cosmological correlators we would ultimately observe from inflation. 
It is worth highlighting again the underlying philosophy behind this approach. 
We know from explicit calculation that the observable in-in correlators are very complicated functions of the spatial momenta. Motivated by the fact that observable cross-sections on Minkowski become much more tractable when phrased in terms of the underlying scattering amplitude as a function of $p_\mu$, here we are searching for an alternative to the in-in correlator which could enjoy a simpler mathematical structure and be easier to compute, while still being connected in a relatively straightforward way to observations. 
The amputated correlators that we study here and their corresponding $S$-matrix elements are one promising candidate for such an alternative.

\section{Mandelstam operators for de Sitter}
\label{sec:mandelstam}

In this section, we define a ``scattering amplitude'' from the amputated correlator and show that it can be written in terms of a basis of differential operators that are direct analogues of the Mandelstam variables. 
For 4-particle scattering, these differential operators are similar in spirit to the ones appearing in cosmological scattering equations \cite{Gomez:2021qfd,Gomez:2021ujt,Armstrong:2022csc}, though they act on the $\{ \omega, \bfk \}$ of the off-shell $\G_n$ rather than the $\{ \bfk\}$ of the on-shell wavefunction or $S$-matrix. 
In the on-shell limit, the Mandelstam $\hat{s}$ and $\hat{t}$ defined below should coincide with those of the cosmological scattering equations.

\subsection{Scattering amplitudes}

On Minkowski, the scattering amplitude is extracted from the amputated correlator via\footnote{
When $\omega^2 = k^2 + m^2$ for each particle, the amputated correlator becomes the Minkowski $S$-matrix and $A$ becomes the usual on-shell scattering amplitude. Here we are considering the \emph{off-shell} amplitude for which $p^2$ is independent of $m^2$. 
}
\begin{align}
 \G_n \left(   \col{\omega_1}{\bfk_1} , ... ,   \col{\omega_n}{\bfk_n}   \right)   =  A ( p_1, ... , p_n ) \; 2 \pi i \, \delta \left( \omega_T \right)  \qquad \text{(Minkowski)}
\end{align}
where $p_b^\mu = ( \omega, \bfk )$ is the energy and momentum of each particle, and $\delta$ functions impose conservation of the total $\omega_T = \sum_{b=1}^n \omega_b$ and $\bfk_T = \sum_{b=1}^n \bfk_b$. 
On de Sitter, we can analogously extract an amplitude from the amputated correlator,
\begin{align}
    \G_n \left(  \col{\omega_1}{\bfk_1} , ... ,   \col{\omega_n}{\bfk_n}   \right)    &=  
    A (p_1, ... , p_n )  \; \G^{\rm con}_n \left( \omega_T \right)   \qquad  \text{(de Sitter)}
     \label{eqn:first_A_def}
\end{align}
where $p_b = ( \omega_b, \bfk_b)$ denotes the conformal energy and spatial momenta of each field, and $\G^{\rm con}_n$ is the simplest invariant allowed by the de Sitter symmetries.

\paragraph{A first example.}
Let us first consider the simplest polynomial interaction, $\lambda \phi^n$. 
In terms of the canonically normalised $\varphi$, the action is explicitly 
\begin{align}
S_{\rm int} =  \int d^{d} \bfx \int_{-\infty}^0 \frac{d \tau}{-\tau}   \;  (-\tau)^{J_n} \; \frac{\lambda}{n!} \varphi^n  \; , 
\end{align}
where $J_n \equiv \frac{d}{2} (n-2)$ depends on the number of fields. 
At first order in small $\lambda$, the time-ordered correlation function $\langle T \hat{\varphi} (\tau_1, \bfk_1 ) ... \hat{\varphi} (\tau_n, \bfk_n ) \rangle'$ is
\begin{align}
 G_n \left( \begin{array}{c}
	\tau_1  \\
	\bfk_1 
\end{array} , ... , \begin{array}{c}
	\tau_n  \\
	\bfk_n 
\end{array}  \right)  =  i \lambda \int_{-\infty}^0  \frac{d \tau}{- \tau} \, ( - \tau )^{J_n}  \prod_{b = 1}^n G_2^{\mu} \left( k_b \tau_b , k_b  \tau 
	\right)  \; , 
\end{align}
where $G_2^\mu ( k \tau, k \tau' ) = \langle T \hat{\varphi} (\tau, \bfk) \hat{\varphi} (\tau', -\bfk) \rangle'$ is the Feynman propagator for a field of mass $m^2 = \mu^2 + \frac{d^2}{4}$. 
Since this propagator is a Green function for the equation of motion\footnote{The overall normalisation can be found by setting ${}_{\rm out} \langle 0 | 0 \rangle_{\rm in} = 1$ and writing the free action as $i S_{\rm free} =  - \frac{1}{2} \int \frac{d\tau}{i \tau} d^d \bfx  \; \varphi  \mathcal{E}_\mu \varphi $ in a Gaussian path integral. 
},
\begin{align}
i \mathcal{E}_\mu [k \tau] G_2^\mu \left(  k \tau , k \tau' \right) =  -\tau \delta \left( \tau - \tau'  \right)  \; , 
	\label{eqn:G2_eom}
\end{align}
amputating the external legs and transforming to the energy-domain gives
\begin{align}
 \G_n \left( \begin{array}{c}
	\omega_1  \\
	\bfk_1 
\end{array} , ... , \begin{array}{c}
	\omega_n  \\
	\bfk_n 
\end{array}  \right)  =  i \lambda \int_{-\infty}^0  \frac{d \tau}{- \tau} \, ( - \tau )^{J_n}  \prod_{b=1}^n e^{ i \omega_b \tau}   
=  i  \lambda  \; \frac{  \Gamma (J_n) }{  ( i \omega_1 + ... +  i \omega_n +  \epsilon  )^{J_n}} \; ,
\label{eqn:A_eg_1}
\end{align}
where the $+i \epsilon$ is a small imaginary deformation that arises from making the $\tau \to -\infty$ limit convergent\footnote{
A nice way to keep track of the $i\epsilon$'s is to notice that since the time integral runs over negative values of $\tau$ only, the result must be an analytic function of $\omega$ in the lower-half of the complex plane.
}. 
Comparing this with the Minkowski answer, $i \lambda 2 \pi \delta ( \omega_1 + ... + \omega_n)$, we see that the energy-conserving $\delta$-function has been replaced by a total-energy singularity (which may be a pole or branch cut since $J_n$ can be either integer or half-integer).   
Just as $\delta^d \left( \bfk_1 + ... + \bfk_n \right)$ is required by translation invariance, the power of $1/(\omega_1+...+\omega_n)^{J_n}$ is fixed by dilation invariance, and so overall \eqref{eqn:A_eg_1} is uniquely fixed by the de Sitter isometries\footnote{
Since both this interaction and diagram topology are invariant under the exchange of any two fields, the resulting $\G_n$ must be a symmetric function of its arguments.
 Note that while in principal other symmetric polynomials, e.g. $\omega_1 \omega_2 + \text{perm.}$ may have appeared in $\G_n$, once the $\bfk$ dependence is fixed then the only symmetric function of the $\omega_b$ compatible with de Sitter boosts is \eqref{eqn:A_eg_1}.  }.

\paragraph{Seed solution.}
Regardless of whether a theory contains a particular $\phi^n$ interaction, it will be useful to make reference to this de Sitter invariant combination, and so we shall denote it by $\G^{\rm con}_n$ (to distinguish it from $\G_n$, which is the actual correlator in whatever theory is under consideration). 
Since it depends only on the total energy, we shall write it with a single argument 
\begin{align}
  \G^{\rm con}_n \left( \omega_T \right) = \frac{ i \, \Gamma \left( J_n \right)  }{ ( i \omega_T + \epsilon )^{J_n}  } \; . 
\label{eqn:hatGn_def}
\end{align}
It is by factoring out this simple invariant that we can extract a scale-invariant amplitude from $\G_n$ (see \eqref{eqn:first_A_def}). 
In the example of $\lambda \phi^n$, this gives simply $A = \lambda$ (which coincides with the Minkowski result). 
These ``seed correlators'' will be the main building blocks of our more general results for arbitrary derivative interactions in this section and for exchange diagrams in the next section.

\paragraph{de Sitter symmetries.}
Since the underlying scalar field $\phi$ must transform in a particular way under the de Sitter isometries, the correlators $G_n$ satisfy various Ward identities (assuming de Sitter invariant interactions). 
The most familiar of these follows from spatial translations, which requires that the total spatial momentum is conserved---we have already factored out the resulting $\delta^d (\bfk_T)$. 
Another follows from dilations, which fixes the overall scaling dimension of each $G_n$ to be $J_n = \frac{d}{2} ( n -2 )$, which is one factor of $d/2$ for each canonically normalised field plus the dimension of the $\delta^d ( \bfk_T)$.
Since the equation-of-motion operator $\mathcal{E}$ represents the quadratic Casimir of the de Sitter algebra, it commutes with dilations and therefore the amputated $\G_n$ has the same scaling dimension. 
The amplitude $A$ is therefore constrained to have scaling weight zero: it must be a dimensionless function of the $(\omega, \bfk)$. 
%
The final Ward identity is from de Sitter boosts---these place an additional constraint on $G_n$ and hence on the amplitude. 
Both the dilation and boost constraint take a simple form when written in terms of $\bfk$ and $\rho$ (which determine the energy via $\omega = k \cosh \rho$), 
\begin{align}
0 &= \left( \sum_{b=1}^n   \hat{D} [ \bfk_b ]   \right) A_n  \; ,  
& 0 &= \sum_{b=1}^n \left(  \hat{\bfK} [ \bfk_b ]  +  \frac{\bfk_b}{k_b^2} \partial_{\rho_b}^2   \right)  A_n   \; , 
\label{eqn:Ward}
\end{align}
where $\hat{D} [ \bfk]$ and $\hat{\bfK} [\bfk]$ are the symmetry generators of the $d$-dimensional conformal symmetries,
\begin{align}
 \hat{D} [ \bfk ] = \bfk \cdot \partial_{\bfk} \; , \;\; \hat{\bfK} [ \bfk ] =  2 \bfk \cdot \partial_\bfk \;  \partial_{\bfk}   -  \bfk \;  \partial_{\bfk} \cdot \partial_{\bfk}  \; . 
\end{align}
These Ward identities provide a useful consistency check of the examples we provide below, and it would be interesting to explore whether our results for the Mandelstam operators could be bootstrapped directly from \eqref{eqn:Ward}.

\subsection{Covariant momenta}

Now we turn to interactions that contain derivatives. 
In this case, it will be useful to write the amplitude in terms of a time-dependent object, $\hat{A}_\tau$, that acts directly\footnote{
We also include the $\delta^d \left( \bfk_T \right)$ since below we express $\hat{A}_\tau$ in terms of derivatives with respect to $\bfk$, and since the total momentum is conserved one must be wary of mistakenly replacing $\partial_{\bfk_1} f ( \bfk_1 )$ with $\partial_{\bfk_1} f ( - \bfk_2 - ... -\bfk_n ) = 0$.
} on the $e^{i \omega_T \tau}$ within $\G_n^{\rm con}$:
\begin{align}
     A (p_1, ... , p_n )  \; \G^{\rm con}_n \left( \omega_T \right) \delta^d \left( \bfk_T \right)  
  &=    \int_{-\infty}^0 \frac{d \tau}{-\tau} \, (-\tau)^{J_n} \; \hat{A}_\tau [ p_1, ... , p_n ] \; e^{ i \omega_T \tau} \, \delta^d \left( \bfk_T \right) \; . 
	\label{eqn:A_def}
\end{align}

\paragraph{A second example.}
To build some intuition for how derivative interactions are expressed in this way, consider the simple covariant interaction, $\int d^{d} \bfx d \tau \; \sqrt{-g} \; \phi^{n-2} ( \nabla  \phi )^2 $. 
Writing $\phi = (-\tau)^{d/2} \varphi$ and using the same identity \eqref{eqn:G2_eom} for amputating the external propagators, this interaction gives the following contribution to the amputated $n$-point correlator\footnote{
Note that here we treat the fields as distinguishable for the sake of shorter expressions: for indistinguishable fields one should sum over all permutations of the momenta. 
}
\begin{align}
 & \G_n \left( 
 \begin{array}{c}
	\omega_1  \\
	\bfk_1 
\end{array} , ... , \begin{array}{c}
	\omega_n  \\
	\bfk_n 
\end{array}  
 \right)  \nonumber \\ 
 &=  i \int_{-\infty}^0 \frac{d \tau}{-\tau} ( - \tau )^{J_n}  e^{i ( \omega_3 + ... + \omega_n ) \tau} \, \tau^2 \left( 
  \tau^{-d} \partial_\tau \left( \tau^{d/2} e^{ i \omega_1 \tau} \right)\partial_\tau \left( \tau^{d/2} e^{ i \omega_2 \tau} \right)
   - \bfk_1 \cdot \bfk_2 e^{i \omega_{12} \tau} \right) 
   \label{eqn:Gamp_eg_1} 
\end{align}
which can be written in the form \eqref{eqn:A_def} with
\begin{align}
	\hat{A}_\tau [ p_1 , ... , p_n ] &= \left( \omega_1 \omega_2 - \bfk_1 \cdot \bfk_2 \right) \tau^2 - \frac{d}{2} i \omega_{12} \tau - \frac{d^2}{4} \; ,   \nonumber \\ 
	A (p_1, ... , p_n) &=  J_n ( J_n +1 ) \frac{ \omega_1 \omega_2 - \bfk_1 \cdot \bfk_2  }{\omega_T^2} - \frac{J_n d}{2} \frac{\omega_{12}}{\omega_T} - \frac{d^2}{4} \; . 
\end{align}
Note that this amplitude contains the Minkowski result as the leading term in the limit of large $\tau$ / small $\omega_T$, i.e. 
\begin{align}
g^{\alpha \beta} \nabla_\alpha \phi_1 \nabla_\beta \phi_2 \;\; \Rightarrow \;\; \hat{A}_\tau [ p_1, ... , p_n] = - \tau^2 \, p_1 \cdot  p_2 \left(1 + \mathcal{O} \left( \frac{1}{ p \tau}  \right)\right) \; . 
\label{eqn:A_subleading}
\end{align}
where $p_1 \cdot p_2 = -\omega_1 \omega_2 + \bfk_1 \cdot \bfk_2$. 
This turns out to be the general structure of all contact amplitudes: an arbitrary contraction of covariant derivatives in the action will produce the corresponding contraction of $p$'s in $\hat{A}_\tau$, plus corrections which have fewer powers of the momenta/conformal time and which account for the background curvature.

\paragraph{Momentum operators.}
We now describe a systematic way to determine the curvature corrections to $\hat{A}_\tau$ and $A$ directly in the energy-domain, without explicitly expanding out the $\nabla_\mu$ derivatives as in the example above. 
The central idea is to find a differential operator $\hat{p}_\mu ( \tau )$ in terms of $\omega$ and $\bfk$ that produces the same correlator as $\nabla_\mu$ in the time-domain. For instance
\begin{align}
  \left[ - \tau \nabla_{\mu}  \right]  (-\tau)^{d/2} e^{i \omega \tau - i \bfk \cdot \bfx} \; = \; 
  (-\tau )^{d/2} e^{-i \bfk \cdot \bfx} \left[ i \hat{p}_{\mu} ( \tau ) \right]   e^{ i \omega \tau}  \; ,
\end{align}
where the factor of $( - \tau)^{d/2}$ is from our normalisation of $\varphi$.
Concretely, we seek a $\hat{p}_\mu (\tau)$ that produces $p_\mu = ( - \omega \tau + i \tfrac{d}{2} ,  \bfk \tau )$ when acting directly on $e^{i \omega \tau }$, and which also respects the same commutation relation as the covariant derivatives, which is
\begin{align}
 [ [ \nabla_\mu , \nabla_\nu ] , \nabla_\alpha ] = g_{\mu \alpha} \nabla_\nu - g_{\nu \alpha} \nabla_\mu
\label{eqn:comm_rel}
 \end{align}
since $R_{\mu \nu \alpha \beta} = g_{\mu \alpha} g_{\nu \beta} - g_{\mu \beta} g_{\nu \alpha} $ on a maximally symmetric spacetime. 

One convenient solution with these properties is
\begin{align}
 \hat{p}_\mu (\tau ) = \left( \begin{array}{c}
    i \hat{D} [p]  \\ 
    \frac{\tau}{2} \bfk_i  + \frac{1}{2 \tau} \hat{\bfK}_i [p]
\end{array} \right) \; . 
\label{eqn:phat_def}
\end{align}
where the operators\footnote{
Note that all spatial indices are raised/lowered using $\delta_{ij}$. 
}
\begin{align}
 \hat{D} [ p ] &=  \omega \partial_{\omega} + \bfk_i \partial_{\bfk_i} + \frac{d}{2} \nonumber \\
 \hat{\bfK}_i [ p ] &= \bfk_i ( - \partial_\omega^2 + \partial_{\bfk_j} \partial_{\bfk^j} ) - 2\hat{D} \left[  p \right] \partial_{\bfk^i}
\end{align}
are related to the generators of the dilation/boost isometries of the de Sitter spacetime, which makes it straightforward to verify that they satisfy the commutation relation \eqref{eqn:comm_rel}. 

The amplitude in the preceding example can then be written very succinctly as\footnote{
To be explicit, $\hat{p}_1^\mu$ is the operator \eqref{eqn:phat_def} acting on the arguments $p_1^\mu = ( \omega_1 , \bfk_1)$, i.e.
\begin{align}
 \hat{p}_1^\mu (\tau ) = \left( \begin{array}{c}
    - i \hat{D} [p_1]  \\ 
    \frac{\tau}{2} \bfk^i_1  + \frac{1}{2 \tau} \hat{\bfK}^i [p_1]
\end{array} \right) \; . 
\end{align}
}
\begin{align}
g^{\alpha \beta} \nabla_\alpha \phi_1 \nabla_\beta \phi_2 \quad \Rightarrow \quad 	\hat{A}_\tau [ p_1 , ... , p_n ] &= - \eta_{\mu \nu} \hat{p}_1^\mu ( \tau ) \hat{p}_2^\nu ( \tau )
\end{align}
where it is the commutation relations between $\hat{D}$ and $\hat{\bfK}$ that fully account for the subleading terms in \eqref{eqn:A_subleading}.

\paragraph{A third example.}
These $\hat{p}_\mu (\tau)$ operators lead to even greater simplifications as the number of derivatives grows. For example, consider the covariant $\int d^{d} \bfx d \tau \, \sqrt{-g} \,  \phi^{n-2} ( \nabla_\alpha \nabla_\beta \phi )^2$. 
Following the above discussion, we can immediately write down the energy-domain amplitude
\begin{align}
 \hat{A}_\tau [ p_1, ... , p_n ]  &= \left( \eta_{\mu\nu} \hat{p}_1^\mu (\tau)  \hat{p}_2^\nu (\tau ) \right)^2 \; .
\label{eqn:Atau_eg_2}
\end{align}
Expanding out the $\hat{D}$ and $\hat{\bfK}$ operators leads to the explicit expression\footnote{
One can verify that \eqref{eqn:Atau_eg_2} is the same result as expanding $\nabla_\mu \nabla_\nu \phi = ( \partial_\mu \partial_\nu + \Gamma_{\mu\nu}^\alpha \partial_\alpha ) (- \tau )^{d/2} \varphi$ and then transforming to the energy-domain like in \eqref{eqn:Gamp_eg_1}.
} 
\begin{align}
 e^{-i \omega_{12} \tau} \hat{A}_\tau [ p_1, ... , p_n ] e^{+i \omega_{12} \tau} &=  ( p_1 \cdot p_2 )^2 \tau^4 
+
\left[  (d +2 ) \omega_{12} p_1 \cdot p_2 -
 \omega_1 k_2^2 - \omega_2 k_1^2  + \omega_{12} \omega_1 \omega_2   \right] i \tau^3 \nonumber \\
&+ \left[ 
 \tfrac{d}{2} s_{12} + 
  \tfrac{d^2 +6d + 4 }{2} \,  p_1 \cdot p_2 - \tfrac{d (d+2)}{4} \omega_{12}^2 +  \omega_1 \omega_2
   \right] \tau^2 \nonumber \\
&+ \frac{d^2 (d+3)}{4} \omega_{12} i \tau  +\frac{1}{16} d^3 ( d + 4 ) \; , 
\label{eqn:Atau_eg_3}
\end{align}
where $p_1 \cdot p_2 = -\omega_1 \omega_2 + \bfk_1 \cdot \bfk_2$, $s_{12} = \omega_{12}^2 - k_s^2$ and $\omega_{12} = \omega_1 + \omega_2$. 
Again we see that the leading term in the large $\tau$ limit is the expected Minkowski result, and this is corrected by terms which are generated by the non-zero commutator between $\hat{p}^0 ( \tau)$ and $\hat{p}^i ( \tau )$. 
The final amplitude $A (p_1, ... , p_n)$ is found by performing the time integral in \eqref{eqn:A_def}: this is straightforward since its only effect is to replace each $(-\tau)^n$ with $\Gamma (J_n + n) / (i \omega_T)^{J_n + n}$ in \eqref{eqn:Atau_eg_3}.

\subsection{A basis of kinematic invariants}

Armed with the covariant $\hat{p}_\mu (\tau )$ operators, we can now build a complete basis for the time-dependent $\hat{A}_\tau$ (and therefore also for the final amplitude $A$).  
To ensure that the basis is not overcomplete, we must account for two redundancies:
\begin{itemize}

\item[(i)] total $\tau$ derivatives in $\hat{A}_\tau$ that do not affect $A$, 

\item[(ii)] total $\rho$ derivatives in $\G_n =  A \G_n^{\rm con}$ that do not affect the on-shell $S$-matrix. 

\end{itemize}

\paragraph{Total spacetime derivatives.}
In addition to simplifying perturbative calculations, 
the representation \eqref{eqn:phat_def} of covariant derivatives makes the vanishing of total derivatives transparent, namely that any interaction $\int d^{d+1}x \sqrt{-g} \, \nabla_\alpha \mathcal{O}^\alpha$ in the action does not contribute to the amputated Green function.
This is because the spacetime isometries effectively set the total $\sum_{a=1}^n \hat{p}_a ( \tau ) = 0$ within $\hat{\mathcal{A}}_\tau$. 
In more detail, total time derivatives give a vanishing contribution to the amputated correlator thanks to the dilation Ward identity, 
\begin{align}
	0 &=  \sum_{a=1}^n \hat{D} [p_a] \, \G_n (p_1, ... , p_n ) \delta^d \left( \bfk_T \right)  \nonumber \\
	&= \int_{-\infty}^0 \frac{d \tau}{- \tau} ( - \tau )^{ J_n } \left( \sum_{a=1}^n  i \hat{p}_a^0 ( \tau )  \right) \, \hat{A}_{\tau} [ p_1, ... , p_n ] e^{i \omega_T \tau} \delta^d \left( \bfk_T \right)  \; .  
	\label{eqn:DA_vanishes}
\end{align}
Total spatial derivatives, $\sum_{a=1}^n \hat{p}_a^i ( \tau) $ vanish thanks to the translation Ward identity (i.e. $\bfk_T = 0$) and also the fact that contact invariants satisfy,
\begin{align}
	\left( \sum_{a=1}^n  \hat{\bfK}_i [ p_a ]  \right) \, \hat{A}_{\tau} [ p_1, ... , p_n ] e^{i \omega_T \tau} \delta^d \left( \bfk_T \right) = 0 
	\label{eqn:KA_vanishes}
\end{align}
for any $\tau$, which implies the boost Ward identity.

\paragraph{Total $\rho$ derivatives.}
In the previous section, we showed that $\hat{p}^2 \approx - m^2$ under the integral transform which puts the external particles on-shell. 
However, note that the time-dependent operator $\hat{p} (\tau) \cdot \hat{p} (\tau ) = \eta^{\alpha \beta} \hat{p}_\alpha (\tau) \hat{p}_\beta (\tau)$ does not commute with $\hat{p}_\alpha (\tau)$, which reflects the fact that a contact interaction containing $\Box \nabla_\mu \phi \nabla^\mu \phi$ is \emph{not} equivalent to $m^2 \nabla_\mu \phi \nabla^\mu \phi$ (since $[\Box , \nabla_\mu ] \neq 0$). 
Accounting for this non-zero commutator, the general on-shell condition for covariant derivatives can be written as
\begin{align}
	\hat{p} ( \tau ) \cdot \hat{p} (\tau ) = \hat{p}^2 - \hat{\mathcal{J}}^2 (\tau) \approx - m^2 - \hat{\mathcal{J}}^2 (\tau) \; ,
	\label{eqn:p2_onshell}
\end{align}
where\footnote{
Note that $\hat{J}_{ij} [p] =  \bfk_i \partial_{\bfk^j} -  \bfk_j \partial_{\bfk^i}$ is the generator of spatial rotations, while  $\frac{\tau}{2} \bfk_i - \frac{1}{2 \tau} \hat{\bfK}_i [p]$ generates a rotation in embedding space.
}
\begin{align}
 	\hat{\mathcal{J}}^2 (\tau) = \frac{1}{2} \left( \hat{J}_{ij} [ p ] \right)^2 +  \left( \frac{\tau}{2} \bfk_i - \frac{1}{2 \tau} \hat{\bfK}_i [p]  \right)^2  
 	\label{eqn:Jhat_def}
\end{align}
obeys $\hat{\mathcal{J}}^2 (\tau) e^{i \omega \tau} = 0$ and $\left[  \hat{\mathcal{J}}^2 (\tau)  , \hat{p}_\mu ( \tau ) \right]$ reproduces $[\Box , \tau \nabla_\mu ]$.
 We stress that although $\hat{\mathcal{J}}^2 ( \tau )$ cannot be written in terms of $\hat{p}_\mu (\tau)$, its commutator with any momentum can be, and so when \eqref{eqn:p2_onshell} is substituted into $\hat{A}_\tau e^{i \omega \tau}$ it always produces terms that are functions of the $\hat{p}_\mu ( \tau )$ only.

\paragraph{A fourth example.}
For example, consider the covariant interaction $\int d^{d+1} x \, \sqrt{-g} \, \phi^{n-2} \Box^2 \phi^2$, where $\Box = g^{\mu\nu} \nabla_\mu \nabla_\nu$. 
The corresponding amplitude is
\begin{align}
 \hat{A}_\tau [ p_1, ... , p_n] = \left( \hat{p}_1 (\tau)  +  \hat{p}_2 (\tau) \right)^4 \; , 
\end{align}
where the four indices are contracted using $\eta_{\alpha \beta}$. 
On Minkowski, it would be straightforward to argue that the on-shell amplitude from this operator is not independent of the two previous examples $\phi^{n-2} (\nabla \phi)^2$ and $\phi^{n-2} (\nabla \nabla \phi )^2$, since $( p_1 + p_2 )^4$ can be written purely in terms of $p_1 \cdot p_2$ and $(p_1 \cdot p_2 )^2$ once $p_1^2 = -m_1^2$ and $p_2^2 = - m_2^2$ are fixed to their on-shell values. 
On de Sitter, we can make this same argument by writing
\begin{align}
\left( \hat{p}_1 ( \tau)  +  \hat{p}_2 ( \tau) \right)^4 &= 
\left( \hat{p}_1^2 ( \tau) + \hat{p}_2^2 ( \tau) \right)^2 
+ 4 \left( \hat{p}_1 ( \tau) \cdot \hat{p}_2 ( \tau) \right)^2   \nonumber \\
&+  2 \{   \hat{p}_1 ( \tau) \cdot \hat{p}_1 ( \tau) + \hat{p}_2 ( \tau) \cdot \hat{p}_2 ( \tau)  ,  \hat{p}_1 ( \tau) \cdot \hat{p}_2 ( \tau) \} 
\end{align}
and then use \eqref{eqn:p2_onshell} to go on-shell, which gives 
\begin{align}
 \left( \hat{p}_1 ( \tau)  +  \hat{p}_2 (\tau) \right)^4 e^{ i \omega_{12} \tau }  &\approx 
\Bigg\{ 
\left( m_1^2 + m_2^2 \right)^2 
- 4  \hat{p}_1 (\tau) \cdot \hat{p}_2 (\tau) \left( m_1^2 + m_2^2 \right)
+ 4 \left( \hat{p}_1 (\tau) \cdot \hat{p}_2 (\tau) \right)^2  \nonumber \\ 
&\qquad -  \left[   \hat{\mathcal{J}}^2_1 (\tau) +  \hat{\mathcal{J}}^2_2 (\tau)    , 2 \hat{p}_1 (\tau) \cdot \hat{p}_2 (\tau) \right]  
\Bigg\} e^{ i \omega_{12} \tau} \; . 
\end{align}
The first line is the Minkowski result, and the commutator on the second line captures the effect of the de Sitter spacetime curvature. 
It is straightforwardly evaluated using
\begin{align}
\left[   \hat{\mathcal{J}}^2_1 (\tau )  , 2 \hat{p}_1 (\tau) \cdot \hat{p}_2 (\tau) \right] e^{ i \omega_{12} \tau}  
=
- 2 d \,  \hat{p}_1 (\tau) \cdot \hat{p}_2 (\tau) e^{ i \omega_{12} \tau}  \; ,
\end{align}
which establishes that the contribution of $\Box^2 \phi^2$ is not independent from $(\nabla \nabla \phi)^2$ and $(\nabla \phi)^2$ once we go on-shell.

\paragraph{Kinematic basis.}
Thanks to the above results, we can now enumerate the independent kinematic operators on which $\hat{\mathcal{A}}_\tau$ can depend in essentially the same way that we count kinematic invariants on Minkowski. 
In particular, a local interaction of $\phi$ and its covariant derivatives in the action will produce a $\hat{\mathcal{A}}_\tau$ built from the $\hat{p}_\alpha (\tau)$ operators subject to the conditions:
\begin{itemize}

\item The $\hat{p}^\alpha (\tau)$ must come in pairs with spacetime indices contracted using $\eta_{\alpha \beta}$, 

\item The total $\hat{p}^\alpha (\tau)$ of all external lines gives a vanishing contribution to the correlator, 

\item All contractions of the form $\hat{p}_b (\tau) \cdot \hat{p}_b (\tau )$ are redundant, since they can be replaced by $-m_b^2$ plus lower order invariants using \eqref{eqn:p2_onshell}. 

\end{itemize}
We will explicitly describe the resulting kinematics for both 3- and 4-point correlators in sections~\ref{sec:3-pt} and \ref{sec:4-pt} below. 

\paragraph{Mandelstam operators.}
Before focussing on a particular $n$, let us give a general characterisation of the allowed kinematic invariants on which any de Sitter amplitude can depend.
By analogy with Minkowski scattering, we define the differential operators $\hat{s}_{12...n} ( \tau)$ as the invariant associated with particles $\{ 1, 2, ... ,n \}$,
\begin{align}
  \hat{s}_{12...n} ( \tau ) = - \eta_{\alpha \beta} \left( \sum_{b=1}^n \hat{p}_b^\alpha (\tau)  \right) \left( \sum_{c=1}^n \hat{p}_c^\beta (\tau)  \right) \; .
  \label{eqn:shat_def_1}
\end{align}
We can again separate this operator into a time-independent part, which we denote as $\hat{s}_{12...n}$ with no $\tau$ argument, and a remainder 
\begin{align}
 \hat{s}_{12...n} ( \tau ) = \hat{s}_{12 ... n} - \hat{\mathcal{J}}^2_{12...n} ( \tau ) 
 \label{eqn:shat_def_2}
\end{align}
which is given in terms of the isometry generators by
\begin{align}
 \hat{s}_{12 ... n} = \left( \sum_{a=1}^n \hat{D} [ p_a ] \right)^2 + \frac{1}{2} \left( \sum_{a=1}^n \bfk_a^i \right) \left( \sum_{a=1}^n \hat{\bfK}^i [ p_a] \right) + \frac{1}{2} \left( \sum_{a=1}^n \hat{\bfK}^i [ p_a] \right) \left( \sum_{a=1}^n \hat{\bfk}_a^i \right) + \frac{1}{2} \left( \sum_{a=1}^n \hat{J}_{ij} [ p_a] \right)^2
 \label{eqn:shat_def_3}
\end{align}
and represents the Casimir associated with transforming the points $1, ..., n$. 
The remainder is then
\begin{align}
 \hat{\mathcal{J}}^2_{12...n} ( \tau ) = \frac{1}{2} \left( \sum_{a=1}^n \hat{J}_{ij} [ p_a] \right)^2 + \left( \frac{\tau}{2} \sum_{a=1}^n \bfk^i_a - \frac{1}{2 \tau} \sum_{a=1}^n \hat{\bfK}^i [ p_a ] \right)^2
\end{align}
and by analogy with \eqref{eqn:Jhat_def} accounts for the non-commutativity of the different Mandelstam operators. 

In the large $\tau$ (or small $\omega_T$) limit, these Mandelstam operators reduce to the usual Mandelstam variables on Minkowski. 
At finite $\tau$ (or finite $\omega_T$), the main effect of the background curvature is to introduce non-commutativity between the different Mandelstam operators. 
In general, the commutator
\begin{align}
 [ \hat{s}_{A} (\tau ) , \hat{s}_B (\tau)  ]  = \hat{\mathcal{R}}_{AB} ( \tau )
\end{align}
is non-zero whenever the lists $A$ and $B$ have a particle in common, and is equal to a $d$-dependent operator which contains strictly fewer powers of the $\hat{p}_\alpha$ than $\hat{s}_A \hat{s}_B$. 
Finally, since the total $\hat{p}_\alpha$ vanishes, we also have identities such as
\begin{align}
\left( \hat{s}_{12 ... b} - \hat{s}_{b+1 ... n} \right) A ( p_1, ... , p_n ) \G^{\rm con}_n (\omega_T )  = 0    
\end{align}
which are familiar from Minkowksi.

\subsection{The unique 3-point correlator}
\label{sec:3-pt}

It was previously argued in \cite{Melville:2023kgd} that the 3-particle $S$-matrix is unique (up to crossing) because it is completely fixed by the de Sitter Ward identities.
The off-shell picture of momenta operators developed above makes this uniqueness manifest, and also clarifies which further structures may appear in the off-shell correlator (and how they vanish when going on-shell).

\paragraph{Uniqueness.}
The argument for uniqueness is straightforward: any bilinear combination of $\hat{p}_1, \hat{p}_2, \hat{p}_3$ subject to the constraint that $\sum_{a=1}^3 \hat{p}_a = 0$ (total derivatives vanish) can be written in terms of the $\hat{p}_a^2$ only.
Once these are fixed by the on-shell relation~\eqref{eqn:p2_onshell}, the resulting three-point amplitude $A$ will always reduce to a constant (function of the masses).
Therefore an arbitrary cubic interaction will give rise to a correlator of the form 
\begin{align}
 \G_3 \left(  \col{\omega_1}{\bfk_1} ,  \col{\omega_2}{\bfk_2} , \col{\omega_3}{\bfk_3} \right) \approx f ( \m_1, \m_2, \m_3 ) \, \G^{\rm con}_3 \left( \omega_T \right)
 \label{eqn:G3amp_unique}
\end{align}
for some function $f$ of the masses and with $\G^{\rm con}_3$ given in \eqref{eqn:hatGn_def}.

\paragraph{Three-particle $S$-matrix.}
Given~\eqref{eqn:G3amp_unique}, the 3-particle $S$-matrix is therefore unique, up to crossing and a dependence on the fixed masses (and other quantum numbers) of the three particles. 
In particular, the LSZ reduction of $\G^{\rm con}_3$ produces an $S$-matrix element 
\begin{align}
 {}_{\rm out} \langle  \bfk_1 \bfk_2 \bfk_3 | 0 \rangle_{\rm in} 
  &= i g  \left[ - \sqrt{\pi} \int_{-\infty}^{+\infty} \frac{d \rho}{2 \pi} \, e^{i \m \rho }  \right] \frac{  \delta^d \left( \bfk_1 + \bfk_2 + \bfk_3 \right) }{ ( \sum_{b=1}^3 i k_b \cosh \rho_b )^{\frac{d}{2}}}  
\end{align}
and its crossing images \eqref{eqn:S3}.
This integral can be written as a combination of Appell $F_4$ functions, given explicitly in \cite{Melville:2023kgd}. 
This particular $F_4$ function can also be derived from the Ward identities as the unique structure compatible with all of the de Sitter isometries \cite{Bzowski:2013sza,Bzowski:2015pba}. 
However, the preceding kinematic argument for uniqueness makes it clear what role is being played by the on-shell condition. If we relax this condition, we find that $\delta^d \left( \bfk_1 + \bfk_2 + \bfk_3 \right) / ( \omega_1 + \omega_2 + \omega_3 )^{d/2}$ is not the only function of the $\omega$ and $\bfk$ which is invariant under the de Sitter Ward identities. A general amputated correlator may additionally depend on an arbitrary function of $\hat{p}^2_1 , \hat{p}_2^2 , \hat{p}_3^2$. 
A local interaction Lagrangian may therefore produce contributions of the form
\begin{align}
    \partial_{\rho_1}^{2n_1} \partial_{\rho_2}^{2n_2} \partial_{\rho_3}^{2n_3} \; \G^{\rm con}_3 ( \omega_T )   
\end{align}
in the amputated correlator, which would indeed reduce to \eqref{eqn:G3amp_unique} once all three external particles are put on-shell. 
For instance, the following combination
\begin{align}
\left( \hat{p}_1^2 + m_1^2 \right) \G_3^{\rm con}  =    \left(  - \omega_1^2 + k_1^2  +  \frac{2 \omega_1 \omega_T}{ d + 2 }   -  \mu_1^2 \frac{ 4 \omega_T^2 }{ d (d+2) }     \right) \frac{ \Gamma \left( \frac{d}{2} + 2\right) }{ ( i \omega_T )^{ \frac{d}{2} + 2} }
\end{align}
is consistent with the de Sitter isometries and will generically appear in $\G_3$ when the action contains derivative interactions, however since it does not contribute to the on-shell $S$-matrix it may be discarded without affecting any physical observable.

\paragraph{Two-particle mode function.}
The 3-particle case highlights how taking each particle on-shell increases the complexity of special functions required. Starting from the off-shell correlator $\G^{\rm con}_3 ( \omega_T )$, which is simply a power of $\omega_T$, we take just one of the particles on-shell:
\begin{align}
\left[ - \sqrt{\pi} \int_{-\infty}^{+\infty} \frac{d \rho}{2\pi} e^{i \mu \rho}   \right]  \G^{\rm con}_3 ( \omega_T  ) \big|_{\omega_3 = k_{s} \cosh \rho} 
%
%
=  \frac{  F_{j_2} \left( \frac{\omega_{12}}{k_{s}} , \mu \right) }{ ( i k_{s} )^{ \frac{d}{2} } } \; . 
\label{eqn:F2_def}
\end{align}
Such integrals can be performed by considering the differential equations they must satisfy. 
For instance, since $\hat{p}_3^2= ( \hat{p}_1 + \hat{p}_2 )^2 =  - \hat{s}_{12} $ when acting on $\G_3^{\rm con}$, the function $F_{j_2}$ must solve
\begin{align}
 \left(   m^2 - \hat{s}_{12} \right) F_{j_2} \left( \tfrac{\omega_{12}}{k_{s}} , \mu \right) = 0 \; .
\end{align}
$F_{j_2}$ is therefore an eigenfunction of $\hat{s}_{12}$ with eigenvalue $m^2 = \mu^2 + \tfrac{d^2}{4}$.
Together with the Bunch-Davies vacuum condition \eqref{eqn:mode_asy_exp} inherited from $f^+$, this uniquely fixes $F_{j_2}$---we give this function explicitly in \eqref{eqn:Fmode} in terms of a Legendre function.  
These eigenfunctions of the Mandelstam operators will play a central role when we construct spectral representations\footnote{
There is a nice parallel between the $F_{j_n}$---an off-correlator in which one leg is put on-shell---and the Berends-Giele current \cite{Berends:1987me}---an on-shell amplitude with one leg taken off-shell. Both can be used to recursively construct higher-point exchange diagrams.
} in section~\ref{sec:spectral} and we will explore this differential equation approach to determining correlators further in section~\ref{sec:bootstrap}.

Physically, $F_{j_2}$ has the interpretation of a two-particle mode function, since it describes the projection of two fields (with energy $\omega_1$ and $\omega_2$) onto the mass eigenstate $\mu$,  
\begin{align}
    \begin{tikzpicture}[scale=0.6,baseline=0.0cm]
			\draw (-2,0) -- (-1,0);
			\draw (0,1) -- (-1,0);
			\draw (0,-1) -- (-1,0);
                \draw (-2,0) node [anchor=east][inner sep=2.0pt] {$\mu$};
                \draw (0,1) node [anchor=west][inner sep=2.0pt] {$\omega_1$};
			\draw (0,-1) node [anchor=west][inner sep=2.0pt] {$\omega_2$};
                \draw (-1.1,-0.1) node [anchor=north][inner sep=3.0pt] {\small \color{gray} $\phi^3$};
        \end{tikzpicture} 
           \; = \;  \frac{  F_{j_2} \left( \frac{\omega_{12}}{k_{s}} , \mu \right) }{ ( i k_{s} )^{ \frac{d}{2} } }
\end{align}
This is analogous to $e^{i ( p_1 + p_2 ) \cdot x}$ on Minkowski, which describes two particles with an effective total mass of $s = - (p_1 + p_2)^2$. 
Just as $\G_3^{\rm con}$ has a branch cut at $\omega_T = 0$, this mode function has a branch cut at $\omega_{12} + k_s \to 0$. The physical interpretation is that $k_s > 0$ corresponds to the ``energy'' of the outgoing on-shell third particle, while $\omega_1 + \omega_2 < 0$ would correspond to particles 1 and 2 incoming: there is therefore a singularity whenever $\omega_{12} < - k_s$ since the total ingoing energy is enough to create the on-shell outgoing particle.

\subsection{All contact 4-point correlators}
\label{sec:4-pt}

Next we focus on the 4-point function. 
Our main observation here is that the on-shell 4-point correlator can be written in terms of $\hat{s} = \hat{s}_{12}$ and $\hat{t} = \hat{s}_{13}$ only, and in the flat-space limit these become the usual Mandelstam $s$ and $t$ variables.   
The argument goes as follows: 
\begin{itemize}
	
	\item The spacetime symmetries require that interactions are built from covariant contractions of $\nabla_\mu$, and hence $\hat{A}_\tau$ is a function of $\hat{s}_{ab} (\tau)$ and $\hat{p}_a^2 ( \tau)$ only. 
	
	\item The Ward identities \eqref{eqn:DA_vanishes} and \eqref{eqn:KA_vanishes} (namely $\sum_{j=1}^n \hat{p}_j^\mu (\tau) = 0$) further require that
	\begin{align}
		\hat{s}_{34} ( \tau)  = \hat{s}_{12} ( \tau) \;\; , \;\;  \hat{s}_{24} ( \tau) = \hat{s}_{13} ( \tau) \;\; ,  \;\; \hat{s}_{14} ( \tau) = \hat{s}_{23} ( \tau) \; ,
	\end{align} 
	once commuted to the outermost operation, and also that\footnote{
 \eqref{eqn:stu} is the usual relation $s+t+u = 4m^2$ on Minkowski. 
    },
	\begin{align}
		\hat{s}_{12} ( \tau) + \hat{s}_{13} (\tau) + \hat{s}_{23} (\tau ) = - \sum_{a=1}^4 \hat{p}_a^2  \; , 
  \label{eqn:stu}
	\end{align}
	which ensures that only $\{ \hat{s}_{12} ( \tau ), \hat{s}_{13} (\tau ) , \hat{p}_1^2 ( \tau ) , \hat{p}_2^2 (\tau ) , \hat{p}_3^2 (\tau ), \hat{p}_4^2 (\tau) \}$ are independent. 	
	
	\item The on-shell relation \eqref{eqn:p2_onshell} can then be used to replace the four $\hat{p}_a^2 ( \tau )$ with the particle masses and (after commuting the $\hat{\mathcal{J}}^2_a ( \tau)$ until they annihilate the mode functions) powers of $\hat{s}_{12} (\tau )$ and $\hat{s}_{13} ( \tau )$. 
	
\end{itemize}
\noindent The most general contact contribution to the 4-point $S$-matrix can therefore be written as
\begin{align}
 \G_4 ( p_1, ... , p_4 ) =  \sum_{a, b = 0} c_{ab} \, \hat{t}^b \hat{s}^a \; \G^{\rm con}_n ( \omega_T  ) \; , 
 \label{eqn:G4cont}
\end{align}
since any $\hat{s}_{13}^b ( \tau ) \hat{s}_{12}^a (\tau )$ term in $\hat{A}_{\tau}$ becomes $\hat{s}_{13}^b \hat{s}_{12}^a$ in $A$, plus lower-order invariants (generated by the commutations of $\hat{\mathcal{J}}_{12}^2$ and $\hat{\mathcal{J}}_{13}^2$) which can be absorbed into a shift of the other $c_{a'b'}$ coefficients.

\paragraph{Evaluating the Mandelstams.}
To explicitly evaluate the terms in \eqref{eqn:G4cont}, recall that the operator $\hat{p}^2$ for a single particle takes a particularly simple form when expressed in terms of the angular variable $\omega = k \cosh \rho$. 
There is an analogous simplification that occurs for the $n$-particle Mandelstam operators. 
Choosing a subset of $n$ energies and parametrising their total as $\sum_{a=1}^n \omega_a  = | \sum_{a = 1}^n \bfk_a | \cosh \rho$, we can write a simple expression for \eqref{eqn:shat_def_3},
\begin{align}
- \hat{s}_{12...n}  e^{-i \omega_T \tau} =  \left( 
\frac{1}{  \text{sinh}^{2 j_n} \rho } \partial_\rho  \; \text{sinh}^{ 2 j_n} \rho  \; \partial_\rho  + j_n^2 - \frac{d^2}{4}   
\right) e^{- i \omega_T \tau} \; . 
\label{eqn:shat_rho_def}
\end{align}
where $j_n = \frac{d}{2} (n-1)$ and $\partial_\rho$ acts on everything to its right. 

Using \eqref{eqn:shat_rho_def} to represent $\hat{s}_{12}$ as an angular derivative makes the evaluation of all $\hat{s}_{12}^n$ invariants particularly straightforward. In fact, starting from the expression, 
\begin{align}
\hat{s}_{12}^n  \G^{\rm con}_4   =  \left( -\partial_{\rho_{12}}^2 - d \coth \rho_{12} \, \partial_{\rho_{12}} \right)^n  \G^{\rm con}_4
\label{eqn:shat_def_5}
\end{align}
it is possible to derive a simple closed-form expression for these invariants in terms of the angles,
\begin{align}
\cosh \rho_{12} = \frac{\omega_{12}}{k_{s}}  \;\;\;\; \text{and} \;\;\;\; \cosh \rho_{34} = \frac{\omega_{34}}{k_{s}} \; .   
\label{eqn:rho12_def} 
\end{align}
To do this, we first note that (as is often the case when dealing with rotational Casimirs) it is more convenient to replace $\hat{s}_{12}^n$ with the shifted product, 
\begin{align}
 \hat{S}_{12}^n = \prod_{j=0}^{n-1} \left( -\hat{s}_{12} - j (j + d ) \right) \; . 
 \label{eqn:Sn_def}
\end{align}
These invariants are particularly simple when written in terms of $2 \rho_+ = \rho_{12} + \rho_{34}$ and $2 \rho_- = \rho_{12} - \rho_{34}$, and take the form of an $(d + n)^{\rm th}$ order polynomial in $1/\cosh^2 \rho_+$ and $1/\cosh^2 \rho_-$. In fact, this polynomial can be written for general $n$ in terms of a hypergeometric function,
\begin{align}
 \hat{S}^n_{12}  \G^{\rm con}_4 = \frac{ i \, \Gamma \left( d + 2 n\right)}{  \left( 2 k_{s} \cosh \rho_+ \cosh \rho_- \right)^{d} } \; \frac{1}{ \left( - 4 \cosh^{2} \rho_+ \right)^n} {}_2 F_1 \left( -n , \frac{d}{2} ; 1 - n - \frac{d}{2} ;  \frac{\cosh^2 \rho_+}{ \cosh^2 \rho_- } \right) 
\label{eqn:sn_invariant_1}
\end{align}

Since $\hat{s}_{12} \G^{\rm con}_4 = \hat{s}_{34}  \G^{\rm con}_4$ (thanks to momentum conservation), we expect that $\hat{s}_{12}^n \G^{\rm con}_4$ should be a symmetric function of $\omega_{12}$ and $\omega_{34}$. 
This symmetry is not manifest in \eqref{eqn:sn_invariant_1}. 
To make it manifest, note that any symmetric polynomial of $1/\cosh^2 \rho_+$ and $1/\cosh^2 \rho_-$ can be written as a polynomial in the two symmetric variables,
\begin{align}
 -4 X =  \frac{1}{ \cosh^2 \rho_+} + \frac{1}{ \cosh^2 \rho_-} \; , \;\;\; 4 Y^2 = \frac{1}{  \cosh^2 \rho_+ \cosh^2 \rho_-} \; . 
\end{align}
Using these, we can re-write \eqref{eqn:sn_invariant_1} as,
\begin{align}
 \hat{S}^n_{12} \G^{\rm con}_4 = \frac{ i\,  \Gamma \left( d + 2 n \right) }{ k_{s}^d } Y^d X^n \;
 {}_2 F_1 \left(  -\frac{n}{2} , \frac{1-n}{2} ; 1 - n - \frac{d}{2}  ; \frac{Y^2}{X^2}  \right)
 \label{eqn:sn_invariant_2}
\end{align}
which is now manifestly invariant under $\omega_{12} \leftrightarrow \omega_{34}$. 
Analogous expressions for the $\hat{s}_{13}^n  \G^{\rm con}_4$ terms follow immediately from relabelling $p_2 \leftrightarrow p_3$.  

The change of variables \eqref{eqn:rho12_def} requires $\omega_{12} \geq k_{s}$ and $\omega_{34} \geq k_{s}$.  In this kinematic region, we have found a simple expression for any $\hat{s}_{12}^n \G^{\rm con}_4$. This would be enough to find the $\SBD_{0 \to 4}$ matrix element, since if $\omega_1 \geq k_1$ and $\omega_2 \geq k_2$ then $\omega_{12} \geq k_{12}$ by the triangle inequality. 
Other kinematic regions can be reached by crossing, i.e. replacing $\rho_{\pm}$ with $\omega_{12}$ and $\omega_{34}$ and then analytically continuing past the branch points at $\omega_{12} = k_{s}$ and $\omega_{34} = k_{s}$. 

\paragraph{Mandelstam eigenbasis.}
Constructing closed form expressions for the mixed terms involving both $\hat{s}_{12}$ and $\hat{s}_{13}$ is more involved. 
The reason is essentially that, since $\hat{s}_{12}$ and $\hat{s}_{13}$ do not commute, it is not possible to find simultaneous eigenfunctions of both operators.
Focussing on the $\hat{s}_{12}^n \G_4^{\rm con}$ terms, \eqref{eqn:sn_invariant_1} can be reproduced very efficiently using the $\hat{s}_{12}$ eigenfunction found in \eqref{eqn:F2_def}.
The trick is to recognise that these eigenfunctions form a complete basis for the $\omega_{12}$ dependence of $\G_4^{\rm con}$, which can therefore be written as\footnote{
The overall $i$ is included so that $\mathcal{N}_\nu$ is positive, given our normalisation of $F_{j_2}$---see Appendix~\ref{app:mode}. This can also be viewed as $\int d \Delta$ over the principal series $\Delta = \frac{d}{2} + i \nu$ with $\nu \geq 0$. 
}
\begin{align}
 \G_4^{\rm con} ( \omega_T) = \frac{i}{ ( i k_s )^{d} } \int_0^{\infty} d \nu \, \mathcal{N}_\nu \;  F_{j_2} \left(  \frac{\omega_{12}}{k_s} , \nu  \right)   F_{j_2} \left(  \frac{\omega_{34}}{k_s} , \nu  \right)  \,  
 \label{eqn:G4con_spec_rep}
\end{align}
for some density $\mathcal{N}_\nu$.
Physically, this decomposition corresponds to inserting a complete set of mass eigenstates like so: 
 \begin{align}
     \begin{tikzpicture}[scale=0.6,baseline=0.0cm]
			\draw (-2,1) -- (-1,0);
			\draw (-2,-1) -- (-1,0);
			\draw (0,1) -- (-1,0);
			\draw (0,-1) -- (-1,0);
                \draw (-2,1) node [anchor=east][inner sep=2.0pt] {$\omega_1$};
                \draw (-2,-1) node [anchor=east][inner sep=2.0pt] {$\omega_2$};
                \draw (0,1) node [anchor=west][inner sep=2.0pt] {$\omega_3$};
			\draw (0,-1) node [anchor=west][inner sep=2.0pt] {$\omega_4$};
                \draw (-1,-0.1) node [anchor=north][inner sep=3.0pt] {\small \color{gray} $\phi^4$};
        \end{tikzpicture} 
&= 
 \int_\nu  \; \left( 
     \begin{tikzpicture}[scale=0.6,baseline=0.0cm]
			\draw (2,0) -- (1,0);
			\draw (0,1) -- (1,0);
			\draw (0,-1) -- (1,0);
                \draw (2,0) node [anchor=west][inner sep=2.0pt] {$\nu$};
                \draw (0,1) node [anchor=east][inner sep=2.0pt] {$\omega_1$};
			\draw (0,-1) node [anchor=east][inner sep=2.0pt] {$\omega_2$};
                \draw (1.1,-0.1) node [anchor=north][inner sep=3.0pt] {\small \color{gray} $\phi^3$};
        \end{tikzpicture} 
        \right) \left( 
           \begin{tikzpicture}[scale=0.6,baseline=0.0cm]
			\draw (-2,0) -- (-1,0);
			\draw (0,1) -- (-1,0);
			\draw (0,-1) -- (-1,0);
                \draw (-2,0) node [anchor=east][inner sep=2.0pt] {$\nu$};
                \draw (0,1) node [anchor=west][inner sep=2.0pt] {$\omega_3$};
			\draw (0,-1) node [anchor=west][inner sep=2.0pt] {$\omega_4$};
                \draw (-1.1,-0.1) node [anchor=north][inner sep=3.0pt] {\small \color{gray} $\phi^3$};
        \end{tikzpicture}         
        \right)
\end{align}
The model-independent $\mathcal{N}_\nu$ is given explicitly by $\mathcal{N}_\nu = \tfrac{2}{\pi} \nu \sinh ( \pi \nu )$ and is notionally a density of states for the de Sitter group.
In this basis of two-particle mode functions, the action of $\hat{s}_{12}$ then becomes multiplication by the effective mass parameter $\nu^2 + \tfrac{d^2}{4}$, 
\begin{align}
  \hat{s}_{12}^n \G_4^{\rm con} ( \omega_T )  = \frac{i}{ ( i k_s )^{d} } \int_0^{\infty} d \nu \, \mathcal{N}_\nu \,  F_{j_2} \left(  \frac{\omega_{12}}{k_s} , \nu  \right)   F_{j_2} \left(  \frac{\omega_{34}}{k_s} , \nu  \right)  \left(  \nu^2 + \tfrac{d^2}{4} \right)^{n} \; . 
\end{align}
Evaluating this integral produces \eqref{eqn:sn_invariant_1}.
The symmetry $\hat{s}_{12} \G_4^{\rm con} = \hat{s}_{34} \G_4^{\rm con}$ is also now manifest.

\paragraph{Separating kinematics and dynamics.}
Finally, this two-particle basis has the advantage that, when we perform the LSZ integral to project this $\G_4$ onto mass eigenstates, there is a clear split between the \emph{dynamical} information (the interaction) and the \emph{kinematical} propagation of the asymptotic states. 
In particular, one may write:
\begin{align}
\text{LSZ} \left[ \chi \left( \hat{s}_{12} \right) \G_4^{\rm con} ( \omega_T ) \right] = 
\frac{ i }{ ( i k_s )^{d} } \int_0^{\infty} d \nu \, \mathcal{N}_\nu \,  \left( 
\col{\mu_1}{\bfk_1} \col{\mu_2}{\bfk_2} \col{\nu}{ \bfk_{34} }
\right)  \left( 
\col{\mu_3}{\bfk_3} \col{\mu_4}{\bfk_4} \col{\nu}{ \bfk_{12} }
\right) \chi \left( \nu^2 + \tfrac{d^2}{4} \right)
\label{eqn:chi_def}
\end{align}
where we have defined a bracket analogous to the $3j$ symbol from angular momentum addition, which captures how two one-particle mass eigenstates combine to form a single two-particle mass eigenstate,
\begin{align}
  \left( 
\col{\mu_1}{\bfk_1} \col{\mu_2}{\bfk_2} \col{\nu}{ \bfk_{34} }
\right) 
= \left[  \pi   
 \int_{-\infty}^{+\infty} \frac{d^2 \rho}{ (2\pi)^2 } \; e^{i \mu_1 \rho_1 } e^{ i \mu_2 \rho_2} \right]  
  F_{j_2} \left(  \frac{\omega_{12}}{k_s} , \nu  \right)  \Bigg|_{\omega_{12} = k_1 \cosh \rho_1 + k_2 \cosh \rho_2 } \; .  
  \label{eqn:3j}
\end{align}
Crucially, this is a model-independent structure constant, uniquely fixed by the de Sitter symmetries. Once these factors are stripped from the $S$-matrix element, the residual density $\chi \left( \nu^2 + \tfrac{d^2}{4} \right)$ captures the derivative interaction vertex that mediates the scattering. 
In pictures:
 \begin{align}
     \begin{tikzpicture}[scale=0.6,baseline=0.0cm]
			\draw (-2,1) -- (-1,0);
			\draw (-2,-1) -- (-1,0);
			\draw (0,1) -- (-1,0);
			\draw (0,-1) -- (-1,0);
                \draw (-2,1) node [anchor=east][inner sep=2.0pt] {$\mu_1$};
                \draw (-2,-1) node [anchor=east][inner sep=2.0pt] {$\mu_2$};
                \draw (0,1) node [anchor=west][inner sep=2.0pt] {$\mu_3$};
			\draw (0,-1) node [anchor=west][inner sep=2.0pt] {$\mu_4$};
        \end{tikzpicture} 
&= 
 \int_\nu \; \left( 
     \begin{tikzpicture}[scale=0.6,baseline=0.0cm]
			\draw (2,0) -- (1,0);
			\draw (0,1) -- (1,0);
			\draw (0,-1) -- (1,0);
                \draw (2,0) node [anchor=west][inner sep=2.0pt] {$\nu$};
                \draw (0,1) node [anchor=east][inner sep=2.0pt] {$\mu_1$};
			\draw (0,-1) node [anchor=east][inner sep=2.0pt] {$\mu_2$};
                   \draw (1.1,-0.1) node [anchor=north][inner sep=3.0pt] {\small \color{gray} $\phi^3$};
        \end{tikzpicture} 
        \right) \chi \left( \nu^2 + \tfrac{d^2}{4} \right) \left( 
           \begin{tikzpicture}[scale=0.6,baseline=0.0cm]
			\draw (-2,0) -- (-1,0);
			\draw (0,1) -- (-1,0);
			\draw (0,-1) -- (-1,0);
                \draw (-2,0) node [anchor=east][inner sep=2.0pt] {$\nu$};
                \draw (0,1) node [anchor=west][inner sep=2.0pt] {$\mu_3$};
			\draw (0,-1) node [anchor=west][inner sep=2.0pt] {$\mu_4$};
                \draw (-1.1,-0.1) node [anchor=north][inner sep=3.0pt] {\small \color{gray} $\phi^3$};
        \end{tikzpicture}         
        \right)
        \label{eqn:chi_picture}
\end{align}
where now the labels denote the mass of each leg.
For example, from the effective Lagrangian
\begin{align}
 \mathcal{L}_{\rm int} = \phi^2 \left( \lambda_0 + \lambda_1 \frac{\Box}{\Lambda^2} + \frac{\Box^2}{\Lambda^4} + ... \right) \phi^2
\end{align}
we can immediately read off:
\begin{align}
 \chi ( m^2 ) = \lambda_0 + \lambda_1 \frac{m^2}{\Lambda^2} + \lambda_2 \frac{m^4}{\Lambda^4} + ...
\end{align}
The on-shell $S$-matrix and corresponding cosmological correlators in this theory will be complicated functions of $\{ \bfk_1, ... \bfk_4 \}$, but \eqref{eqn:chi_picture} makes it clear that these complications stem entirely from how we treat the kinematical combination and propagation of free fields on de Sitter. 

\paragraph{Beyond 4-particle scattering.}
The above generalises straightforwardly to contact diagrams with any number of particles. 
In particular, we can find the ``$n$-particle mode function'' by considering $\G_{n+1}$ with a single leg put on-shell,
\begin{align}
\left[ - \sqrt{\pi} \int_{-\infty}^{+\infty} \frac{d \rho}{2\pi} e^{i \mu \rho }   \right]  \G^{\rm con}_{n+1} ( \omega_T + \omega  ) \big|_{\omega = | \bfk_{T} | \cosh \rho } =  \frac{  F_{j_n} \left( \frac{\omega_{T}}{ | \bfk_{T} | } , \mu \right) }{ ( i | \bfk_{T} | )^{ j_n } }
\end{align}
where $\omega_T = \sum_{b=1}^n \omega_n$ and $\bfk_T = \sum_{b=1}^n \bfk_b$. This is the eigenfunction of $\hat{s}_{12...n}$, since when acting on $\G_{n+1}^{\rm con}$ we have $\hat{s}_{12...n} = -\hat{p}_{n+1}^2 \approx  \mu^2 + \tfrac{d^2}{4}$ when on-shell. 
A useful corollary is that,
\begin{align}
    \frac{  F_{j_n} \left(  \frac{\omega}{ k } , \mu \right) }{ ( i k )^{ j_n } }
    =
    \left[ \int_{-\infty}^{0} \frac{ d \tau }{  -\tau  } e^{i \omega \tau }   \right]  ( - \tau )^{j_n}  i f^{+} ( k \tau , \mu )  \; 
    \label{eqn:f_to_F}
\end{align}
which illustrates how the $n$-particle mode function is related to the time-domain mode function.  
These $n$-particle mode functions can be used to decompose any correlator and are particularly useful for correlators that depend on only a single Mandelstam operator.
Suppose we split the $n$ particles into a subset of size $n_1$ and its complement (of size $n_2 
= n -n_1$), with partial energies $\omega_L$ and $\omega_R$ respectively (so $\omega_L + \omega_R = \omega_T$) and a partial momentum $\bfk$ (i.e. $| \bfk_L | = | \bfk_R | = k$). 
Then, 
\begin{align}
 \G_n^{\rm con} ( \omega_T )  =  \frac{i}{ ( i k )^{J_n}} \int_0^{\infty} d \nu \, \mathcal{N}_\nu \, F_{j_{n_1}} \left( \frac{\omega_{L}}{k} , \nu \right) F_{j_{n_2}} \left( \frac{\omega_{R}}{k} , \nu \right) \; . 
 \label{eqn:Gn_spec_rep}
\end{align}
This representation makes it straightforward to treat any interaction that contains only a single Mandelstam operator, for instance:
\begin{align}
\hat{s}_{1...n_1}^a \G_n^{\rm con} ( \omega_T )  = \frac{ i }{ ( i k )^{J_n}} \int_0^{\infty} d \nu \, \mathcal{N}_\nu \, F_{j_{n_1}} \left( \frac{\omega_{L}}{k} , \nu \right) F_{j_{n_2}} \left( \frac{\omega_{R}}{k} , \nu \right)  (  \nu^2 + \tfrac{d^2}{4} )^{a} \; . 
\end{align}

~\\

To sum up, we have defined a differential operator $\hat{p}_\mu$ that captures the insertion of $\nabla_\mu$ into any contact interaction vertex, and found a complete basis of invariant Mandelstam operators in which to expand $\G_n$. These operators do not commute (because $\nabla_\mu$ do not commute), but for interactions that depend on only a single Mandelstam invariant we can expand the resulting correlator in a basis of the corresponding eigenfunction and this leads to very simple integral expressions and a natural way to separate the dynamical and kinematical information contained within the $S$-matrix elements.

\section{New representations of exchange diagrams}
\label{sec:spectral}

Now we turn our addition to exchange diagram contributions to the amputated correlator / off-shell $S$-matrix. 
We will start by considering the simplest diagram from two cubic vertices with no derivatives, which is given in time-ordered perturbation theory by
\begin{align}
   \left[ \int_{-\infty}^0 \frac{d \tau_L \, e^{i \omega_{12} \tau_L}}{ ( - \tau_L)^{1 - j_2}}  \right] \left[ \int_{-\infty}^0 \frac{d \tau_R \, e^{i \omega_{34} \tau_R} }{ ( - \tau_R )^{1 - j_2}}  \right]  G_2^{\mu_\sigma} \left( k_{s} \tau_L , k_{s} \tau_R  \right)  \equiv   \G_4^{\mu_{\sigma}} \left( \omega_{12}, \omega_{34}, k_s  \right)
   \label{eqn:G4exch_def}
\end{align}
where $G_2$ is the propagator of the internal field which has mass $m_\sigma^2 = \mu_\sigma^2 + d^2/4$. 
As argued above, the tree-level exchange contribution to $\G_4$ from \emph{any} pair of cubic interactions can be written in terms of this exchange structure, plus a remainder that is indistinguishable from a contact interaction. 
It can already be anticipated, from the results of the previous section, that this exchange structure ought to correspond to
\begin{align}
 \chi ( m_\sigma^2 ) \sim \frac{1}{ \nu^2 - \mu_\sigma^2 } \; 
\end{align}
in \eqref{eqn:chi_def}. 
We will now derive this result, show how it relates to previous results in the literature for the exchange diagram, and for the first time determine a consistent $+i \epsilon$ prescription which encodes the causal time-ordering of the $S$-matrix.

\subsection{Exchange differential equation}

Equation \eqref{eqn:G4exch_def} can be viewed as a Fourier transform of the propagator, replacing the conformal time of each vertex with the total conformal energy flowing into that vertex from the boundary. 
Since the propagator obeys the simple differential equation~\eqref{eqn:G2_eom}, one might ask what differential equation is satisfied by this Fourier transform? 
Using the relation
\begin{align}
 \left[ \int_{-\infty}^0 \frac{d \tau_L}{ ( -\tau_L )^{1-j_2} } \,  e^{i \omega_{12} \tau_L} \right]  \mathcal{E}_{\mu_\sigma} \left[ k_s \tau_L  \right] = \left( m_\sigma^2 - \hat{s}_{12} \right)  \left[ \int_{-\infty}^0 \frac{d \tau_L}{ ( -\tau_L )^{1-j_2} } \,  e^{i \omega_{12} \tau_L} \right] \, ,
\end{align}
we see that the action of $( m_\sigma^2 - \hat{s}_{12} )$ is to replace the internal $G_2$ with a $\delta$ function, effectively collapsing the internal line. 
The 4-particle exchange structure, $\G_4^{\mu_\sigma}$, therefore satisfies the defining relation:
\begin{align}
    \left( m^2_\sigma - \hat{s}_{12} \right)  \G_4^{\mu_\sigma}   =  - \G^{\rm con}_4 \left(  \omega_T \right)  \; 
 \label{eqn:exch_eqn}
\end{align}
which we refer to as ``the exchange equation''. 

\paragraph{Conformally coupled example.}
As usual, the simplest example is the exchange of a scalar with conformally coupled mass, $m^2_\sigma = 2$ in $d=3$ (so $\mu_\sigma = i/2$). 
In that case the propagator is simple,
\begin{align}
 G_2^{\mu_\sigma = i/2} ( k \tau , k \tau'  ) =  \frac{ e^{ - i k | \tau - \tau' |}   }{2 k \sqrt{\tau \tau'} } 
\end{align}
and the time integrals in \eqref{eqn:G4exch_def} can be performed explicitly, giving the correlator:
\begin{align}
\G_4^{\mu_\sigma = i/2}   =  - \frac{  \omega_T + 2 k_s  }{ 2  k_s  \omega_T ( \omega_{12} + k_s ) ( \omega_{34} + k_s ) } \; . 
\end{align}
This indeed satisfies the exchange equation \eqref{eqn:exch_eqn}. 
Notice that it has so-called ``partial energy'' singularities at $\omega_{12} = -k_s$ and $\omega_{34} = -k_s$, as well as a total energy singularity at $\omega_T = 0$. 
The residue of the total energy pole is the Minkowski 4-point amplitude,
\begin{align}
 \lim_{\omega_T \to 0}  \omega_T  \, \G_4^{\mu_\sigma = i/2}  = \frac{1}{s} \;\; \text{ with }\;\; s = \omega_{12}^2 - k_s^2
\end{align}
and the correlator factorises on the partial energy poles into a product of Minkowski and de Sitter 3-point amplitudes
\begin{align}
 \lim_{\omega_{12} \to -k_{s}} \left(  \omega_{12} + k_s \right)  \G_4^{\mu_\sigma = i/2}   = 
 \frac{ 1 }{ 2 k_s ( k_s - \omega_{34} )  } \; . 
 \label{eqn:cc_factor}
\end{align}
This turns out to be a general feature: as $\omega_{12} + k_s \to 0$, the corresponding vertex becomes long-lived and persists to very early times where the spacetime curvature is unimportant and thus it produces the Minkowski 3-point amplitude (which in this case is just a constant).
See the cartoon in Figure~\ref{fig:factorisation}. 
Notice that in terms of the 2-partlce mode function \eqref{eqn:F2_def}
\begin{align}
  F_{j_2} \left( \frac{\omega}{k} , \mu_\sigma  = \frac{i}{2} \right)  = \frac{k}{ \sqrt{2} ( \omega_{12} + k ) }
\end{align}
the limiting behaviour \eqref{eqn:cc_factor} can be written as
\begin{align}
\lim_{\omega_{12} \to - k_s}  i \G_4^{ \mu_\sigma} = \frac{1}{  ( i k_s )^3 } F_{j_2} \left( \tfrac{\omega_{12}}{k_s} , \mu_\sigma \right) F_{j_2}^* \left( -\tfrac{\omega_{34}}{k_s} , \mu_\sigma \right) \; .  
\label{eqn:cc_factorisation}
\end{align}
This factorisation provides the required boundary condition for solving the exchange equation \eqref{eqn:exch_eqn}, leading to the exchange bootstrap we develop in section \ref{sec:bootstrap} along the lines of \cite{Arkani-Hamed:2018kmz}.
Ultimately, this boundary condition is equivalent to choosing a particular $i \epsilon$ prescription when inverting the $( m_\sigma^2  - \hat{s}_{12} )$ operator, much like Feynman's $1/(m^2 - s - i \epsilon)$ from the causal propagator on Minkowski. 

\paragraph{Large mass expansion.}
One approach to solving the exchange equation \eqref{eqn:exch_eqn} for general masses is to expand in large $m_\sigma$, giving a power series solution,
\begin{align}
  \G_4^{\mu_\sigma}  = - \sum_{n=0}^\infty \frac{ \hat{s}_{12}^n  }{ m_{\sigma}^{2+2n}}  \G_{4}^{\rm con} ( \omega_T )  
\label{eqn:G4exch_large_m}
\end{align}
which is analogous to the usual Minkowski expansion.
This has the advantage that no $+i \epsilon$ prescription is required, i.e. any choice of propagator will produce the same expansion since the large $m_\sigma$ limit is dominated by space-like momenta (so there is no operator-ordering ambiguity to resolve). 
The propagator $+i \epsilon$ only becomes necessary when one resums the series \eqref{eqn:G4exch_large_m} (i.e. analytically continues beyond the radius of convergence at $| \hat{s}_{12} | \sim m_\sigma^2$).   

As discussed above, the invariant $\hat{s}_{12}^n$ is can be simplified by introducing the shifted products $\hat{S}_{12}^n$ defined in \eqref{eqn:Sn_def}.
In terms of these, the exchange invariant can be expanded as,
\begin{align}
  \G_4^{\mu_\sigma} = - \sum_{n=0}^{\infty} \frac{\hat{S}_{12}^n \G^{\rm con}_4 }{ m_\sigma^2 ( 1 + \frac{d}{2} + i \mu_\sigma )_n ( 1 + \frac{d}{2} - i \mu_\sigma )_n } \; . 
 \label{eqn:exch_large_mu}
\end{align}
Using \eqref{eqn:sn_invariant_1} or \eqref{eqn:sn_invariant_2}, this is an infinite sum over hypergeometric ${}_2 F_1$'s, which can be written in terms of a generalised hypergeometric function of two variables (a Kamp\'{e} de F\'{e}riet function) if so desired. 

\subsection{Propagator and the $i \epsilon$ prescription}

To determine the correct $+i\epsilon$ prescription and prove that the factorisation observed above for conformally coupled fields is indeed a general phenomenon, we now turn our attention to the propagator $G_2$ appearing in \eqref{eqn:G4exch_def}. 
Inspired by our previous integral representation of the contact correlators \eqref{eqn:chi_def}, notice that the defining differential equation for the propagator~\eqref{eqn:G2_eom} is formally solved by
\begin{align}
 i  G_2^{\mu}  \left(  k \tau_1,  k \tau_2  \right)
 =
 \int_0^{\infty} d \nu \, \mathcal{N}_\nu   \; \frac{ f^+ ( k \tau_1 , \nu) f^+ (k \tau_2 , \nu) }{  \nu^2 - \mu^2  } \;  \qquad \text{(de Sitter)} \;
 \label{eqn:G2_t_spec_rep}
\end{align}
where $\mathcal{N}_\nu  = \frac{2}{\pi} \, \nu \sinh ( \pi \nu)$ is the de Sitter density of states and the $f^{\pm}$ mode functions are given in \eqref{eqn:fmode}. This is the straightforward analogue of the familiar Minkowski result
\begin{align}
i G_k (t_1, t_2)  = \int_{-\infty}^{+\infty} \frac{d \omega}{2\pi } \, \frac{ e^{i \omega ( t_1 - t_2 ) } }{ \omega^2 - k^2 - m^2 } \;   \qquad \text{(Minkowski)} \; . 
 \label{eqn:G2_Mink}
\end{align}

\paragraph{Boundary conditions.}
However, \eqref{eqn:G2_t_spec_rep} and \eqref{eqn:G2_Mink} are not well-defined as written because we must specify how to integrate past the poles on the real axis. On Minkowski, whether we deform the integration contour above or below these poles corresponds to specifying Feynman, retarded or advanced boundary conditions for the propagator. 
For the Bunch-Davies $S$-matrix on de Sitter, we need a prescription that selects the time-ordered boundary condition:
\begin{align}
 G_2^\mu ( k \tau_1 , k \tau_2 ) &=  {}_{\rm out} \langle 0 | T \hat{\varphi} ( \tau_1 , \bfk ) \hat{\varphi} ( \tau_2 , - \bfk )   | 0 \rangle'_{\rm in}   \\ 
 &=  f^- ( k \tau_1 , \mu ) f^+ ( k \tau_2 , \mu ) \Theta ( \tau_1 - \tau_2 )   +  f^+ ( k \tau_1 , \mu ) f^- ( k \tau_2 , \mu ) \Theta ( \tau_2 - \tau_1 ) \nonumber 
\end{align}
in the free theory. 
The $i \epsilon$ prescription that implements this Bunch-Davies boundary condition is
\begin{align}
\frac{1}{ ( \nu^2 - \mu^2 )_{i\epsilon} }  =  \lim_{\epsilon \to 0^+} \frac{1}{2 \sinh ( \mu \pi )}  \left( \frac{e^{+ \mu \pi}}{\nu^2 - \mu^2 + i \epsilon} -  \frac{e^{-\mu \pi}}{\nu^2 - \mu^2 - i \epsilon} \right) \; . 
\label{eqn:our_prescription}
\end{align}

\paragraph{Derivation.}
Let's consider $\tau_1$ to be to the future of $\tau_2$, so that the desired $G_2^\mu ( k \tau_1, k \tau_2 ) = f^- ( k \tau_1 , \mu ) f^+ ( k\tau_2 , \mu ) $. 
To show that \eqref{eqn:G2_t_spec_rep} indeed produces this, we first use the identity \eqref{eqn:Hankel_to_Bessel}
to write the integral as
\begin{align}
  i  G_2^\mu  \left(  k \tau_1,  k \tau_2  \right) = \lim_{\epsilon \to 0^+} \frac{1}{2}  \int_{-\infty}^{+\infty}  d \nu \frac{ \nu J_{i \nu} ( - k \tau_1 ) H^{(2)}_{i \nu} ( - k \tau_2 )  }{ ( \nu^2 - \mu^2 ) _{i \epsilon} } \; . 
\end{align}
The numerator $\nu J_{i \nu} ( - k \tau_1 ) H^{(2)}_{i \nu} ( - k \tau_2 )$ is analytic in the complex $\nu$ plane. It was shown in \cite{10.1093imamat26.2.133, GUTIERREZTOVAR2007359} that such integrals can be closed in the lower half of the complex $\nu$ plane when $- k \tau_2 > - k \tau_1$, since
\begin{align}
\nu  J_{i \nu} ( z_1 ) H_{i \nu}^{(2)} (z_2 ) &\sim \frac{ i \, e^{ + i \nu \log \left( z_1 /  z_2  \right) } }{  \pi  } \; . 
\end{align}
By Cauchy's residue theorem, the closed contour integral of the $2 \nu/(\nu^2 - \mu^2 \pm i \epsilon)$ term then contributes a pole residue at $\nu = \pm \mu$. 
This gives
\begin{align}
i G_2^\mu  \left(  k \tau_1,  k \tau_2  \right) &= \frac{ 2 \pi i }{8 \sinh ( \mu \pi ) } \left(    
 e^{+ \mu \pi}  J_{i \mu} ( - k \tau_1 ) H^{(2)}_{i \mu} ( - k \tau_2 )
 - 
e^{- \mu \pi}  J_{-i \mu} ( - k \tau_1 ) H^{(2)}_{- i \mu} ( - k \tau_2 ) 
\right) \nonumber \\ 
&= i f^- ( k \tau_1, \mu ) f^+ ( k \tau_2, \mu )
\end{align}
as desired, since $H^{(2)}_{-i \mu} ( z ) = e^{+ \mu \pi} H^{(2)}_{+i \mu} (z)$. 
The other case, $\tau_1 < \tau_2$, can be tackled in the same way (by replacing $f^+ ( k\tau_2, \nu)$ with a Bessel $J_{i \nu}$ and then closing the contour).

\paragraph{Particle production.}
At first sight, \eqref{eqn:our_prescription} may seem surprising. 
Since the Bunch-Davies $S$-matrix reduces to the usual Minkowski $S$-matrix in the flat space limit, one may have expected simply the usual Feynman prescription $1/( \nu^2 - \mu^2 + i \epsilon)$. However, note that since $\mu \sim m/H \to \infty$ in the Minkowski limit $H \to 0$, \eqref{eqn:our_prescription} \emph{does} reduce to the usual prescription. The appearance of the additional term at finite $H$ is physically related to particle production. 
In fact, we show in Appendix~\ref{app:mode} that using the naive prescription $1/(\nu^2 - \mu^2 + i\epsilon)$ produces the propagator of the \emph{Unruh-de Witt} $S$-matrix: this is the basis of asymptotic states in which particle production appears explicitly as off-diagonal $S$-matrix elements in the free theory \cite{Melville:2023kgd}. 
By contrast, the Bunch-Davies $S$-matrix is simply the identity in the free theory, in which case particle production is accounted for in the definition of the asymptotic states. We see in \eqref{eqn:our_prescription} that this choice of asymptotic states introduces an additional $- i \epsilon$ term into the propagator, which is accounting for particle production effects.

\paragraph{Crossing.}
Another feature of our split representation \eqref{eqn:G2_t_spec_rep} which is perhaps unexpected (and seems to differ from the Minkowski result) is that we use $f^+ f^+$ rather than say $f^+ f^-$.
This is surprising because the $S$-matrix for $2$-particle scattering in the free theory is
\begin{align}
& \langle 0 | \bfk_1 \bfk \rangle = \langle \bfk_1 \bfk_2 | 0 \rangle = 0 \; & \langle \bfk_2 | \bfk_1 \rangle &= (2 \pi )^d \delta \left( \bfk_1 - \bfk_2 \right) \; ,
\label{eqn:S2_free}
\end{align}
and so we might expect that $G_2$ should contain an $f^+$ for one outgoing particle and an $f^-$ for one ingoing particle. 
The resolution of this confusion is the following identity:
\begin{align}
 \int_0^\infty d \nu \, \mathcal{N}_\nu \; \frac{ f^+ ( k \tau, \nu )  }{ ( \nu^2 - \mu^2 )_{i \epsilon} } = i f^- ( k \tau, \nu ) \; 
\end{align}
which shows that the $i \epsilon$ prescription in \eqref{eqn:our_prescription} secretly implements a \emph{crossing transformation} from outgoing to ingoing---we derive this in Appendix~\ref{app:mode}.
As a result, once one of the two fields is projected onto an outgoing mass eigenstate via LSZ\footnote{
Notice that the derivation of \eqref{eqn:LSZ_intro} neglects contributions proportional to $\delta ( \bfk_b - \bfk_{b'} )$ from collinear particles: these correspond to disconnected diagrams \emph{except} in the case of the two-point function, where they give rise to the non-zero contribution responsible for \eqref{eqn:S2_free}.  
} 
\begin{align}
  - i f^- ( k \tau , \mu ) ( \overset{\leftrightarrow}{ \tau \partial_\tau} )  G_2^\mu ( k \tau, k \tau' ) =  - f^- ( k \tau', \mu )
  \label{eqn:G2_LSZ}
\end{align}
we see that projecting the remaining field (which now corresponds to an ingoing particle) onto a mass eigenstate indeed gives \eqref{eqn:S2_free}. 
In fact, this means that the continuation $k \to -k$ in \eqref{eqn:cross_kbar} may be implemented by instead integrating over all mass values:
\begin{align}
 \SBD_{0 \to n} |_{\bfk_b \to \bar{\bfk}_b }
 = \left[ \int_0^{\infty} d \nu \; \mathcal{N}_\nu \; 
 \frac{  - i   }{ (\nu^2 - \mu_b^2 )_{i \epsilon} }  \right] \SBD_{0 \to n} |_{ \substack{ \bfk_b \, \to - \bfk_b \\  \mu_b \, \to \; \nu \,\;~ }}
\end{align}
We note in passing the similarity with the so-called shadow transformation that interchanges $\frac{d}{2} \pm i \mu$ irreps in the principal series.
This can also be expressed as an integral over the propagator (see e.g. \cite{Karateev:2018oml} for a pedagogical introduction).
Here we are effectively describing a shadow transform for the Hankel mode functions, which interchanges ingoing and outgoing particles.

\paragraph{Energy-domain.}
Upon Fourier transforming each $\tau$ to $\omega$,  we arrive at an analogous split representation for the propagator:
\begin{align}
i G_2^\mu \left( \frac{\omega_1}{k} , \frac{\omega_2}{k}  \right) 
=
- \int_0^{\infty} d \nu  \; \mathcal{N}_\nu \,  \frac{ F_{j_1} \left( \frac{\omega_1}{k} , \nu \right) F_{j_1} \left( \frac{\omega_2}{k} , \nu \right)  }{ (\nu^2 - \mu^2 )_{i \epsilon} }
\end{align}
Notice that in the factorisation limit,  
\begin{align}
 \lim_{\omega_1 \to - k} i G_2^\mu \left( \frac{\omega_1}{k} , \frac{\omega_2}{k}  \right)  \propto  - \int_0^{\infty} d \nu \, \mathcal{N}_\nu \, \frac{ F_1 \left( \frac{\omega_2}{k} , \nu \right) }{ (\mu^2 - \nu^2 )_{\epsilon} } =  F_1^* \left( - \frac{\omega_2}{k} , \mu \right) \; ,
\label{eqn:F_cross}
\end{align}
which is the analogue of the LSZ reduction~\eqref{eqn:G2_LSZ}, since $F_1$ is the mode function for an ingoing mass eigenstate. Furthermore, the only branch point of $F_1 \left( - \frac{\omega_2}{k} , \mu \right)$ is at $\omega_2 = + k$---this is in contrast to other propagators (e.g. the bulk-to-bulk propagator or Schwinger-Keldysh propagators for wavefunction or in-in correlator) which also have branch points at $\omega_2 = - k$. We will see below that this is in fact a general property: the $S$-matrix (amputated correlator) may not have any $\omega_R \to -k$ singularity in the $\omega_L \to -k$ limit for any pair of partial energies $\omega_L$ and $\omega_R$. 


\paragraph{Position-domain.}
It turns out that \eqref{eqn:our_prescription} is the unique\footnote{
More precisely, \eqref{eqn:our_prescription} is the unique replacement of $1/(\nu^2 - \mu^2)$ which involves only $\mu$. Other prescriptions involving $\nu$ are possible, but these can change the analytic structure of $G_2$.
} way of shifting the poles at $\nu = \pm \mu$ infinitesimally into the complex $\nu$-plane which satisfies the following two conditions:
\begin{itemize}

\item[(i)] in the flat space limit (which sends $\mu \to \infty$), it recovers the usual Feynman prescription,

\item[(ii)] when continued from the expanding Poincar\'{e} patch to the whole of global de Sitter, it leads to a singularity only at coincident spacetime points (and those connected by null geodesics). In contrast, the naive prescription $1/(\mu^2 - \nu^2 + 
i \epsilon)$ is also singular at antipodal points (and any point connected to its antipode by a null geodesic).   
 
\end{itemize}
These are analogous to the two conditions used to select the Bunch-Davies vacuum from the general $\alpha$-vacuum of de Sitter \cite{Allen:1985ux}. 
To see this, transform the propagator for $\phi$ to $(\tau, \bfx)$ variables, 
\begin{align}
 G_2 ( \cosh \sigma ) =  \left[  \int \frac{d^d \bfk }{ (2\pi)^d}  e^{i \bfk \cdot ( \bfx_1 - \bfx_2 ) }\right] (- \tau_1 )^{d/2} ( - \tau_2 )^{d/2} G_2 ( k \tau_1 , k \tau_2 ) 
\end{align}
Thanks to the de Sitter symmetries, the result is a function of the invariant chordal distance,
\begin{align}
 \cosh \, \sigma  \equiv 1 + \frac{ ( \tau_1 - \tau_2 )^2  - | \bfx_1 - \bfx_2 |^2  }{2 \tau_1 \tau_2} \; . 
\end{align}
The desired Bunch-Davies propagator is, 
\begin{align}
 G_2 \left(  \cosh \sigma \right) 
&= \frac{1}{ 2^{\frac{d+3}{2}}  \pi^{\frac{d-1}{2}} \cosh ( \pi \mu )} \frac{ P^{\frac{d}{2} -\frac{1}{2}}_{i \mu - \frac{1}{2}} ( - \cosh \sigma )  }{ (  \sqrt{ \sinh \sigma } )^{\frac{d}{2} - \frac{1}{2} } } \; ,
\label{eqn:G2inin}
\end{align}
The only singularity in $G_2$ is at coincident points $x_1^\mu = x_2^\mu$ or at null separations, and the corresponding branch cut from $\cosh \sigma = 1$ to $\infty$ encodes the ordering ambiguity when $x_1$ and $x_2$ are time-like separated.
The Feynman prescription for a time-ordered product corresponds to approaching this cut from below. 
Using the crossing identity \eqref{eqn:F_cross} for the Legendre functions, we find that,
\begin{align}
i G_2 \left(  \cosh \sigma \right) 
&=  \int_0^{\infty} d \nu \,  \mathcal{N} ( \nu)  \frac{ G_2 ( - \cosh \sigma ) }{ \left( \nu^2 - \mu^2 \right)_{i\epsilon}  }
\end{align}
where $\cosh \sigma \to - \cosh\sigma$ sends $x_2$ to its antipode, $\tau_2 \to - \tau_2$.  
If we now invert the spatial Fourier transform and return to $(\tau, \bfk)$, 
\begin{align}
i G_2 \left(  k \tau, k \tau'  \right) 
&=  \int_0^{\infty} d \nu \,  \mathcal{N} ( \nu)  \frac{ f^- ( k \tau , \nu ) f^- ( k \tau' , \nu ) \Theta ( - \tau - \tau' ) +  f^+ ( k \tau , \nu ) f^+( k \tau' , \nu ) \Theta ( \tau + \tau' ) }{ \left( \nu^2 - \mu^2 \right)_{i \epsilon}  } \; . 
\end{align}
When restricted to the expanding Poincar\'{e} patch, this coincides with \eqref{eqn:G2_t_spec_rep}. 
This calculation highlights how our split representation \eqref{eqn:G2_t_spec_rep} is related to the usual propagator in position space, and could be used as a starting point for future work that studies the extension to the whole of de Sitter, e.g. in global co-ordinates.

\subsection{Spectral representation}

Inserting the split representation for the propagator \eqref{eqn:G2_t_spec_rep} into the exchange integral \eqref{eqn:G4exch_def} allows us to immediately perform the time integrals,
\begin{align}
 	\G_{4}^{\mu_\sigma}   = 
 	\frac{i}{ ( i k_s )^d } \int_0^{\infty} d \nu  \; \mathcal{N}_\nu \,  \frac{ F_{j_2} \left( \frac{\omega_{12}}{k_s} , \nu \right) F_{j_2} \left( \frac{\omega_{34}}{k_s} , \nu \right)  }{ ( \nu^2 - \mu_\sigma^2  )_{i \epsilon} } \; . 
 	\label{eqn:G4exch_spec}
\end{align}
It is clear that this satisfies the exchange equation~\eqref{eqn:exch_eqn}, since $F_{j_2}$ is an eigenstate of $\hat{s}_{12}$ and therefore the effect of $\left( m_\sigma^2 - \hat{s}_{12} \right)$ is to remove the pole and collapse the internal line (producing the integral representation \eqref{eqn:G4con_spec_rep} of $\G_4^{\rm con}$). 
Physically, we can recognise this as inserting two complete sets of states in the $s$-channel diagram:
\begin{align}
  \begin{tikzpicture}[scale=0.6,baseline=0.0cm]
			\draw (-4,1) -- (-3,0);
			\draw (-4,-1) -- (-3,0);
			\draw (-3,0) -- (-1,0);
			\draw (0,1) -- (-1,0);
			\draw (0,-1) -- (-1,0);
                \draw (-4,1) node [anchor=east][inner sep=2.0pt] {$\omega_1$};
                \draw (-4,-1) node [anchor=east][inner sep=2.0pt] {$\omega_2$};
                \draw (-2,0) node [anchor=south][inner sep=2.0pt] {$\mu_\sigma$};
                \draw (0,1) node [anchor=west][inner sep=2.0pt] {$\omega_3$};
			\draw (0,-1) node [anchor=west][inner sep=2.0pt] {$\omega_4$};
        \end{tikzpicture} 
&= 
 \int_{\nu_L} \int_{\nu_R} \, \left( 
     \begin{tikzpicture}[scale=0.6,baseline=0.0cm]
			\draw (2,0) -- (1,0);
			\draw (0,1) -- (1,0);
			\draw (0,-1) -- (1,0);
                \draw (2,0) node [anchor=west][inner sep=2.0pt] {$\nu_L$};
                \draw (0,1) node [anchor=east][inner sep=2.0pt] {$\omega_1$};
			\draw (0,-1) node [anchor=east][inner sep=2.0pt] {$\omega_2$};
        \end{tikzpicture} 
        \right) \left( 
             \begin{tikzpicture}[scale=0.6,baseline=0.0cm]
			\draw (2,0) -- (0,0);
                \draw (2,0) node [anchor=west][inner sep=2.0pt] {$\nu_R$};
                \draw (0,0) node [anchor=east][inner sep=2.0pt] {$\nu_L$};
			\draw (1,0) node [anchor=south][inner sep=2.0pt] {$\mu_\sigma$};
        \end{tikzpicture} 
        \right) \left( 
           \begin{tikzpicture}[scale=0.6,baseline=0.0cm]
			\draw (-2,0) -- (-1,0);
			\draw (0,1) -- (-1,0);
			\draw (0,-1) -- (-1,0);
                \draw (-2,0) node [anchor=east][inner sep=2.0pt] {$\nu_R$};
                \draw (0,1) node [anchor=west][inner sep=2.0pt] {$\omega_3$};
			\draw (0,-1) node [anchor=west][inner sep=2.0pt] {$\omega_4$};
        \end{tikzpicture}         
        \right)
\end{align}
which makes it clear that the internal line is off-shell and may take any mass value. 
Here we use $\int_{\nu_L} = \int_0^{\infty} d \nu_L \, \mathcal{N}_{\nu_L}$ for the integral over de Sitter principal series states, and the following diagram for the free propagator
\begin{align}
             \begin{tikzpicture}[scale=0.6,baseline=0.0cm]
			\draw (2,0) -- (0,0);
                \draw (2,0) node [anchor=west][inner sep=2.0pt] {$\nu'$};
                \draw (0,0) node [anchor=east][inner sep=2.0pt] {$\nu$};
			\draw (1,0) node [anchor=south][inner sep=2.0pt] {$\mu$};
        \end{tikzpicture} 
 = 
 \frac{ i \, \delta ( \nu - \nu' ) }{ \mathcal{N}_{\nu} \,  ( \nu^2 - \mu^2 )_{i \epsilon} } \; . 
\end{align}
Rather pleasingly, we see that free propagation preserves the angular momentum $\nu$.  
One way to think of the spectral representation \eqref{eqn:G4exch_spec} is that \emph{two} integrals in the time domain have been reduced to \emph{one} integral in the $\nu$-domain since $\nu$ is conserved between the vertices.
On Minkowski, there is also a conservation law at each vertex (which reduces the number of integrals to $-1$, i.e. an energy-conserving $\delta ( \omega_T)$ function)---on de Sitter, even once we project the external particles onto $\nu$ eigenstates as in \eqref{eqn:chi_def} there is no strict conservation at the vertices since angular momentum addition allows for multiple possibilities (i.e. the symbols \eqref{eqn:3j} are effectively the Clebsch-Gordan coefficients for this angular momentum addition).

\begin{figure}
\centering
\includegraphics[width=0.62\textwidth]{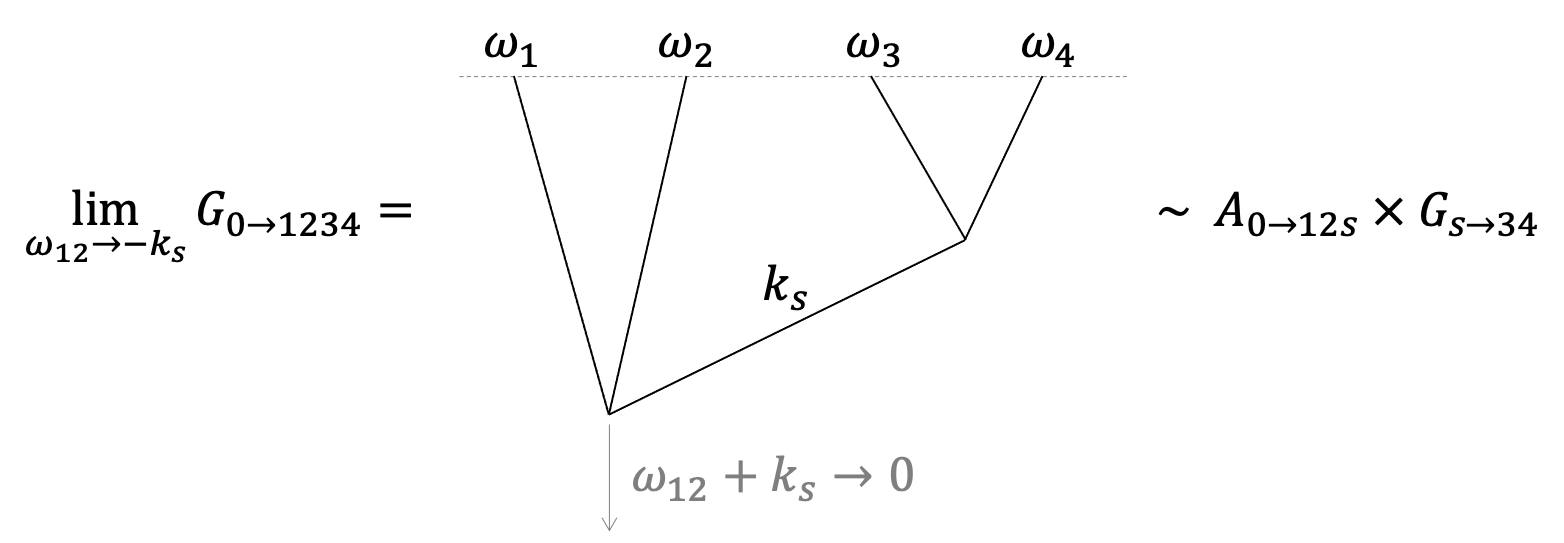}
\caption{In the factorisation limit $\omega_{12} \to k_s$, the left-vertex is pushed to early times and decouples from the right-vertex.}
\label{fig:factorisation}
\end{figure}

\paragraph{Factorisation.}
One of the most important features of the representation~\eqref{eqn:G4exch_spec} is that it makes factorisation manifest. 
In the limit $\omega_{12} \to -k_s$,  the exchange integral becomes proportional to
\begin{align}
\lim_{\omega_{12} \to -k_s } i \G_4^{\mu_\sigma} 
\left( \frac{\omega_{12}}{k_{s}} , \frac{\omega_{34}}{k_s} \right)  
\propto  -
 	\int_0^{\infty} d \nu  \; \mathcal{N}_\nu \,  \frac{ F_{j_2} \left( \frac{\omega_{34}}{k_s} , \nu \right)  }{ (\nu^2 - \mu_\sigma^2  )_{i \epsilon} } =  F_{j_2} \left( - \frac{\omega_{34}}{k_s} , \mu_\sigma \right) \; . 
\end{align}
This confirms the factorisation observed in the conformally coupled example above \eqref{eqn:cc_factorisation}.
Physically, taking the partial energy $\omega_{12} + k_s \to 0$ corresponds to this interaction taking place at early times (since there is no longer any $e^{i ( \omega_{12} + k_s) \tau}$ suppression), which produces the Minkowski 3-point amplitude (a constant). The remaining vertex is now a 3-point interaction between one incoming particle, of mass $\mu_\sigma$ and two outgoing fields with energies $\omega_3$ and $\omega_4$. 
This is depicted in Figure~\ref{fig:factorisation}, and can be written as the condition
\begin{align}
\lim_{\omega_{12} \to -k_s } \G_4
\left(  \col{\omega_1}{\bfk_1} , \col{\omega_2}{\bfk_2} , \col{\omega_3}{\bfk_3}  , \col{\omega_4}{\bfk_4} \right)  
\propto
\underset{ \omega_s \to \mu_\sigma}{ \text{LSZ} } \left[ \G_3 \left(  \col{\omega_1}{\bfk_1} , \col{\omega_2}{\bfk_2}  , \col{\omega_s}{ \bfk_{s} } \right) \right]  \times  \underset{ \omega_s \to \mu_\sigma}{ \text{LSZ} } \left[  \G_3 \left( \col{-\omega_s}{-\bfk_s} , \col{\omega_3}{\bfk_3} , \col{\omega_4}{\bfk_4}    \right) \right]  \; . 
\end{align}
In the next section, we will use this factorisation as a boundary condition for solving the exchange equation~\eqref{eqn:exch_eqn}, allowing us to efficiently bootstrap exchange diagrams.

\paragraph{Large mass expansion.}
As a consistency check of our results so far, we can evaluate the spectral representation as an expansion in large $\mu_\sigma$. 
Focussing on $d=1$, we can massage the integral into,
\begin{align}
    \G_{4}^{\mu_\sigma}   &= \frac{\pi \Gamma (d) }{2 \cosh ( \mu \pi )}  \int_1^\infty d z  \frac{ P_{i \nu - \frac{1}{2}} ( - z ) }{ \sqrt{ (z + \cosh (2 \rho_- ) ) ( z + \cosh ( 2 \rho_+ ) } }  \nonumber \\ 
    &= \frac{\pi \Gamma (d) }{2 \cosh ( \mu \pi )}  \int_1^\infty d y  \frac{ {}_2 F_1 \left( \frac{1}{2} - i \mu , \frac{1}{2} + i \mu , 1 ; y   \right) }{ \sqrt{ (y + \sinh^2 ( \rho_- ) ) ( y + \sinh^2 (  \rho_+ ) ) } } 
\end{align}
where $2 \rho_{\pm} = \rho_{12} \pm \rho_{34}$, and $\omega_{ij} = k_{s} \cosh \rho_{ij}$.
Note that expanding in powers of $1/\cosh^2 ( \rho_\pm )$, 
gives the series expression,
\begin{align}
& \G_{4}^{\mu_\sigma}  \nonumber \\ 
 &= \frac{1}{2 \pi k_{s}} \sum_{a,b = 0}^{\infty} \frac{  \Gamma ( a + b + 1 )^2  }{ m^2_\sigma \left( \frac{3}{2}  + i \mu \right)_{a+b}  \left( \frac{3}{2}  - i \mu \right)_{a+b} }  \frac{ \Gamma \left( a + \frac{1}{2} \right) \Gamma \left( b + \frac{1}{2} \right) }{a! b!}  \frac{1}{\cosh^{1+2a} ( \rho_+ ) \cosh^{1+2 b} ( \rho_- )}
\end{align}
Doing the summation over $b$ at fixed $a+b$ gives,
\begin{align}
 \G_{4}^{\mu_\sigma}  &= \sum_{n = 0}^{\infty} \frac{  1  }{ m_\sigma^2 \left( \frac{3}{2}  + i \mu \right)_{n}  \left( \frac{3}{2}  - i \mu \right)_{n} } \nonumber \\ 
&\qquad \times \left(  
\frac{ \Gamma ( 1 + 2n ) }{ 2 k_s \cosh \rho_+ \cosh \rho_- } \, \frac{1}{ ( - 4 \cosh^2 \rho_+ )^n } \; {}_2 F_1 \left( -n , \frac{1}{2} ; \frac{1}{2} - n ; \frac{\cosh^2 \rho_+}{\cosh^2 \rho_-} \right) 
\right)
\end{align}
which precisely reproduces the large-$\mu$ expansion given in \eqref{eqn:exch_large_mu}.

\paragraph{Derivative interactions}
We have thus far considered polynomial interactions with no derivatives. 
Including derivatives is straightforward, thanks to the $\hat{p}_\mu$ operators of section~\ref{sec:mandelstam}. 
For instance, suppose we had $(\nabla \phi)^2 \sigma$ in the interaction Lagrangian. Since the 3-point correlator is
\begin{align}
 \G_3 \left( \col{\omega_1}{\bfk_1} , \col{\omega_2}{\bfk_2} , \col{\omega_3}{\bfk_3}  \right)  = \hat{p}_1 \cdot \hat{p}_2 \,  \G_3^{\rm con} \left( \omega_T  \right)
\end{align} 
the 4-point exchange diagram from $(\nabla \phi)^2 \sigma \times \phi^2 \sigma$ is,
\begin{align}
 \G_4 \left( \col{\omega_1}{\bfk_1} , \col{\omega_2}{\bfk_2} , \col{\omega_3}{\bfk_3}, \col{\omega_4}{\bfk_4}  \right)  =  
 	i \int_0^{\infty} d \nu  \; \mathcal{N}_\nu \,  \frac{ \left[ \hat{p}_1 \cdot \hat{p}_2\,  F_{j_2} \left( \frac{\omega_{12}}{k_s} , \nu \right) \right] \left[ F_{j_2} \left( \frac{\omega_{34}}{k_s} , \nu \right) \right]  }{ ( \nu^2 - \mu_\sigma^2  )_{i \epsilon} } \; . 
\end{align} 
This obeys the exchange equation \eqref{eqn:exch_eqn} with an updated source that contains the derivatives. It has the right factorisation behaviour as $\omega_{12} \to -k_s$ or $\omega_{34} \to -k_s$. 
And finally, since we can write $2 \hat{p}_1 \cdot \hat{p}_2 = \hat{p}_3^2 - \hat{p}_1^2 - \hat{p}_2^2$, we have
\begin{align}
 \G_4 \approx \frac{ m_1^2 + m_2^2 - m_\sigma^2}{2} \, \G_4^{\mu_\sigma} + \frac{1}{2} \G_4^{\rm con}  
\end{align}
and therefore this exchange diagram is equivalent, when on-shell, to the exchange structure above plus contact correlators from the previous section.

\subsection{Exchange bootstrap}
\label{sec:bootstrap}

We will now show how exchange correlators can be efficiently ``bootstrapped'', following the approach of \cite{Arkani-Hamed:2018kmz}. 
As shown above, the unique exchange contribution is given by solving the differential equation:
\begin{align}
    \left( m_\sigma^2 - \hat{s}_{12}  \right) \G_4^{\mu_\sigma}  =  - \G_4^{\rm con} \left( \omega_T  \right)
    \label{eqn:exch_eqn_2}
\end{align}
subject to an appropriate boundary condition.
Our explicit integral solution contains an $i \epsilon$ prescription which implements the Bunch-Davies boundary conditions relevant for $\SBD_{n \to n'}$ 
At general kinematics this integral is difficult to evaluate analytically, but it is clear that the only singularities in this function are at $\omega_{12} = -k_s$ and $\omega_{34} = -k_s$. 

In this subsection, we show that the $i \epsilon$ prescription for the exchange integral can be replaced by the conditions, 
\begin{itemize}
    \item The correlator (for all outgoing energies) contains singularities only when the total energy at any vertex vanishes, and not in other ``folded'' configurations, 

    \item at these singular points, the residue may only contain non-overlapping singularities, since when any vertex is pulled to early times (partial energy $\to 0$) the correlator must factorises as in \eqref{eqn:factorisation}, 
    
\end{itemize}
which are sufficient to uniquely fix the solution to the exchange equation.

This is in the same spirit as the ``cosmological bootstrap'' of \cite{Arkani-Hamed:2018kmz}, in which a particular analytic structure was used to bootstrap solutions to the exchange equation for in-in correlators of conformally coupled fields.
To parallel the notation in that paper, for the remainder of this section we will introduce the ratios\footnote{
Note that our $u$ and $v$ are \emph{not} the same variables as in \cite{Arkani-Hamed:2018kmz}: although they do coincide upon setting $\omega_b = |\bfk_b |$ (i.e. $\cosh \rho_b = 1$) for each particle. 
},
\begin{align}
    u = \frac{k_s}{\omega_{12}} \; , \;\; v = \frac{k_s}{\omega_{34}} \; .
    \label{eqn:uv_def}
\end{align}
Somewhat remarkably, in terms of these variables the $\hat{s}_{12}$ operator \eqref{eqn:shat_def_5} in $d$ dimensions coincides with the $\Delta_u$ operator of \cite{Arkani-Hamed:2018kmz} in $d+2$ dimensions, 
\begin{align}
   - \hat{s}_{12} |_{d} = u^2(1-u^2)\partial^2_u-2u\left(u^2+\frac{d-1}{2}\right)\partial_u = \Delta_u |_{d+2} \; . 
\end{align}
Many of the earlier cosmological bootstrap results can therefore be carried over to the de Sitter $S$-matrix---the only differences are (i) the relevant boundary condition is now given by the factorisation property \eqref{eqn:factorisation}, and (ii) when computing correlators in which the external legs no longer have a definite mass value, the dimension is effectively two lower than for conformally coupled external scalars. 

To be explicit, we can write the exchange structure \eqref{eqn:G4exch_def} in terms of a scale-invariant function $F$, 
\begin{align}
   i \G_4^{\mu} =  \frac{ F ( u ,v ) }{ ( i k_s )^d } \; , 
\end{align}
which is forbidden from having singularities as $u \to 1$ or $v \to 1$, and in the factorisation limit
\begin{align}
    \lim_{ u \to - 1} F (u , v )  = \lim_{u \to -1} F_{j_2} \left( \tfrac{1}{u}  , \mu_\sigma \right) F_{j_2}^* \left( - \tfrac{1}{v} , \mu_\sigma \right)  \; .
    \label{eqn:factorisation}
\end{align}
In the limit $\omega_{12} \to -k_s$, the correlator $F_{j_2} ( \frac{\omega_{12}}{k_s} , \mu_\sigma )$ reduces to a Minkowski scattering amplitude (multiplied by a singular factor of $(\omega_{12} - k_s )^j$, where $j$ is fixed by dilation invariance).
This factorisation of the dS correlator $G_4^\mu$ into a product of a Minkowski amplitude and a dS correlator is shown pictorially in Figure~\ref{fig:factorisation}.
Notice that the residue of the $\omega_{12} = -k_s$ singularity may only have singularities at $\omega_{34} =  k_s$ (but not at $\omega_{34} = - k_s$). That certain overlapping singularities are forbidden in the $S$-matrix is analogous to the Steinman relations on Minkowski.

\paragraph{Exchange of a conformally coupled field.}
For example, consider the exchange of a conformally coupled scalar in $d=3$. The source for the exchange equation is
\begin{align}
 \left( m_\sigma^2 - \hat{s}_{12} |_{d=3} \right) F (u,v)  =  \frac{  2 (u v)^3}{ (u+v)^3 }
 \label{eqn:exch_eqn_3}
\end{align}
which is nothing but the $\Gamma(3)/(i \omega_T)^3$ from above. 
Solving both \eqref{eqn:exch_eqn_3} and the analogous $\hat{s}_{34}$ equation and imposing the absence of folded singularities at $u=1$ and $v=1$ determines $G_4^{\mu_\sigma}$ up to a single constant of integration,
\begin{align}
    F (u, v ) =  \frac{  u v (  (u + v) C - u v) }{ ( u + 1 ) ( v + 1 ) (u + v) } \; . 
    \label{eqn:G4s_eg_C}
\end{align}
In the factorisation limit $\omega_{12} \to - k_s$,
\begin{align}
   \lim_{u \to -1} F (u, v ) = -  \frac{1}{2 ( u+ 1 )} \left[ 2 + 2 C + \frac{1}{v-1} - \frac{1 + 2 C}{v+1} \right] \; . 
\end{align}
To implement the in-out boundary conditions of the $S$-matrix, we must set $C=-1/2$ to remove the overlapping singularity at $v=-1$. 
This uniquely determines $\G_4$, and indeed this value matches a brute-force computation using bulk time integrals. 
In contrast, the wavefunction coefficient\footnote{
Recall that $\psi_4 \propto 1/ \left[ \omega_T (\omega_{12} + k) (\omega_{34} + k) \right]$ for the exchange of a conformally coupled scalar in $d=3$ dimensions.
} corresponds to setting $C= 0$. 
Since the coefficient $C$ captures the freedom to add any disconnected product $\G_3 \times \G_3$ to the exchange solution, \eqref{eqn:G4s_eg_C} can be read as the exchange Feynman diagram in which the internal line represents the propagator,
\begin{align}
   \Theta ( \tau_1 - \tau_2 ) f_1^- f^+_2  + \Theta ( \tau_2 - \tau_1 ) f_1^- f^+_2  - \left( 1 +  2 C  \right) f_1^+ f_2^+ \; 
\end{align}
where $f_j^{\pm} = f^{\pm} ( k_s \tau_j, \mu_\sigma)$.
The choices above ($C = -1/2$ and $C = 0$) correspond to the Feynman and bulk-to-bulk propagators respectively.

\paragraph{Shifting the spacetime dimension.}
In other spacetime dimensions, we could bootstrap solutions in the same way. 
But given a solution in $d=3$, an efficient method of generating solutions in $d = 3+ 2n$ is to repeatedly use the identity
\begin{equation}
    \hat{s}_{12} |_{d+2} u^2\partial_u = u^2 \partial_u ( d + 1 + \hat{s}_{12}|_{d}  ) \; . 
\end{equation}
This means that if there is a Green's function in $d$ dimensions that solves $\hat{s}_{12}|_{d} {\cal G}=\hat{s}_{34}|_{d} {\cal G}$ then $(u v)^2 \partial_{u} \partial_v {\cal G}$ will be a kinematically allowed Green's function in $d+2$ dimensions\footnote{
$(uv)^2 \partial_u \partial_v$ is the ``spin-raising'' operator of \cite{Arkani-Hamed:2018kmz}. 
}. 
For instance, the contact invariants $\G_4^{(n)} = (  d + 1 -\hat{s}_{12}  )^n \G_4^{\rm con}$ in $d=3$ dimensions can be written as 
\begin{equation}
    {\cal G}^{(n)}_{4} |_{d=3} =(u v)^2 \partial_{u} \partial_v \; ( - \hat{s}_{12} )^n \G^{(0)}_{4} |_{d=1}  = (u v)^2 \partial_{u} \partial_v  \Delta_{u}^n \left(\frac{u v }{u+v}\right) \; , 
\end{equation}
where $\Delta^{n}_{u}\left(\frac{u v }{u+v}\right)$ are precisely the contact invariants previously computed in \cite{Arkani-Hamed:2018kmz}. However, notice that now $u$ and $v$ correspond to functions of the energy variables $\omega$ which label the \emph{off-shell} external lines, so these are qualitatively different physical objects. 
Applying this same trick to the exchange equation, we have that any solution $\G_4^{\mu_\sigma}$ to \eqref{eqn:exch_eqn_2} in $d$ dimensions will also be a solution of: 
\begin{equation}
    (   m_\sigma^2 + d + 1 - \hat{s}_{12} |_{d+2} )(u v)^2 \partial_{u} \partial_v \G_4^{\mu_\sigma} = ( i k_s )^{d+2} \G_4^{\rm con} |_{d+2} =  \Gamma (d+2) \left(  \frac{  u v  }{ u+v } \right)^{d+2} \; .
\end{equation}
$(u v)^2 \partial_{u} \partial_v \G_4^{\mu_\sigma}$ therefore corresponds to the exchange of a particle with mass $(m_\sigma^2 + d + 1)$ in $d+2$ dimensions\footnote{
Note that the spin-raising operator does not affect the singularity structure and therefore preserves the required boundary conditions of $\G_4^{\mu_\sigma}$.
}.

\paragraph{Exchange of a general massive field.}
To connect with previous literature on the cosmological collider signal from the exchange of heavy scalar, let us now consider solving the exchange equation \eqref{eqn:exch_eqn_2} in $d=1$ dimensions for a general mass. This can be written in $u, v$ variables as,
\begin{align}
    \left(  \mu^2 + \frac{1}{4}  - \hat{s}_{12} |_{d=1} \right) F (u, v) =  \frac{u v}{u + v} \; .
\end{align}
Demanding no folded singularities (regularity at $u, v \to +1$) produces a general solution for $F (u,v)$ that depends on a single undetermined constant, which we denote by $\beta$. 
Noting that $- \hat{s}_{12} |_{d=1} = \Delta_u |_{d=3}$, an explicit series representation for this function when $|u| < |v|$ was found in \cite{Arkani-Hamed:2018kmz}
\begin{align}
    F ( u, v) = \sum_{m,n = 0}^{\infty} c_{mn} (\mu ) u^{2m+1} \left( \frac{u}{v} \right)^n  + \frac{\pi}{2 \cosh ( \pi \mu_\sigma )} g (u ,v) \; \;
    \label{eqn:Fuv_def}
\end{align}
where the coefficients $c_{mn}$ are
\begin{align}
    c_{mn} (\mu) = \frac{ (-1)^n (n+1) (n+2 ) ... (n+2m) }{  \left( (n+\frac{1}{2} )^2 + \mu^2  \right) \left( (n+\frac{5}{2} )^2 + \mu^2  \right) ... \left( (n+\frac{1}{2} + 2m )^2 + \mu^2  \right)   }
\end{align}
and $g (u,v)$ is a solution to the homogeneous equation\footnote{
The first term is fixed by the requirement that it cancels the singularity at $u \to 1$ from the infinite sum,
\begin{align}
    \lim_{u \to 1} \left[ \sum_{m,n} c_{m,n} u^{2m +1} \left( \frac{u}{v} \right)^n \right]  =  \frac{\pi}{2 \cosh (\pi \mu )} P_+ (v)  \log ( 1 - u )  = \lim_{u \to 1} \left[   \frac{ - \pi^2}{2 \cosh^2 (\pi \mu )} P_- (u) P_+ (v) \right] \; .
\end{align}
},
\begin{align}
  g (u,v) =  \frac{ \pi }{ \cosh ( \pi \mu ) } P_+ (v) P_- (u)  -  \beta P_+ (u)  P_+ \left( v \right) 
\label{eqn:guv_def}
\end{align}
where $P_\pm (u) = P_{i \mu - \frac{1}{2}} \left( \pm \frac{1}{u} \right)$.
The complementary region $|v| > |u|$ then follows from the symmetry $F (u,v) = F (v,u)$. 

The coefficient $\beta$ again captures which propagator should be used for the internal line\footnote{
Note that the Legendre function in the second term of \eqref{eqn:guv_def} has a branch cut for the physical region $0 < u < 1$. We evaluate this function by summing its value just above and just below this cut, which preserves the identity $P_n^* ( z) = P_{n^*} ( z^* )$ on the real axis. Since the branch cut discontinuity of $P_n ( -z)$ is proportional to $P_n ( +z)$, a different convention would correspond to a shift in $\beta$. For instance, \cite{Goodhew:2021oqg} defines the Legendre functions on the real axis as being approached from one side only, and as a result their $\beta_+$ has an additional contribution. Explicitly, our $\beta = \beta_+ - i \pi $ in the notation of \cite{Goodhew:2021oqg}, and $\beta = - 
 \beta_0 \pi \sinh ( \pi \mu) / \cosh ( \pi \mu ) $ in the notation of \cite{Arkani-Hamed:2018kmz}. 
}. Different choices correspond to different physical objects: correlator, wavefunction, $S$-matrix, etc. 
We can fix the value of $\beta$ by matching to the so-called ``factorisation limit'' in which $u \to -1$ and $v \to -1$, for instance:  
\begin{align}
    \langle \phi^4 \rangle &\sim &\frac{1}{2} \times &\log \left( 1 + u \right) \log \left( 1 + v \right) \;  &\Rightarrow&  \;\; &\beta &= \frac{\pi}{ \cosh ( \pi \mu )}  ,  \nonumber \\ 
    \psi_4  &\sim &- \frac{ f^-_{\mu}   }{ 2 f^+_{\mu}   } \times &\log \left( 1 + u \right) \log \left( 1 + v \right) \;  &\Rightarrow& \;\; &\beta &= - \frac{\pi}{\cosh ( \pi \mu ) } \frac{ f^-_{\mu}  }{ f^+_{\mu}   }  ,  \label{eqn:beta_fixing} \\ 
    \SBD_{0\to 4} &\sim  &0 \times &\log \left( 1 + u \right) \log \left( 1 + v \right) \; &\Rightarrow& \;\; &\beta &= 0 ,  \nonumber 
\end{align}
where $f^{\pm}_{\mu} = f^{\pm} ( k_s \tau_0 , \mu )$ with $\tau_0 \to 0$ a regulator for the late-time behaviour oscillatory behaviour of the mode function. 
This is the same method described in \cite{Arkani-Hamed:2018kmz}, where our simpler factorisation condition for the $S$-matrix means that $\beta$ can be fixed straightforwardly without any bulk calculation.

\paragraph{Particle production.}
The effects of particle production are most clearly visible in the collapsed limit $k_s \to 0$, which can be implemented by taking $v \to 0$ with $\xi = u/v$ held fixed. 
The general solution to the exchange equation $F (u,v)$ in this limit takes the form,
\begin{align}
 F (u,v) = v \left(  F_{\rm EFT} (\xi) +  F_{\rm collider} (\xi) + 2 \text{Re} \left[  v^{2 i \mu} F_{\rm OPE} (\xi) \right] \right) + \mathcal{O} ( v^2 )
\end{align}
where $F_{\rm collider}$ are the first non-analyticities in $\xi$, and $F_{\rm OPE}$ are the first non-analytic terms in $k_s$.  
Explicitly, expanding the Legendre functions gives, 
\begin{align}
 F_{\rm EFT} (\xi) &= \sum_{n} c_{0n} (\mu) \xi^n  \;  \nonumber \\ 
  F_{\rm collider} (\xi) &= \frac{ i \pi  + \beta \coth ( \pi \mu   )   }{4 \mu \, \cosh ( \pi \mu  )} \xi^{ \frac{1}{2} + i \mu_\sigma}   + \text{c.c.}  \\ 
   F_{\rm OPE} (\xi) &= \frac{ \sinh ( \pi \mu_\sigma )}{2\pi^2 } \left( \frac{\xi}{4} \right)^{ \frac{1}{2} + i \mu } \left( i \pi + \beta \coth ( \pi \mu  )  \right)
   \Gamma \left( \tfrac{1}{2} + i \mu  \right)^2 \Gamma \left( - i \mu  \right)^2    \nonumber 
\end{align}
The $F_{\rm collider}$ signal dominates in the squeezed limit and produces characteristic oscillations with a periodicity set by the mass of the heavy field. 
Beyond the EFT series, both $F_{\rm collider}$ and $F_{\rm OPE}$ depend on $\beta$ and will differ for different physical observables (e.g. wavefunction, correlator, $S$-matrix). 
However, in the large mass limit this changes the overall amplitude of the signal by a simple numerical factor and does not affect the Boltzmann $\sim e^{- \pi \mu_\sigma}$ suppression.

\paragraph{Beyond 4-particle scattering.}
One of the virtues of our off-shell formalism is that it can accommodate any number of external particles. 
To illustrate this, let us now consider the tree-level exchange diagram for $n_1 \to n_2$ scattering in $d$ dimensions. 
If we denote the total energy and momentum of the $n_1$ ingoing particles as $-\omega_1$ and $-\bfk$, and the total energy and momentum of the $n_2$ outgoing particles as $\omega_2$ and $\bfk$, then we can express this diagram in terms of the two dimensionless ratios 
\begin{align}
    u &= \frac{k}{\omega_1} \; ,  & v &= \frac{k}{\omega_2} \; . 
\end{align}
which are the straightforward generalisation of \eqref{eqn:uv_def} to multiple particles. 
The relevant exchange equation to be solved is:
\begin{align}
    \left( \mu^2  + \frac{d^2}{4} - \hat{s}_{12... n_1} \right) F (u, v) =  \Gamma \left( j_{n_1} + j_{n_2} \right)  \left( \frac{ u v }{ u + v} \right)^{j_{n_1} + j_{n_2}}
    \label{eqn:exch_eqn_n1n2}
\end{align}
where the $n$-particle Mandelstam operator \eqref{eqn:shat_def_3} can be written as
\begin{align}
- \hat{s}_{12 ... n} = u^2 ( 1- u^2 ) \partial_u^2 - u \left( 2 u^2 - 1 + 2 j_n \right) \partial_u + j_n^2 - \frac{d^2}{4} \; . 
\end{align}
Proceeding as before, we can express the general solution to the exchange equation in the region $|u| < |v|$ as
\begin{align}
    F(u,v)  &=  \sum_{m,n = 0}^{\infty} c_{mn} u^{j_{n_1} + j_{n_2} + 2m} \left( \frac{u}{v} \right)^{n}  
    + \sum_{\sigma_1 , \sigma_2 = \pm} \beta_{\sigma_1 \sigma_2} F^{\sigma_1}_{n_1} (u) F^{\sigma_2}_{n_2} ( v) \; 
    \label{eqn:Gn1n2_soln}
\end{align}
where the $\beta$ coefficients reflect the freedom to add any homogeneous solution of both \eqref{eqn:exch_eqn_n1n2} and the complementary equation in $v$, which we have written in terms of the mode functions~\eqref{eqn:Fmode} as\footnote{
Note that $F^{\pm}_n \left( \frac{k}{\omega} \right)$ corresponds to the overlap of $n$ particles with total energy $\omega$ with an on-shell particle of mass $\mu$ which is either outgoing $(F^+_n)$ or ingoing $(F^-_n)$. In particular, if the $\omega$ energies are then also projected onto outgoing mass eigenstates using \eqref{eqn:intro_LSZ_rho}, we would have $\text{LSZ} \left[ F^+_n \right] = \SBD_{0 \to n+1}$ and $\text{LSZ} \left[ F^-_n \right] = \SBD_{1 \to n}$. 
}
\begin{align}
    F^+_n ( u ) &= F_{j_n} \left( \frac{1}{u} \right) \; , &F^-_n (u) = F^*_{j_n} \left( - \frac{1}{u} \right) \; .
    \label{eqn:Fpm_def}
\end{align}
The $c_{mn}$ are fixed by matching to the analogous expansion of the source
\begin{align}
  \Gamma ( J )  \left( \frac{uv}{u+v} \right)^{J} = \sum_{n=0}^{\infty} \frac{ (-1)^n }{n!}  \Gamma ( J + n ) u^{J} \left( \frac{u}{v} \right)^n
\end{align}
which gives
\begin{align*}
    c_{mn}  = \frac{ (-1)^n \Gamma ( n + J_{n_1 + n_2} )}{ \Gamma ( n + 1 )} \frac{ ( n + J_{n_1 + n_2} ) ( n + J_{n_1 + n_2} + 1 ) ... ( n + J_{n_1 + n_2} + 2m - 1 )  }{ \left[ \mu^2 + ( n + j_{n_2} )^2 \right] \left[ \mu^2 + ( n + j_{n_2} +2 )^2 \right] ... \left[ \mu^2 + ( n + j_{n_2} + 2 m )^2 \right] }
\end{align*}
where $J_{n_1 + n_2} = j_{n_1} + j_{n_2} = \frac{d}{2} ( n_1 + n_2 - 2 )$. 
To fix the $\beta$ coefficients, we must specify which object we would like to compute (wavefunction, in-in correlator, $S$-matrix, etc)---this is equivalent to providing the $i \epsilon$ prescription for the bulk propagator in \eqref{eqn:G4exch_def}. 
Thhe relevant conditions for the $S$-matrix or amputated correlator is
\begin{itemize}

\item[(i)] \emph{No singularity as $v \to 1$}. Since the $c_{mn}$ series is regular as $v \to 1$ for generic values of $u$, this condition requires $\beta_{+-} = \beta_{--} = 0$ in order to remove the singularity in $F^-_{n_2} (v)$. 

\item[(ii)] \emph{No singularity as $u \to 1$}. 
The $c_{mn}$ series diverges as $u \to 1$
\begin{align}
    \lim_{u \to 1} \sum_{m,n=0}^{\infty} c_{mn} u^{2m + J_{n_1 + n_2}} \left( \frac{u}{v} \right)^n =  - \frac{ \Gamma \left( j_{n_1} - \frac{1}{2} \right) }{i \sqrt{2} ( 1 - u )^{ j_{n_1} - \frac{1}{2} } }  F^+_{n_2} ( v )
\end{align}
and so comparing with the singularity in $F^-_{n_1} ( u) $---see \eqref{eqn:F_u1_limit}---we see that $\beta_{-+} = 1$ is needed to remove this singularity.

\item[(iii)] \emph{No overlapping singularity as $u \to-1$ and $v \to -1$.} The preceeding conditions leave $\beta_{++}$ as the only undetermined coefficient, just as in the simpler example of $n_1 = n_2 = 2$ and $d=1$ we considered above. 
However, the corresponding $F^+_{n_1} (u) F^+_{n_2} (v)$ contains an overlapping singularity that $\sim ( 1 + u )^{\frac{1}{2} - j_{n_1}} ( 1 + v )^{ \frac{1}{2} - j_{n_2}}$ as both $u$ and $v$ approach $-1$. 
Such singularities are forbidden by the factorisation condition that 
\begin{align}
    \lim_{u \to -1} \, F (u,v)  =  F_{n_1}^+ \left( u \right) F^-_{n_2} \left( v \right)  
\end{align}
and therefore we must set $\beta_{++} = 0$. 

\end{itemize}
Altogether, the amputated correlator for this $n_1 \to n_2$ scattering via the $s$-channel exchange of a heavy scalar is given by \eqref{eqn:Gn1n2_soln} with
\begin{align}
    \beta_{++} = \beta_{--} =  \beta_{+-} =  0 \; , \qquad \beta_{-+} =  1\; . 
\end{align}
For the complementary region $|v| > |u|$, one can proceed in the same fashion and would arrive at the analogous series solution with both $u \leftrightarrow v$ and $n_1 \leftrightarrow n_2$. 
The spectral representation that unifies both regions is
\begin{align}
   F (u,v)  = \int_{0}^{\infty} d \nu \; \mathcal{N}_\nu \; \frac{ F_{j_{n_1}} \left( \frac{1}{u} \right) F_{j_{n_2}} \left( \frac{1}{v} \right) }{ \left( \nu^2 - \mu^2 \right)_{i \epsilon} }
\end{align}
where $\mathcal{N}_\nu = \tfrac{2}{\pi} \nu \sinh ( \pi \nu )$ and $F_{j_n}$ is given in \eqref{eqn:Fmode}. This manifestly satisfies both the exchange equation (thanks to~\eqref{eqn:Gn_spec_rep}) and the required factorisation (thanks to~\eqref{eqn:F_cross}).

\subsection{Unitarity and positivity}
\label{sec:unitarity}

Another way to fix the coefficient $C$ in \eqref{eqn:G4s_eg_C} and $\beta$ in \eqref{eqn:guv_def}, first described in \cite{Goodhew:2021oqg}, is to use the cutting rules which follow from perturbative unitarity. 
In that approach, one considers the so-called ``discontinuity'', which for the exchange of a conformally coupled scalar is 
\begin{align}
    &F (u, v) + F^* ( - u^*, - v^* ) \nonumber \\
    &=  g_3 ( u) g_3 (-v) + g_3 ( - u ) g_3 ( v) + \left( 1 + 2 C  \right) \left( g_3 (u) g_3 (v) + g_3 (-u) g_3 (-v) \right) \; 
\end{align}
where $g_{3} ( u ) = \tfrac{1}{\sqrt{2}} u / (u + 1)$, 
and for the exchange of a general massive scalar,
\begin{align}
   &F (u , v ) +  F^* ( - u^*, - v^* )  \nonumber \\
   &= \frac{\pi}{2 \cosh \left( \pi \mu_\sigma \right) } \left[ 
   \frac{\pi P_- (u) P_+ (v) }{\cosh ( \pi \mu_\sigma )}  +  \frac{\pi P_+ (u) P_- (v) }{\cosh ( \pi \mu_\sigma )} - 
   \beta  P_+ (u) P_+ (v) - \beta^* P_- (u) P_- (v)    
   \right] \;  \\ 
   &=  F^-_2 \left(u \right) F_{2}^+ \left( v \right) + F_{2}^+ \left( u \right) F^-_2 \left( v \right) + \tfrac{1}{\pi} \cosh (\pi \mu ) \left(   
   \beta  F^+_2 \left( u \right) F^+_{2} \left( v \right)  + \beta ^* F_{2}^- \left( u \right) F^-_2 \left( v \right)
   \right)
   \nonumber 
   \label{eqn:cutting_1}
\end{align}
where we have given the result in both our $F_2^{\pm}$ mode function \eqref{eqn:Fpm_def} and in the notation $P_\pm (u) = P_{i \mu - \frac{1}{2}} \left( \pm \frac{1}{u} \right)$ of \cite{Goodhew:2021oqg} for easy comparison.
The different $C$ and $\beta$ for wavefunction versus $S$-matrix reflects their different unitarity cutting rules: the usual Cutkosky rules for the $S$-matrix imply that $C = - \frac{1}{2}$ and $\beta =0$, while the Cosmological Cutting Rules of \cite{Melville:2021lst, Goodhew:2021oqg} imply that $C = 0$ and $\beta$ is given by \eqref{eqn:beta_fixing}. 
In this subsection, we describe unitarity and the associated cutting rules in more detail.

\paragraph{Optical theorem.}
As on Minkowski, the de Sitter $S$-matrix can be written as
\begin{align}
\SBD_{n_1 \to n_2} = \langle n_2 |  \hat{U}  | n_1 \rangle 
\end{align}
where $|n\rangle$ are the free theory particle eigenstates and $\hat{U}$ is a unitary operator related to time translations\footnote{
If $U ( \tau_1, \tau_2)$ and $U_{\rm free} ( \tau_1, \tau_2)$ implement time evolution in the interacting and free theories respectively, then $\hat{U} = U_{\rm free}^{\dagger} (- \infty, 0) U( -\infty, 0)$.
}.
Splitting the $S$-matrix elements into a free disconnected part and an interacting part corresponds at the operator level to writing $ \hat{U} = 1 + i \hat{T}$.
Unitarity of $\hat{U}$ then implies the usual relation $\hat{T} - \hat{T}^\dagger = i \hat{T}^\dagger \hat{T}$. 
Inserting a complete set of states (e.g. the free particle eigenstates~\eqref{eqn:n_def}) between the $\hat{T}^\dagger \hat{T}$ then implies the usual optical theorem for the connected part of the de Sitter $S$-matrix,
\begin{align}
 \SBD_{n_1 \to n_2} + \SBD_{n_2 \to n_1}^* =  \sum_I \SBD_{n_1 \to I} \SBD_{n_2 \to I}^* \; . 
 \label{eqn:optical}
\end{align}
This result is non-perturbative and immediately implies many important properties. The two we will highlight here are:
\begin{itemize}

\item[(i)] the discontinuity in the forward limit is positive, namely $2 \text{Im} \SBD_{n \to n} = \sum_I | \SBD_{n \to I} |^2 > 0$,

\item[(ii)] in perturbation theory, individual diagrams for both $\SBD$ and $\G$ can be cut into simpler diagrams by taking a suitable discontinuity.

\end{itemize}

\paragraph{Cutting rules.}
Let us start with the perturbative cutting rules (ii). 
These can be derived from the fact that the propagator obeys
\begin{align}
 G_2 ( k \tau, k \tau ) +   G_2^* ( k \tau, k \tau' )  &= f^+ ( k \tau ) f^- ( k \tau' ) + f^- ( k \tau ) f^+ ( k \tau' )  
\end{align}
i.e. that the real part does not contain any $\Theta$ functions, and therefore each term factorises into a product of two mode functions. 
Since the spectral representation for the exchange diagram has the same form as the propagator, we immediately have that
\begin{align}
\G_4^{\mu_\sigma} \left( - \frac{\omega_{12} }{k_s},  + \frac{\omega_{34}}{k_s} \right) +   \G_4^{\mu_\sigma *} \left(  + \frac{\omega_{12}}{k_s}, - \frac{\omega_{34}}{k_s} \right)  =  F^+_{j_2} ( \rho_{12} ) F^+_{j_2} ( \rho_{34} ) +  F^-_{j_2} ( \rho_{12} ) F^-_{j_2} ( \rho_{34} ) \; .
\label{eqn:cutting_2}
\end{align}
This extends \eqref{eqn:cutting_1} to any dimension and can also be applied to the $n_1 \to n_2$ example above, for which:
\begin{align}
    F ( u,v ) + F^* ( -u^* , -v^* ) = F^+_{n_1} ( u ) F^-_{n_1} ( v ) + F^-_{n_1} ( u ) F^+_{n_2} ( v ) 
\end{align}
In fact, performing the LSZ integral~\eqref{eqn:intro_LSZ_rho} to project onto mass eigenstates, we have that
\begin{align}
 \SBD_{12 \to 34} + \SBD_{34 \to 12}^*=    \SBD_{12 \to \mu_\sigma} \SBD_{34 \to \mu_\sigma}^* +  \SBD_{12 \mu_\sigma \to 0} \SBD_{34 \mu_\sigma \to 0}^*
\end{align}
Diagrammatically, this cutting rule is
\begin{align}
    \begin{tikzpicture}[scale=0.6,baseline=0.0cm]
			\draw (-3,1) -- (-2,0);
			\draw (-3,-1) -- (-2,0);
   			\draw (-2,0) -- (-1,0);
			\draw (0,1) -- (-1,0);
			\draw (0,-1) -- (-1,0);
                \draw (-3,1) node [anchor=east][inner sep=2.0pt] {$\omega_1$};
                \draw (-3,-1) node [anchor=east][inner sep=2.0pt] {$\omega_2$};
                \draw (-1.5,0) node [anchor=south][inner sep=2.0pt] {$\mu$};
                \draw (0,1) node [anchor=west][inner sep=2.0pt] {$\omega_3$};
			\draw (0,-1) node [anchor=west][inner sep=2.0pt] {$\omega_4$};
        \end{tikzpicture}  
         + 
         \left(     \begin{tikzpicture}[scale=0.6,baseline=0.0cm]
			\draw (-3,1) -- (-2,0);
			\draw (-3,-1) -- (-2,0);
   			\draw (-2,0) -- (-1,0);
			\draw (0,1) -- (-1,0);
			\draw (0,-1) -- (-1,0);
                \draw (-3,1) node [anchor=east][inner sep=2.0pt] {$\omega_3$};
                \draw (-3,-1) node [anchor=east][inner sep=2.0pt] {$\omega_4$};
                \draw (-1.5,0) node [anchor=south][inner sep=2.0pt] {$\mu$};
                \draw (0,1) node [anchor=west][inner sep=2.0pt] {$\omega_1$};
			\draw (0,-1) node [anchor=west][inner sep=2.0pt] {$\omega_2$};
        \end{tikzpicture}  \right)^*  
        =      
        \begin{tikzpicture}[scale=0.6,baseline=0.0cm]
			\draw (-4,1) -- (-3,0);
			\draw (-4,-1) -- (-3,0);
			\draw (-3,0) -- (-1,0);
			\draw (0,1) -- (-1,0);
			\draw (0,-1) -- (-1,0);
			\draw[dashed] (-2,1) -- (-2,-1);
                \draw (-4,1) node [anchor=east][inner sep=2.0pt] {$\omega_1$};
                \draw (-4,-1) node [anchor=east][inner sep=2.0pt] {$\omega_2$};
                \draw (-2.5,0) node [anchor=south][inner sep=2.0pt] {$\mu$};
                \draw (-1.5,0) node [anchor=south][inner sep=2.0pt] {$\mu$};
                \draw (0,1) node [anchor=west][inner sep=2.0pt] {$\omega_3$};
			\draw (0,-1) node [anchor=west][inner sep=2.0pt] {$\omega_4$};
        \end{tikzpicture} 
+
        \begin{tikzpicture}[scale=0.6,baseline=0.0cm]
			\draw (-4,1) -- (-1,0.5);
			\draw (-4,0) -- (-1,0.5);
			\draw (-1,0.5) -- (-3,-0.5);
			\draw (0,0) -- (-3,-0.5);
			\draw (0,-1) -- (-3,-0.5);
			\draw[dashed] (-2,1) -- (-2,-1);
                \draw (-4,1) node [anchor=east][inner sep=2.0pt] {$\omega_1$};
                \draw (-4,0) node [anchor=east][inner sep=2.0pt] {$\omega_2$};
                \draw (0,0) node [anchor=west][inner sep=2.0pt] {$\omega_3$};
			\draw (0,-1) node [anchor=west][inner sep=2.0pt] {$\omega_4$};
        \end{tikzpicture} 
\end{align}
Notice that, on Minkowski, the final term would vanish due to energy conservation---it is present on de Sitter because there is no longer any requirement for the total energy at a vertex to vanish.
A systematic description of these cutting rules and their uses will be developed elsewhere.

\paragraph{Positivity and beyond.}
Notice that since the mode functions $F_n^{\pm}$ are real, the disc in \eqref{eqn:cutting_2} is positive in the forward limit. 
This is the perturbative avatar of the optical theorem \eqref{eqn:optical}.
This positivity follows from the unitarity time evolution of quantum mechanics: it is therefore potentially distinct from \cite{Green:2023ids} (positivity of Fisher information) and \cite{Hogervorst:2021uvp, Penedones:2023uqc, Loparco:2023rug} (positivity of inner product on Hilbert space). 
Interestingly, this disc is also bounded from above: to proceed as on Minkowski would require a partial wave expansion. 
We leave this interesting direction for future work.

\section{Discussion}
\label{sec:disc}

\noindent In this work, we have put forward an extension of the de Sitter $S$-matrix of \cite{Melville:2023kgd} which allows for the scattering of off-shell states in which the particles are not confined to physical mass eigenstates. 
Going off-shell has allowed us to connect with amputated correlators in the energy-domain, which share many useful properties of the off-shell amplitudes in Minkowski.
In particular, the unique 3-particle $S$-matrix takes the particularly simple form of a total-energy pole, and the most general 4-particle $S$-matrix at tree-level can be written as a power series expansion in $\hat{s}$ and $\hat{t}$ (which exactly captures the EFT derivative expansion) plus exchange contributions which have a simple spectral representation. 
Altogether, these results determine the off-shell correlators for a wide range of different theories. 
The on-shell $S$-matrix that determines cosmological observables can then be obtained by projecting onto particular mass eigenstates. \\

To sum up in a single equation: from the tree-level exchange of a scalar field, the most general amputated correlator (up to terms that vanish on-shell) is given by: 
\begin{align}
&\G_4 \left( \col{\omega_1}{\bfk_1} , ... , \col{\omega_4}{\bfk_4} \right) =  \sum_{a,b } c_{ab} \, \hat{t}^b \,\hat{s}^a \; \G^{\rm con}_4 ( \omega_T  )  \\
&+\sum_{\sigma} \left(  \frac{ \lambda_{12\sigma} \lambda_{34\sigma} }{  k_s^d } \G_4^{\mu_\sigma} \left( \frac{\omega_{12}}{ k_{s} } , \frac{\omega_{34}}{k_{s}}  \right)   
+ \frac{ \lambda_{13\sigma} \lambda_{24\sigma} }{ k_t^d }  \G_4^{\mu_\sigma} \left( \frac{\omega_{13}}{ k_{t} } , \frac{\omega_{24}}{k_{t}}  \right)   
+ \frac{ \lambda_{14\sigma} \lambda_{23\sigma} }{ k_u^d } \G_4^{\mu_\sigma} \left( \frac{\omega_{14}}{k_{u}} , \frac{ \omega_{23}}{k_{u}}  \right)    \nonumber 
\right)
\end{align}
where $\{ c_{ab} , \lambda_{abc} \}$ are constant model-dependent couplings, $\{ \G_4^{\rm con} ,  \G_4^{\mu} \}$ are model-independent kinematic structures fixed by the de Sitter symmetries, and the kinematic variables are $\omega_{ab} = \omega_a + \omega_b$, $k_s = | \bfk_1 + \bfk_2 |$, $k_t = | \bfk_1 + \bfk_3 |$ and $k_u = | \bfk_1 + \bfk_4 |$.  \\

There are a number of interesting directions to be pursued further:
\begin{itemize}

\item \emph{Loops}. While our examples have been at tree-level, many of our results are true at any loop order---for instance the optical theorem and consequent positivity, the Mandelstam operators (built from de Sitter generators) and the integral representations for correlators in terms of Mandelstam eigenstates. It would therefore be interesting to explicitly compute loop diagrams using this technology, to compare with the recent computations in the wavefunction and in-in formalisms \cite{Salcedo:2022aal, Lee:2023jby, Xianyu:2022jwk, Qin:2023nhv, Qin:2023bjk}.

\item \emph{IR divergences}. For fields with particular mass values, the time integrals may diverge as $\tau \to 0$. The divergences can be renormalised \cite{Bzowski:2015pba, Bzowski:2023nef} using the framework of holographic renormalisation \cite{Skenderis:2002wp}. They are known to give anomalous contributions to the Ward identities \cite{Bzowski:2015pba, Cespedes:2020xqq, Wang:2022eop} and their resummation leads to the semiclassical stochastic description of inflation \cite{Gorbenko:2019rza,Baumgart:2019clc,Green:2020txs,Cohen:2020php,Cespedes:2023aal}. Here we have focussed on scalar fields with masses and interactions for which there are no such divergences (this includes all principal series fields and also light fields with derivative interactions, such as those in the EFT of inflation \cite{Cheung:2007st}). 

\item \emph{Including spin}. We have focussed on scalar fields largely for simplicity. In subsequent work, we hope to include also spinning fields for the external and internal lines. There is no conceptual obstacle to doing this, since the analogous problem has been solved for the wavefunction: for instance the non-Gaussianity for gravitons is well-defined and has been recently bootstrapped \cite{Cabass:2021fnw,Bonifacio:2022vwa}.

\item \emph{Going on-shell.} We have focussed on the amputated correlator because it cleanly separates the dynamical part of the problem from the kinematic complications of the external asymptotic states. Ultimately, we would also like a way to take these amputated objects on-shell in order to construct $S$-matrices and observable in-in correlators. This produces various generalisations of the hypergeometric function, and there has been a lot of recent progress in understanding their structure \cite{Bzowski:2022rlz,Caloro:2023cep, Qin:2022fbv, Qin:2023ejc, Xianyu:2023ytd, Fan:2024iek} particularly in momentum space \cite{Bzowski:2013sza,Bzowski:2019kwd,Bzowski:2020kfw,Caloro:2022zuy}.

\item \emph{Canonical forms.} It was recently shown that the same differential equation technology from Minkowski amplitudes can be applied to cosmological correlators and the wavefunction~\cite{Arkani-Hamed:2023bsv, Arkani-Hamed:2023kig}. Since the $S$-matrix elements we have discussed here belong to the same class of integrals (we have seen they solve the same exchange equation but with different boundary conditions), the ``kinematic flow'' approach could also be used to efficiently determine their system of differential equations.
This would likely come at the cost of the de Sitter isometries, but could allow for an object similar to the $S$-matrix to be defined and computed for other power law cosmologies. 

\item \emph{Fewer symmetries}. We have focussed on de Sitter invariant interactions, although the standard $S$-matrix construction that we presented could equally be applied on other time-dependent backgrounds (albeit with more complicated mode functions). Following the same boostless bootstrap procedure developed for the wavefunction \cite{Green:2020ebl,Pajer:2020wnj,Pajer:2020wxk, DuasoPueyo:2020ywa,Pimentel:2022fsc, Du:2024hol} should lead to analogous results for the $S$-matrix elements (which, after all, are convenient combinations of wavefunction coefficients). 

\item \emph{Other variables}. As highlighted in the Introduction, there is now a variety of options for what kinematic variables to use and which object to compute when analysing cosmological correlators. 
We would argue that the $\{ \omega, \bfk \}$ variables used here have a transparent flat-space limit and a simple spectral representation, though other properties may be more manifest with a different set of kinematic variables. 
For instance, AdS correlators correspond to remarkably simple amplitudes in Mellin space 
\cite{Penedones:2010ue,Fitzpatrick:2011ia,Raju:2012zr,Penedones:2019tng} and much of this simplification can be directly imported to de Sitter 
\cite{Sleight:2019mgd,Sleight:2019hfp,Sleight:2020obc,Sleight:2021plv}. 
Previous spectral representations that integrate over the principal series with fields labelled by spatial position \cite{DiPietro:2021sjt, DiPietro:2023inn}, 
spatial momentum \cite{Sleight:2020obc},
or spatial angular momentum \cite{Marolf:2012kh} also enjoy a simple analytic structure in $\nu$ in which the poles have concrete physical meaning.
Ultimately, on Minkowski there are a number of desirable properties that coincide in the covariant momenta $p_\mu$---including a simple LSZ reduction, simple analytic structure and simple linearisation formula---but on de Sitter we have yet to find variables that can enjoy all of these properties simultaneously.

\item \emph{Subhorizon limit}. 
Many previous applications of $S$-matrix techniques to cosmology have considered subhorizon modes that do not feel the time-dependence of the background and therefore within a WKB approximation scattering via approximately energy conserving amplitudes. This has led to a variety of constraints on field theories of inflation \cite{Baumann:2011su,Baumann:2015nta,Grall:2020tqc,Grall:2021xxm,Green:2022slj,Freytsis:2022aho} and dark energy \cite{Melville:2019wyy,deRham:2021fpu,Davis:2021oce,Melville:2022ykg,deBoe:2024gpf}. We have shown only that the $S$-matrix considered here reduces to the Minkowski $S$-matrix when the Hubble rate $H\to 0$, but it would be interesting to instead perform a subhorizon expansion in $\omega, k \gg H$ to see how these $S$-matrix elements are related to the subhorizon amplitudes (and whether their positivity can impose any further constraints in the subhorizon regime, where analyticity is better understood).

\item \emph{Dispersion relation}. Finally, perhaps the most pressing question raised by this work is: what is the analogue of deforming the amplitude into the complex $s$ plane at fixed $t$? Since the Mandelstam operators no longer commute on de Sitter, it is not possible to simultaneously fix $t$ and $s$. It could be that another set of variables is better suited for exploiting analyticity, for instance complex angular momenta adapted to the full de Sitter group. 
We have referred to the combination appearing in the optical theorem as a ``discontinuity'', but we have not specified what this is a discontinuity with respect to. That remains an open question.  

\end{itemize}

\paragraph{Acknowledgements.}
It is a pleasure to thank the participants of the workshop ``Correlators in Cortona'' (Sep 2023) for many useful discussions, as well as Dionysios Anninos, Santiago Ag\"{u}\'{i} Salcedo, Tarek Anous, Nima Arkani-Hamed, Daniel Baumann, Paolo Benincasa, Carlos Duaso Pueyo, Austin Joyce, Hayden Lee, Juan Maldacena, Enrico Pajer, Sasha Polyakov, and Dong-Gang Wang for useful discussions during this work. 
SM is supported by a UKRI Stephen Hawking Fellowship (EP/T017481/1). 
GLP is supported by Scuola Normale, by a Rita-Levi Montalcini fellowship from the Italian Ministry of Universities and Research (MUR), and by INFN (IS GSS-Pi).

\appendix
\section{Mode function details}
\label{app:mode}

In this appendix we collect various mathematical identities that are useful when manipulating the mode functions for a massive scalar field on de Sitter.
In particular, we use
\begin{align}
 &f^- ( k \tau , \m ) = \frac{ i \sqrt{\pi} }{2}  e^{- \frac{\pi}{2} \m } H^{(1)}_{i \m} ( - k \tau )  \; , \;\;\;
 f^{+} ( k \tau , \m ) =  \frac{ \sqrt{\pi} }{2i} e^{+ \frac{\pi}{2} \m } H^{(2)}_{i \m} ( - k \tau ) \; ,
 \label{eqn:fmode}
\end{align}
for the Bunch-Davies mode functions in the time-domain, and 
\begin{align}
   F_{j_n} (z , \nu)  = \frac{1}{  \sqrt{2} } \;\; | \Gamma \left( j_n + i \nu  \right) |^2  \frac{ P^{ \frac{1}{2} - j_n }_{i \nu - \frac{1}{2}} (z) }{ \left( \sqrt{z^2 - 1} \right)^{j_n - \frac{1}{2}}  }
     \label{eqn:Fmode}
\end{align}
for the $n$-particle mode functions in the energy domain, where $j_n = \frac{d}{2} ( n- 1)$ in the main text but for this Appendix $j_n$ may take any positive real value.

\paragraph{Time-domain.}
The mode functions \eqref{eqn:fmode} satisfy the free equations of motion \eqref{eqn:E_def} with the Bunch-Davies vacuum condition $f^{\pm} \sim e^{\pm i k \tau}$ in the far past. 
Notice that $f^+$ should only be analytically continued to $\text{Im} ( - k \tau ) < 0$ and $f^-$ should only be continued to $\text{Im} (k \tau) > 0$ to ensure convergence of time integrals as $\tau \to -\infty$. 
They are related by complex conjugation,
\begin{align}
 \left[ H^{(1)}_{i \mu} (z )\right]^* &= 
 e^{\mu \pi} H_{i \mu}^{(2)} ( z^* ) \qquad \Rightarrow \qquad \left[ f^- ( k \tau, \mu ) \right]^* = f^+ ( k \tau, \mu ) \; . 
 \label{eqn:f_cc}
\end{align}
Since the Wronskian of two Hankel functions is
\begin{align}
H_{i \mu}^{(2)} ( - k \tau ) \overset{\leftrightarrow}{ \left( \tau \partial_\tau \right) } H_{i\mu}^{(1)} ( - k \tau ) &= 
\frac{4 i }{\pi}  \; .
\label{eqn:HankelWronskian}
\end{align}
we have normalised $f^{\pm}$ so that
\begin{align}
f^- ( k \tau , \mu ) \overset{\leftrightarrow}{ \left( i \tau \partial_\tau \right) } f^+ ( k \tau , \mu ) &= 1 \; 
\end{align}
which ensures a canonical commutation relation $[ \hat{a}_{\bfk} , \hat{a}_{\bfk'}^\dagger] = (2\pi)^3 \delta^3 ( \bfk - \bfk' )$.
Upon time reversal,
\begin{align}
 H_{-i \mu}^{(1)} ( e^{i \pi} z ) &= 
  - H_{i \mu}^{(2)} ( z)
&H_{-i \mu}^{(2)} ( e^{-i \pi} z  ) &=
  - H_{i \mu}^{(1)} ( z)
\end{align}
and the chosen phase of $f^{\pm}$ corresponds to a trivial CPT phase,
\begin{align}
 f^{\mp} ( z , \mu ) =  f^{\pm} ( e^{\mp i \pi} z , - \mu ) 
 \label{eqn:CPT_f}
\end{align}
in addition to the useful relation $ f^{\pm} (k \tau , -\m ) = f^{\pm} (k \tau, + \m)$. 
Note also that our normalisation corresponds to $f^{\pm} ( k \tau , \m ) \to e^{\pm i k \tau}/\sqrt{\mp 2 i k \tau}$ in the far past.

\paragraph{Energy-domain.}
The mode functions \eqref{eqn:Fmode} satisfy the free equations of motion $\hat{p}^2 + m^2 = 0$, with boundary conditions such that they are related to $f^{\pm}$ by a Fourier transform. 
Explicitly, our conventions for the Legendre functions are\footnote{
These corresponds to $P_\nu^\mu ( z) = \text{LegendreP}[\nu,\mu,3,z]$ and $Q_\nu^\mu (z) = \text{LegendreQ} [\nu,\mu,3,z]$ in Mathematica.  
}
\begin{align}
	\frac{ P^{ \frac{1}{2} - j}_{i \nu - \frac{1}{2}} (z) }{ \left( \sqrt{z^2-1} \right)^{j - \frac{1}{2}} } &= \frac{1}{\Gamma \left( j + \frac{1}{2}  \right) } \left( 1 + z \right)^{\frac{1}{2} - j}  {}_2 F_1 \left( \frac{1}{2} - i \nu , \frac{1}{2} + i \nu ;
	j + \frac{1}{2} ; \frac{1 - z}{2} \right) \\
\frac{ e^{- i \pi ( j - \frac{1}{2} ) }  Q^{j - \frac{1}{2}}_{\nu - \frac{1}{2}} (z) }{  \left( \sqrt{z^2 - 1} \right)^{j - \frac{1}{2}} }   &= \sqrt{ \frac{\pi}{2} } \frac{ \Gamma \left( \nu + j \right) }{ \Gamma ( \nu + 1 ) } \; \frac{1}{2^\nu z^{\nu+j}  } \;  {}_2 F_1\left(\frac{\nu + j + 1}{2} ,\frac{  \nu
   + j }{2} ; \nu +1;\frac{1}{z^2}\right) \; . 
   \nonumber 
\end{align}
For $z > 1$, the Legendre $P$ is a \emph{real} function and therefore 
\begin{align}
    F_j^* (z, \mu ) =  F_j ( z, \mu ) \; .
\end{align}

\paragraph{Connection formula.}
The Hankel mode functions can be expanded in terms of Bessel functions using
\begin{align}
\frac{e^{-\mu \pi/2}}{2} H_{i \mu}^{(1)} (z) &= \frac{ e^{+ \mu \pi/2} J_{+i\mu} (z) - e^{- \mu \pi/2} J_{-i \mu} (z) }{ e^{+\mu \pi} - e^{- \mu \pi} }  \; , \nonumber \\
\frac{e^{+\mu \pi/2}}{2} H_{i \mu}^{(2)} (z) &= \frac{ e^{+ \mu \pi/2} J_{-i\mu} (z) - e^{- \mu \pi/2} J_{+i \mu} (z) }{ e^{+\mu \pi} - e^{- \mu \pi} } \; . 
\label{eqn:Hankel_to_Bessel}
\end{align}
The inverse is simply $J_{i \nu} (z) = \frac{1}{2} \left( H^{(1)}_{i\nu} (z) + H^{(2)}_{i\nu} (z) \right)$. 
The Legendre mode functions can be similarly related:
\begin{align}
-  | \Gamma \left( j + i \nu \right) |^2 P^{\frac{1}{2} - j}_{i \nu - \frac{1}{2}} ( x) 
  = \frac{ e^{-i \pi j} }{ \sinh ( \pi \nu ) } \left(  
Q^{j - \frac{1}{2}}_{+ i \nu - \frac{1}{2}} (x)  - Q^{j - \frac{1}{2}}_{-i \nu - \frac{1}{2}} (x)   
   \right) \; . 
   \label{eqn:P_to_Q}
\end{align}
These relations are useful because it is the $J$ ($P$) function that has a simple scaling behaviour at $z \approx 0$ ($\approx 1$), while it is the $H$ and $Q$ functions that have a simple behaviour at large $z$. Explicitly:
\begin{align}
J_{i \nu} ( z ) &\sim  \left( \frac{z}{2} \right)^{+ i \nu} \frac{1}{ \Gamma ( 1 + i \nu ) }   \; , 
& \frac{e^{+ \nu \pi/2}}{2} H^{(2)}_{i \nu} ( z ) &\sim \sqrt{\frac{i}{ 2 \pi z}}  e^{- i z}   \nonumber \\ 
e^{-i \pi ( j - \frac{1}{2} )} Q_{i \nu - \frac{1}{2}}^{j - \frac{1}{2}} (z) &\sim   \sqrt{\frac{\pi}{2 z}} \left( \frac{1}{2 z} \right)^{+i \nu}  \frac{ \Gamma ( j + i \nu )}{ \Gamma ( 1 + i \nu )  }   \; , 
&
P_{i \nu - \frac{1}{2}}^{\frac{1}{2}-j } (z) &\sim  \frac{ 
\left( \sqrt{z^2 - 1} \right)^{j - \frac{1}{2}} }{ 2^{j-\frac{1}{2}} \, \Gamma \left( j + \frac{1}{2} \right) }
\label{eqn:mode_asy_exp}
\end{align}

\paragraph{Fourier transform.}
The mode functions in the two domains are related by a Fourier transform \cite[p394]{watson1922treatise},
\begin{align}
\left[  \int_{-\infty}^{0} \frac{d \tau}{ - \tau} e^{i \omega \tau} \right] ( - \tau )^j  \; e^{- \nu \pi/2} J_{i\nu} ( - k \tau ) \; 
 = 
\frac{1}{ ( i k)^j }  \sqrt{ \frac{2}{\pi} } \;\; \frac{ e^{- i \pi j} Q^{j - \frac{1}{2}}_{i \nu - \frac{1}{2}} (z) }{ \left( \sqrt{z^2 - 1} \right)^{j - \frac{1}{2}}  }
\end{align}
where $z = \omega/k$, $\text{Re} (j + i \nu) > 0$ and $\omega$ must have a small $\text{Im} \, \omega < 0$ for the integral to converge in the far past. 
Given the above connection formula \eqref{eqn:P_to_Q}, the corresponding Fourier transform of a Hankel function is\footnote{
Notice that $| \Gamma ( j + i \nu ) |^2 = \Gamma ( j + i \nu ) \Gamma ( j - i \nu )$ since we assume here that $\nu$ is real. 
}: 
\begin{align}
\left[ \int_{-\infty}^{0} \frac{d \tau}{ - \tau} e^{+ i \omega \tau} \right] ( - \tau )^j  \;  e^{+ \nu \pi/2}  H^{(2)}_{i\nu} ( - k \tau ) \; 
 = 
\frac{  1 }{ ( i k)^j }  \sqrt{ \frac{2}{\pi} } \;\; | \Gamma \left( j + i \nu  \right) |^2  \frac{ P^{ \frac{1}{2} - j }_{i \nu - \frac{1}{2}} (z) }{ \left( \sqrt{z^2 - 1} \right)^{j - \frac{1}{2}}  } \; . 
\label{eqn:Hankel_FT}
\end{align}
The integrals satisfy the same differential equation, and we can fix their overall normalisation by comparing their behaviour at low and large $z$, namely \eqref{eqn:mode_asy_exp}\footnote{
\eqref{eqn:Hankel_FT} can also be performed by rotating the contour into the complex plane and using the identity \cite[p398]{watson1922treatise}
\begin{align}
\int_0^\infty \frac{dy}{y} \,  y^j \; e^{- y \cosh \rho} K_{i \nu} ( y )  = \sqrt{\frac{\pi}{2}} \Gamma ( j - i \nu ) \Gamma ( j + i \nu ) \frac{ P^{\tfrac{1}{2} - j}_{i \nu - \tfrac{1}{2}} ( \cosh \rho ) }{ \sinh^{j - \tfrac{1}{2}} \rho}
\label{eqn:K_to_P}
\end{align}
for $\text{Re} ( j ) > | \text{Im} ( \nu ) |$ and $\text{Re} ( \cosh \rho ) > -1$, where $K_{i\nu} (y) = 
-\frac{i \pi}{2} e^{ \nu \pi / 2} H_{i\nu}^{(2)} ( y e^{- i \pi /2} ) $ in upper-half of complex $y$-plane.
}. 
In terms of the mode functions, \eqref{eqn:Hankel_FT} implies
\begin{align}
\left[ \int_{-\infty}^{0} \frac{d \tau}{ - \tau} e^{+ i \omega \tau} \right] ( - \tau )^j  \; i f^{+} ( k \tau , \nu )  \; 
 = 
\frac{  1 }{ ( i k)^j }  \, F_j \left(  \frac{\omega}{k}  , \mu \right)  \; . 
\end{align}
Using the conjugation property~\eqref{eqn:f_cc}, we also have
\begin{align}
\left[ \int_{-\infty}^{0} \frac{d \tau}{ - \tau} e^{+ i \omega \tau} \right] ( - \tau )^j  \; i f^{-} ( k \tau , \nu )  \; 
 = 
\frac{ -1 }{ ( - i k)^j }  \, F_j^* \left( - \frac{\omega}{k}  , \mu \right)  \; . 
\end{align}

We also have the inverse transformation,
\begin{align}
  e^{+ \nu \pi/2}  H_{i \nu}^{(2)} ( - k \tau ) = - i \int_{-\infty}^{+\infty} \frac{d \rho}{ \pi } e^{ i \nu \rho} \, e^{+ i k \tau \cosh \rho}  \\ 
  e^{- \nu \pi/2}  H_{i \nu}^{(1)} ( - k \tau ) = + i \int_{-\infty}^{+\infty} \frac{d \rho}{ \pi } e^{ i \nu \rho} \, e^{- i k \tau \cosh \rho}  
\end{align}
where we implicitly assume $- k \tau$ has a small positive (negative) imaginary part in $f^-$ ($f^+$) to ensure converge in the far past.

\paragraph{Analytic continuation in $z$.}
The Legendre $P$ function has a branch point at $z=-1$ and we take the branch cut along the negative real axis. 
This $Q$ functions has branch points at both $z = + 1$ and $-1$, and we take the branch cuts along the positive and negative real axis respectively. 
Approaching the cut from above or below corresponds to: 
\begin{align}
- | \Gamma \left( j + i \nu \right) |^2 \, P_{i \nu - \frac{1}{2}}^{j - \frac{1}{2}} ( e^{i \pi n} x) = \frac{ e^{i \pi (j- n/2 )} }{ \sinh ( \nu \pi) } \left( 
e^{n \nu \pi } Q^{j - \frac{1}{2}}_{+ i \nu - \frac{1}{2}} (x)  - e^{- n \nu \pi} Q^{j - \frac{1}{2}}_{-i \nu - \frac{1}{2}} (x) 
 \right)
 \label{eqn:P_cont}
\end{align}
since if we continue along an elliptical arc with $x > 1$ we get,
\begin{align}
 Q_{i\nu - \frac{1}{2}}^{j - \frac{1}{2}} ( e^{i \pi n} x ) = e^{- i \pi n/2}  e^{ + n \nu \pi} Q_{i\nu - \frac{1}{2}}^{j - \frac{1}{2}} (z) \; .
\end{align}
The analogous result for the Hankel function is
\begin{align}
\frac{e^{+\nu \pi/2}}{2}  H_{i \nu}^{(2)} ( e^{- i \pi n} z  ) =  \frac{ e^{-\nu \pi/2 (1 - 2n)} J_{+ i \nu} ( z ) - e^{+ \nu \pi/2 (1 - 2n)} J_{-i \nu} ( z )   }{ e^{\nu \pi} - e^{-\nu \pi} }
\end{align}
since,
\begin{align}
 J_{i \nu} (  e^{ i \pi n} z ) = e^{- n \nu \pi} J_{i \nu} ( z ) \; . 
\end{align}

\paragraph{LSZ reduction.}
Using \eqref{eqn:P_cont}, the branch cut discontinuity of the mode function can therefore be written as,
\begin{align}
 F_j \left(  e^{+ i \pi}  \cosh \rho , \mu \right) -  F_j \left(  e^{- i \pi}  \cosh \rho  , \mu \right) 
 = 
 \sqrt{2} \pi i \,  e^{ i \pi j } \, \frac{ P_{i \nu
   -\frac{1}{2}}^{j-\frac{1}{2}} ( \cosh \rho) }{ ( \sinh \rho )^{j - \frac{1}{2}} } \; . 
\end{align}
One interesting special case is when $j=1$,
\begin{align}
 F_1 \left(  e^{+ i \pi}  \cosh \rho , \mu \right) -  F_1 \left(  e^{- i \pi}  \cosh \rho  , \mu \right) 
 = 
- 2 i \sqrt{\pi }  \,  \frac{ \cos ( \nu \rho ) }{ \sinh \rho }  \; .
 \label{eqn:one_particle_Disc_F}
\end{align}
This gives an alternative way to derive the LSZ formula~\eqref{eqn:intro_LSZ_rho}, since
\begin{align}
\left[ \int_{-\infty}^0 \frac{d \tau}{-\tau} f^+ ( k \tau, \mu ) 
\right] \G_n \left( \col{\tau}{\bfk} \right) 
&= 
 \int_{-\infty}^{+\infty} \frac{d \omega}{2\pi}  \left[ \int_{-\infty}^0 d \tau  \,  f^+ ( k \tau, \mu )  \, e^{- i \omega \tau}
\right]    \G_n \left( \col{\omega}{\bfk} \right) \nonumber \\ 
&=  \int_{-\infty}^{+\infty} \frac{d \omega}{2\pi i k}  \; F_1 \left( - \frac{\omega}{k} , \mu \right)    \; \G_n \left( \col{\omega}{\bfk} \right) \nonumber \\
&=  \int_k^{\infty} \frac{d\omega}{2 \pi i k} \, \text{Disc} \left[ F_1 \left( - \frac{\omega}{k} , \mu \right) \right] \G_n \left( \col{\omega}{\bfk} \right)
\end{align}
where we have used analyticity of $\G_n ( \omega, \bfk)$ in the lower-half of the complex $\omega$ plane to close the integration contour around the branch cut discontinuity of $F_1$.
The identity \eqref{eqn:one_particle_Disc_F} for this discontinuity then immediately implies \eqref{eqn:intro_LSZ_rho} upon changing integration variables to $\omega = k \cosh \rho$. 
For an ingoing particle, repeating these steps produces a $\tfrac{1}{-i k} F_1 ( + \tfrac{\omega}{k} , \mu )$ and we instead close the contour around the branch cut along $\omega < -k$: this produces \eqref{eqn:intro_LSZ_rho} with $\omega = - k \cosh \rho$.

\paragraph{Analytic continuation in $\nu$.}
At fixed $z$, the Bessel, Hankel and Legendre functions are all analytic functions of $\nu$. Integrals over $\nu$ can therefore usually be done by closing the contour in the complex plane. For this, the following asymptotic behaviours are useful. 
For the Bessel/Hankel functions,
\begin{align}
e^{- \frac{\pi}{2} \nu} J_{+i \nu} ( z) &\sim \frac{ e^{+i \nu - i \nu \log \left( \frac{2\nu}{z} \right) } }{\sqrt{2 \pi \nu}}  
& e^{+ \frac{\pi}{2} \nu} H_{i \nu}^{(2)} (z) &\sim \frac{2 i \, e^{- i \nu + i \nu \log \left( \frac{2 \nu}{z}  \right) } }{ \sqrt{2 \pi \nu} }
\end{align}
for $|\nu|$ much greater than $1$ or $|z|$ and with a small negative imaginary part \cite{10.1093imamat26.2.133, GUTIERREZTOVAR2007359}.
Similarly, for $\nu \to \infty$ at fixed $j$ and $\rho \geq 0$, 
\begin{align}
 P_{i \nu - \frac{1}{2}}^{j - \frac{1}{2}} ( \cosh \rho )  &\sim \left( \frac{\rho}{ \sinh \rho }  \right)^{1/2} \; \nu^{\frac{1}{2} - j}  e^{i \frac{\pi}{2} ( j - \frac{1}{2} ) } J_{j - \frac{1}{2}} ( \nu \rho )  \nonumber \\ 
e^{- i \pi ( j - \frac{1}{2} ) } Q_{i \nu - \frac{1}{2}}^{j - \frac{1}{2}} ( \cosh \rho )  &\sim  \frac{\pi}{2i}   \left( \frac{\rho}{ \sinh \rho }  \right)^{1/2} \;  \nu^{j - \frac{1}{2} }  e^{ - i \frac{\pi}{2} ( j - \frac{1}{2} )}   H^{(2)}_{j - \frac{1}{2}} (  \nu \rho ) 
\label{eqn:P_cont2}
\end{align}
which can be simplified using \eqref{eqn:mode_asy_exp}, namely
\begin{align}
e^{+ i \frac{\pi}{2} ( j -\frac{1}{2} ) }  J_{j - \frac{1}{2}} ( \nu \rho ) &\sim  \frac{1}{ \sqrt{ 2 \pi i \nu \rho } }   \left( e^{ i \nu \rho } + e^{ - i  \nu \rho} e^{  i \pi j }   \right) ,  
&e^{- i \frac{\pi}{2} ( j -\frac{1}{2} ) } H_{j - \frac{1}{2}}^{(2)} ( \nu \rho ) &\sim  \sqrt{\frac{2 i}{\pi \nu \rho} } \,   e^{- i \nu \rho} .  
\end{align}
In practice, this often means that $Q_{i \nu - \frac{1}{2}}^{j - \frac{1}{2}}$ ($Q_{-i \nu - \frac{1}{2}}^{j - \frac{1}{2}}$) can be closed in the lower-half (upper-half) of the complex $\nu$ plane.

\paragraph{Crossing.}
An interesting way to implement crossing is to combine the analytic continuation in $z$ \eqref{eqn:P_cont} and the analytic continuation in $\nu$ \eqref{eqn:P_cont2}.
If we define the combination\footnote{
As usual, $2 \nu /( \nu^2 - \mu^2 \mp i \epsilon)$ contains poles at $\nu = +\mu \pm  i \epsilon$ and $-\mu \mp i \epsilon$ with residue $+1$, where $\epsilon \to 0^+$. 
}
\begin{align}
 \frac{1}{ ( \nu^2 - \mu^2 )_{i \epsilon, n} } = \frac{1}{ 2 \sinh ( \pi \mu )  } \left( \frac{ e^{ n \mu \pi } }{\nu^2 - \mu^2 + i \epsilon} - \frac{ e^{- n \mu \pi} }{\nu^2 - \mu^2 - i \epsilon} \right) \; , 
 \label{eqn:denom_def}
\end{align}
this implements a rotation of $n$ half-turns counter-clockwise in the complex $z$-plane when integrated over $\nu$, 
\begin{align}
 \int_0^{\infty} \frac{d \nu}{i \pi } \, 2 \nu \sinh ( \pi \nu )  \frac{  | \Gamma ( j + i \nu ) |^2 \, P^{j - \frac{1}{2}}_{i \nu - \frac{1}{2}} ( x ) }{ ( \nu^2 - \mu^2 )_{i \epsilon, n} } = e^{- i \pi n/2}   | \Gamma ( j + i \mu ) |^2 P^{j-\frac{1}{2}}_{i \mu - \frac{1}{2}} ( e^{i \pi n} x)   \; 
\end{align}
since we can replace $P$ with $Q$ using \eqref{eqn:P_to_Q} and then close the contour in complex $\nu$ plane, which picks up a single residue from each term in \eqref{eqn:denom_def}. 
Transforming a mode function from outgoing to ingoing uses $n= +1$,
\begin{align}
   \int_0^{\infty} d \nu \; \mathcal{N}_\nu \; \frac{ F_{j_2} ( z , \nu ) }{ ( \nu^2 - \mu^2 )_{i \epsilon, +1} }  =  - F_{j_2} ( e^{i \pi} z , \nu ) \; .
\end{align}
The inverse transformation from ingoing to outgoing would use $n=-1$. The same is true in the time-domain, where $1/(\nu^2 - \mu^2 )_{i \epsilon, n}$ implements a rotation of $n$ half-turns clockwise in the complex $- k \tau$ plane:
\begin{align}
   \int_0^{\infty} d \nu \; \mathcal{N}_\nu \; \frac{ f^+ ( k \tau , \nu ) }{ ( \nu^2 - \mu^2 )_{i \epsilon, +1} }  =  i  f^+ ( e^{- i \pi} k \tau , \nu )  =  i f^- ( k \tau, \nu ) \;  . 
\end{align}

\paragraph{Spectral representation.}
The key identity on which our spectral representations are based is:
\begin{align}
  \int_0^{\infty} \frac{d \nu}{\pi} \;  \nu \sinh ( \pi \nu )   \;  \frac{ | \Gamma ( j - i \nu ) |^2 P^{\frac{1}{2} - j}_{i \nu - \frac{1}{2}} ( z ) }{ \left( \sqrt{z^2-1} \right)^{j- \frac{1}{2}} }   
  \;
  \frac{ | \Gamma ( j' - i \nu ) |^2 P^{\frac{1}{2} - j'}_{i \nu - \frac{1}{2}} ( z' ) }{ \left( \sqrt{z'^2-1} \right)^{j'- \frac{1}{2}} } 
=
\frac{ \Gamma \left( j + j' \right) }{ \left( z + z' \right)^{j + j'}   }
\end{align}
which holds for $j, j' > 0$. 
In terms of the mode functions \eqref{eqn:Fmode},
\begin{align}
  \int_0^{\infty} d \nu  \; \mathcal{N}_\nu \; F_{j_1} (z_1)  \; F_{j_2} (z_2) 
=
- \frac{ \Gamma \left( j_1 + j_2 \right) }{ \left( z_1 + z_2 \right)^{j_1 + j_2}   }
\end{align}
where we have introduced the shorthand
\begin{align}
 \mathcal{N}_\nu = \frac{2}{\pi} \, \nu \sinh ( \pi \nu ) \; .
\end{align}

\paragraph{Useful limits.}
Whenever $j$ takes an integer or half-integer value $>1/2$, as we approach the branch point at $z = -1$ the Legendre function diverges as
\begin{align}
  \frac{ P_{i \nu - \frac{1}{2}}^{\frac{1}{2} - j} (z)  }{ \left( \sqrt{z^2 - 1 } \right)^{j - \frac{1}{2}} }  \sim \frac{ \Gamma ( j - \frac{1}{2} ) }{ | \Gamma ( j + i \nu ) |^2 } \; \frac{1 }{ \left( z + 1 \right)^{j - \frac{1}{2}} }
\end{align}
or in terms of the mode function \eqref{eqn:Fmode} and the $u$ variable of section \ref{sec:bootstrap},
\begin{align}
      \lim_{u \to 1^-} F_j \left( - \frac{1}{u} \right) = \frac{ \Gamma \left( j - \tfrac{1}{2} \right)  }{ i \sqrt{2}} \left( 1 - u \right)^{\frac{1}{2} - j} \; .
      \label{eqn:F_u1_limit}
\end{align}
At the value $j = 1/2$, the divergence becomes logarithmic:
\begin{align}
          \lim_{u \to 1^-} F_j \left( - \frac{1}{u} \right) =  \frac{i}{\sqrt{2}} \log \left( 1 - u \right) \; . 
\end{align}

Conversely, near the origin the Legendre function approaches a constant value
\begin{align}
   \lim_{z \to 0}   \frac{ P_{i \nu - \frac{1}{2}}^{\frac{1}{2} - j} (z)  }{ \left( \sqrt{z^2 - 1 } \right)^{j - \frac{1}{2}} } 
    =  \frac{ 2^{-j} \sqrt{2 \pi} }{  \left| \Gamma \left( \frac{1 + j - i \nu}{2}  \right) \right|^2  }
\end{align}
and so thanks to an identity for the $\Gamma$ function we can write
\begin{align}
    F_j ( 0 ) = \frac{ 2^{j-2} }{i \sqrt{\pi} }  \; \left|  \Gamma \left(  \frac{j - i \nu}{2} \right) \right|^2  \; . 
\end{align}

\bibliographystyle{utphys}
\bibliography{dS_S_Matrix.bib}
\end{document}